%% file: EXO-19-019_temp.tex
\pdfoutput=1
\documentclass[11pt,twoside,a4paper,cmspaper,final,collab]{cms-tdr}
\def\svnVersion{851f265-D}\def\svnDate{2021/08/03}\def\cmsCernNoTag{CERN-EP-2021-026}\def\cmsCernDate{\today}\def\cmsMessage{"Published in the Journal of High Energy Physics as \href{http://dx.doi.org/10.1007/JHEP07(2021)208}{\doi{10.1007/JHEP07(2021)208}.}"}
\begin{document}\cmsNoteHeader{EXO-19-019}

\newlength\cmsTabSkip\setlength{\cmsTabSkip}{1ex}

\newcommand{\Rmumuee}{\ensuremath{R_{\MM/\EE}}\xspace}
\newcommand{\costheta}{\ensuremath{\cos\theta^\ast}\xspace}
\newcommand{\cPZg}{\ensuremath{\PZ/\gamma^{*}}\xspace}
\newcommand{\GKK}{\ensuremath{\mathrm{G}_\mathrm{KK}}\xspace}

\newcommand{\ZPSSM}{\ensuremath{\PZpr_\mathrm{SSM}}\xspace}
\newcommand{\ZPPSI}{\ensuremath{\PZpr_\psi}\xspace}
\newcommand{\ZPQ}{\ensuremath{\PZpr_\mathrm{Q}}\xspace}

\newcommand{\massADD}[1]{\ensuremath{M_{\mathrm{#1}}}\xspace}
\newcommand{\lambdaCIADD}[1]{\ensuremath{\Lambda_{\mathrm{#1}}}\xspace}
\newcommand{\etaCIADD}[1]{\ensuremath{\eta_{\mathrm{#1}}}\xspace}
\newcommand{\mll}{\ensuremath{m_{\Pell\Pell}}\xspace}
\newcommand{\mee}{\ensuremath{m_{\Pe\Pe}}\xspace}
\newcommand{\muu}{\ensuremath{m_{\PGm\PGm}}\xspace}

\hyphenation{sear-ches}

\cmsNoteHeader{EXO-19-019}
\title{Search for resonant and nonresonant new phenomena in high-mass dilepton final states at \texorpdfstring{$\sqrt{s} = 13\TeV$}{sqrt(s) = 13 TeV}}

\date{\today}

\abstract{
A search is presented for physics beyond the standard model (SM) using electron or muon pairs with high invariant mass. A data set of proton-proton collisions collected by the CMS experiment at the LHC at $\sqrt{s}=13\TeV$ from 2016 to 2018 corresponding to a total integrated luminosity of up to 140\fbinv is analyzed. No significant deviation is observed with respect to the SM background expectations. Upper limits are presented on the ratio of the product of the production cross section and the branching fraction to dileptons of a new narrow resonance to that of the \PZ boson. These provide the most stringent lower limits to date on the masses for various spin-1 particles, spin-2 gravitons in the Randall--Sundrum model, as well as spin-1 mediators between the SM and dark matter particles. Lower limits on the ultraviolet cutoff parameter are set both for four-fermion contact interactions and for the Arkani-Hamed, Dimopoulos, and Dvali model with large extra dimensions. Lepton flavor universality is tested at the \TeV scale for the first time by comparing the dimuon and dielectron mass spectra. No significant deviation from the SM expectation of unity is observed.
}

\hypersetup{
pdfauthor={CMS Collaboration},
pdftitle={Search for resonant and nonresonant new phenomena in high-mass dilepton final states at sqrt(s)= 13 TeV},
pdfsubject={CMS},
pdfkeywords={CMS,  dileptons, resonance}}

\maketitle 

\section{Introduction}
The presence of new phenomena, both resonant and nonresonant, in the high-mass dilepton final state is predicted by various theoretical models aiming to extend the standard model (SM) of particle physics. In this paper, the production of new spin-1 or spin-2 resonances, as well as the nonresonant production of high-mass lepton pairs, is considered. 

The existence of new neutral gauge bosons is a possible signature of grand unified theories, such as superstring and left-right-symmetric (LRS) models, that include a unification of the three forces at a high energy scale~\cite{Leike:1998wr,Zp_SSM_1}.  

One persistent puzzle in modern particle physics is the large difference in the energy scale of electroweak symmetry breaking and the energy scale of gravitation.  This could be explained in theories including spatial extra dimensions, where the gravitational force can propagate into additional dimensions. In models by Arkani-Hamed, Dimopoulos, and Dvali (ADD)~\cite{arkani98:hlz,arkani99:hlz}, the SM particles are confined to the traditional four dimensions of space and time, while in models proposed by Randall and Sundrum (RS)~\cite{Randall:1999ee} the SM particles could also propagate into the additional dimensions. Possible signatures of these theories at LHC energies are graviton excitations, either as distinct spin-2 high-mass resonances in the RS model, or as a series of nearly mass-degenerate excitations that result in an overall nonresonant excess of events at high mass in the ADD model. 

Based on many astrophysical and cosmological observations, it is assumed that dark matter (DM) accounts for the majority of matter in the universe~\cite{Ade:2015xua}. Models have been proposed in which the DM consists of particles that can interact with those of the SM via high-mass, weakly coupled mediator particles~\cite{Alves:2011wf}. These mediators could then be observed via their decay into SM particles, including the dilepton final state.  

It has long been speculated that the presence of three generations of quarks and leptons is a sign that these particles are not fundamental but rather composed of constituent particles commonly called ``preons''~\cite{Eichten:1984eu}. At energies observable at collider experiments, the preons would be confined into bound states by a new interaction analogous to quantum chromodynamics (QCD). This new interaction is characterized by an energy scale $\Lambda$, at which its effects would be directly observable. At center-of-mass energies far below $\Lambda$, the presence of the preon bound states would manifest itself as a flavor-diagonal ``contact interaction'' (CI)~\cite{Eichten:1983hw}, resulting in a nonresonant excess of events at high mass. 

Hints for lepton flavor universality violation in several measurements recently reported by the LHCb Collaboration~\cite{Aaij:2017vbb,Aaij:2019wad,Aaij:2021vac}, together with other flavor anomalies in \PB-meson decays summarized in Ref.~\cite{Bifani:2018zmi}, have sparked interest in models for physics beyond the SM that could explain these effects. These include models with heavy neutral gauge bosons~\cite{Zp_SSM_1} or leptoquarks~\cite{Becirevic:2016oho}. Some of these models would result in a significant deviation from unity of the ratio of the dimuon to dielectron differential cross section as a function of dilepton mass \mll,
\begin{linenomath} 
\begin{equation}
\Rmumuee=\frac{\rd \sigma(\PQq\PAQq \to \MM)/\rd m_{\Pell\Pell}}{\rd \sigma(\PQq\PAQq \to \EE)/\rd m_{\Pell\Pell}},
\end{equation}
\end{linenomath}
at high \mll~\cite{Greljo:2017vvb}. 

Searches for high-mass \PZpr gauge bosons in dilepton final states have a long history at the LHC, with the CMS Collaboration having reported results using proton-proton (pp) collision data from the LHC Run 1 (2010--2012) at $\sqrt{s} = 7\TeV$~\cite{Chatrchyan:2011wq,Chatrchyan:2012it} and 8\TeV~\cite{Chatrchyan:2012oaa,Khachatryan:2014fba}, and more recently from the beginning of the LHC Run 2 (2015--2018) at 13\TeV~\cite{Khachatryan:2016zqb,Sirunyan:2018exx} using early data corresponding to an integrated luminosity of 36\fbinv collected in 2016. Similar searches have also been performed by the ATLAS Collaboration with data collected at 7\TeV~\cite{Aad:2011xp,Aad:2012hf}, 8\TeV~\cite{Aad:2014cka}, and 13\TeV~\cite{Aaboud:2017buh}. Most recently, the ATLAS Collaboration reported results obtained using data recorded at $\sqrt{s} = 13\TeV$, corresponding to an integrated luminosity of 139\fbinv~\cite{Aad:2019fac}.

Results of searches for spin-2 dilepton resonances, as well as constraints on DM models, have been reported by the ATLAS and CMS Collaborations~\cite{Chatrchyan:2012oaa,Sirunyan:2018exx,Aad:2014cka,Aaboud:2019yqu}.

For the case of nonresonant signatures, the CMS Collaboration has reported results at 8~\cite{Khachatryan:2014fba} and 13\TeV~\cite{Sirunyan:2018ipj}.
Additional constraints on these models have been reported by the CMS Collaboration from diphoton and dijet final states~\cite{Sirunyan:2018wnk,Sirunyan:2018wcm} and by the ATLAS Collaboration from dijet final states~\cite{Aaboud:2017yvp}.
The ATLAS Collaboration has presented similar results for these models in the dilepton final state, the most recent using data at 8\TeV~\cite{Aad:2014wca} for the ADD model and at 13\TeV~\cite{Aad:2020otl} for the CI model.

This paper significantly extends the previous CMS results by using the full data set recorded at $\sqrt{s} = 13\TeV$ during the years 2016--2018, corresponding to 137 (140) \fbinv in the dielectron (dimuon) channel~\cite{CMS-PAS-LUM-17-001,CMS-PAS-LUM-17-004,CMS-PAS-LUM-18-002}. The additional data in the dimuon channel are recorded with the detector in conditions unsuitable for electron reconstruction.  Deviations from the SM predictions are searched for in the dilepton invariant mass spectrum for both dielectron and dimuon events. The contribution from dileptonic decays of tau lepton pairs to the signal is small and neglected. The shapes of contributions from SM processes are estimated from simulation, except for backgrounds containing leptons produced inside jets or jets misidentified as leptons, which are estimated from control regions in data. The simulated events describing the dominant Drell--Yan (DY) background are corrected to the highest order calculations available. The resulting  background shape is normalized to the observed data yields in a mass window of 60--120\GeV around the \PZ boson peak, separately for the dielectron and dimuon channels. 

Two analysis strategies are followed, targeting resonant and nonresonant signatures. For resonant signatures, the search is performed in a mass window around the assumed resonance mass, whose size depends on the assumed intrinsic decay width of the resonance and the mass-dependent detector resolution.  A range of masses and widths is scanned to provide results covering a wide selection of signal models. 
Unbinned maximum likelihood fits are performed inside the mass windows, allowing the background normalization to be determined from data. The background parametrizations are obtained from fits to the \mll distribution of the background estimates.

Upper limits are set on the ratio of the product of the production cross section and the branching fraction of a new narrow dilepton resonance to that of the SM \PZ boson. 
Thus, many experimental and theoretical uncertainties common to both measurements cancel out or are reduced, leaving only uncertainties in the ratio that vary with the dilepton mass to be considered.  

In the case of nonresonant signatures,  the event sample is divided into several bins in invariant mass and, improving upon previous CMS analyses, the scattering angle \costheta in the Collins--Soper frame~\cite{Collins:1977iv}. The result is then obtained from a combination of the individual counting experiments within these bins. While most of the sensitivity to new physics stems from the highest mass bins where the signal-to-background ratio is most favorable, the bins at lower masses still contain valuable information, for example on possible interference between the signal and SM backgrounds. In this approach, the signal is not normalized to the \PZ boson cross section, and the background estimate cannot be obtained from the data. Therefore, this part of the analysis is more affected by uncertainties in the signal and background modeling. 

In addition to these two search strategies, lepton flavor universality is tested for the first time in these final states by measuring the dimuon to dielectron ratio \Rmumuee and comparing it to the SM expectation of unity, including corrections for detector effects, lepton acceptances, and lepton efficiencies.

This paper is structured as follows. The CMS detector is briefly described in Section~\ref{sec:detector}. A description of the signal models is given in Section~\ref{sec:models}, followed by a description of simulated event samples used in the analysis in Section~\ref{sec:samples}. The event reconstruction and selection is given in Section~\ref{sec:objsel} and the estimation of SM backgrounds is described in Section~\ref{sec:backgrounds}. Systematic uncertainties are described in Section~\ref{sec:systematics}, followed by the results and their statistical interpretation in Sections~\ref{sec:results} and~\ref{sec:interpretation}, respectively. The paper is summarized in Section~\ref{sec:summary}. Tabulated results are provided in HEPDATA~\cite{hepdata}.

\section{The CMS detector}
\label{sec:detector}

The central feature of the CMS detector is a superconducting solenoid providing an axial magnetic field of 3.8\unit{T} and enclosing an inner tracker, an electromagnetic calorimeter (ECAL), and a hadron calorimeter (HCAL).
The inner tracker is composed of a silicon pixel detector and a silicon strip tracker, and measures charged particle trajectories in the pseudorapidity range $\abs{\eta}<2.5$. Before the data taking period in 2017, an upgraded pixel detector was installed, adding an additional  barrel layer closer to the interaction point and additional disks in the two forward parts of the detector~\cite{Dominguez:1481838}.
The ECAL and HCAL, each composed of a barrel and two endcap sections, extend over the range $\abs{\eta} < 3.0$.
The finely segmented ECAL consists of nearly 76\,000 lead tungstate crystals, while the HCAL is constructed from alternating layers of brass and scintillator.
Forward hadron calorimeters encompass $3.0<\abs{\eta}<5.0$.
The muon detection system covers $\abs{\eta}<2.4$ with up to four layers of gas-ionization detectors installed outside the solenoid and sandwiched between the layers of the steel flux-return yoke.
A more detailed description of the CMS detector, together with a definition of the coordinate system used and the relevant kinematic variables, can be found in Ref.~\cite{Chatrchyan:2008zzk}.

Events of interest are selected using a two-tiered trigger system~\cite{Khachatryan:2016bia}. The first level (L1), composed of custom hardware processors, uses information from the calorimeters and muon detectors to select events at a rate of around 100\unit{kHz} within a fixed time interval of less than 4\mus. The second level, known as the high-level trigger (HLT), consists of a farm of processors running a version of the full event reconstruction software optimized for fast processing, and reduces the event rate to around 1\unit{kHz} before data storage. 

\section{Signal models}
\label{sec:models}
A large number of models predict new phenomena resulting in high-mass dilepton signatures. The available models may be classified depending on whether they predict resonant or non-resonant production.

\subsection{Models with resonant signatures}
We consider three types of models resulting in resonances at high dilepton masses: \PZpr particles in a class of $U^\prime(1)$ gauge group models, spin-2 gravitons in the RS model of extra dimensions, and spin-1 dark matter mediators. The most prevalent are the theories that introduce new  $U^\prime(1)$ gauge groups. The symmetry properties of these groups, or a combination of them, are broken at some energy scale so that the SM group structure emerges as the low-energy limit of the new theory. If this breaking happens at the TeV scale, new gauge bosons \PZpr with masses accessible at the LHC could exist. The properties of the new \PZpr bosons depend on the mixing of  the $U^\prime(1)$ generators, which can be described by a continuously varying angle within a class of models.  Commonly considered classes are the generalized sequential model (GSM)~\cite{Accomando:2010fz}, containing the sequential SM boson \ZPSSM that has SM-like couplings to SM fermions~\cite{Altar:1989}; LRS extensions of the SM based on the $SU(2)_\text{L} \otimes SU(2)_\text{R} \otimes U(1)_\text{B$-$L}$ gauge group, where B$-$L refers to the difference between the baryon and lepton numbers~\cite{Accomando:2010fz}, which predict high-mass neutral bosons; and grand unified theories based on the $E_6$ gauge group, containing the \ZPPSI boson~\cite{Leike:1998wr,Zp_PSI_3}. 

In the narrow width approximation~\cite{Accomando:2013sfa}, the \PZpr production cross section can be expressed as $c_\PQu w_\PQu + c_\PQd w_\PQd$, where  $c_\PQu$ ($c_\PQd$) are its couplings to up-type (down-type) quarks and $w_\PQu$ ($w_\PQd$) are the parton distribution functions (PDFs) for these quarks~\cite{Carena:2004xs,Accomando:2010fz}.  As the values of the model-dependent couplings depend on the aforementioned mixing of  the $U^\prime(1)$ generators, each class of models can be represented by a unique contour in the ($c_\PQd, c_\PQu$) plane as a function of the mixing angle.  The properties of a variety of benchmark models in the three model classes mentioned above are shown in Table~\ref{tab:BenchmarkModels}. Ref.~\cite{Accomando:2010fz} shows that the finite width of the resonance, which can reach up to 12\% for the \ZPQ, has negligible effects on the translation of experimental limits obtained in the narrow width approximation into the  ($c_\PQd, c_\PQu$) plane. Similarly, the effect of interference between the signal and SM backgrounds can be neglected as long as the signal cross section is considered in a sufficiently narrow window around the \PZpr mass~\cite{Accomando:2013sfa}.

\begin{table*}[htb!]
\centering
\topcaption{
\label{tab:BenchmarkModels}
Various benchmark models in the GSM~\cite{Accomando:2010fz}, LRS~\cite{Accomando:2010fz}, and $E_\text{6}$~\cite{Leike:1998wr,Zp_PSI_3} model classes, with their corresponding mixing angles, their branching fraction ($\mathcal{B}$) to dileptons, the $c_{\PQu}$, $c_{\PQd}$ parameters and their ratio, and the width-to-mass ratio of the associated \PZpr boson.}
\begin{tabular}{lcccccc}
\hline
$U^\prime(1)$ model & Mixing angle  & $\mathcal{B}(\Pell^{+}\Pell^{-}$)     & $c_{\PQu}$               &  $c_{\PQd}$              &  $c_{\PQu}$/$c_{\PQd}$     &  $\Gamma_\text{\PZpr}$/M$_\text{\PZpr}$ \\ \hline
GSM                 &               &                      &		               &			         &		                &	 \\
U(1)$_\text{SM}$    &  $-0.072\pi$  &     0.031            & $2.43\times10^{-3}$ & $3.13\times10^{-3}$ &  0.78                & 0.0297 			\\
U(1)$_\text{T3L}$   &    0          &     0.042            & $6.02\times10^{-3}$ & $6.02\times10^{-3}$ &  1.00                & 0.0450 			\\
U(1)$_\text{Q}$     &     0.5$\pi$  &     0.125            & $6.42\times10^{-2}$ & $1.60\times10^{-2}$ &  4.01                & 0.1225 			\\[2ex]
LRS         	        &               &               	     &			           &			         &			            & 			\\
U(1)$_\text{R}$     &     0         &     0.048            & $4.21\times10^{-3}$ & $4.21\times10^{-3}$ &  1.00                & 0.0247 			\\
U(1)$_\text{B-L}$   &     0.5$\pi$  &     0.154            & $3.02\times10^{-3}$ & $3.02\times10^{-3}$ &  1.00                & 0.0150 			\\
U(1)$_\text{LR}$    &  $-0.128\pi$  &     0.025            & $1.39\times10^{-3}$ & $2.44\times10^{-3}$ &  0.57                & 0.0207 			\\
U(1)$_\text{Y}$     &    0.25$\pi$  &     0.125            & $1.04\times10^{-2}$ & $3.07\times10^{-3}$ &  3.39                & 0.0235 			\\[2ex]
$E_\text{6}$        &               &                  	 &			&			&						&	\\
U(1)$_{\chi}$       &     0         &     0.061            & $6.46\times10^{-4}$ & $3.23\times10^{-3}$ &  0.20                & 0.0117 			\\
U(1)$_{\psi}$       &     $0.5\pi$  &     0.044	         & $7.90\times10^{-4}$ & $7.90\times10^{-4}$ &  1.00                & 0.0053 			\\
U(1)$_{\eta}$       &   $-0.29\pi$  &     0.037            & $1.05\times10^{-3}$ & $6.59\times10^{-4}$ &  1.59                & 0.0064 		\\
U(1)$_\text{S}$     &   $0.129\pi$  &     0.066            & $1.18\times10^{-4}$ & $3.79\times10^{-3}$ &  0.31                & 0.0117 			\\
U(1)$_\text{N}$     &    $0.42\pi$  &     0.056            & $5.94\times10^{-4}$ & $1.48\times10^{-3}$ &  0.40                & 0.0064 		\\
\hline
\end{tabular}
\end{table*}

The RS model of extra dimensions~\cite{Randall:1999vf, Randall:1999ee}  introduces an additional warped spatial dimension where gravity can propagate through all of the five-dimensional bulk. This leads to the prediction of spin-2 Kaluza--Klein (KK) excitations of the graviton (\GKK) that could be resonantly produced at the LHC and observed in their decay into lepton pairs. These new resonances are characterized by their masses and the coupling $k/\overline{M}_\mathrm{Pl}$, where $k$ is the warp factor of the five-dimensional anti-de Sitter space and $\overline{M}_\mathrm{Pl}$ is the reduced Planck mass. The ratios of the intrinsic widths of the first excitation of the graviton to its mass for the coupling parameters $k/\overline{M}_\mathrm{Pl}$ of 0.01, 0.05, and 0.10 are 0.01\%, 0.36\%, and 1.42\%, respectively.

To ensure a consistent search strategy and comparability between the results of different searches,  the LHC Dark Matter Working Group has defined a variety of simplified models to be used as benchmarks for DM searches at the LHC~\cite{Albert:2017onk,Backovic:2015soa}. In this paper, we consider a model that assumes the existence of a single DM particle that interacts with the SM particles through a spin-1 mediator, which can be either a vector or axial-vector boson.  The model is fully described by the following parameters: the DM mass $m_\text{DM}$, the mediator mass $m_\text{med}$, the coupling $g_\text{DM}$ between the mediator and the DM particle, and the universal couplings $g_{\Pell}$ and $g_{\PQq}$ between the mediator and the SM charged leptons and quarks, respectively. Two signal scenarios have been defined that feature nonzero values of $g_{\Pell}$, so that the mediator could be directly observed in its decay into the dilepton final state. The first choice represents a relative strength of the couplings to quarks and charged leptons typical of some models with a pure vector mediator, while the second choice is a representative case found in the simplest complete models with axial-vector \PZpr bosons~\cite{Albert:2017onk}:
\begin{itemize}
\item   Vector mediator with small couplings to leptons: $g_\text{DM}=1.0$, $g_{\PQq}=0.1$, $g_{\Pell}=0.01$;
\item   Axial-vector mediator with equal couplings to quark and leptons:  $g_\text{DM}=1.0$,\\ $g_{\PQq}=g_{\Pell}=0.1$.
\end{itemize}
Interference between the mediator exchange and the SM processes resulting in the dilepton final state has been studied and found to be negligible for experimental searches~\cite{Albert:2017onk}.

\subsection{Models with nonresonant signatures}
We consider two types of nonresonant excess at high \mll: a series of virtual spin-2 graviton excitations in the ADD model of large extra dimensions~\cite{arkani98:hlz}, and a four-fermion CI caused by fermion substructure~\cite{Eichten:1984eu}. In these models, the differential  cross section for dilepton pair production can be expressed as:
\begin{linenomath}
\begin{equation}
  \dd{\sigma_{\text{X}\to\Pell\Pell}}{\mll}=
  \dd{\sigma_{\text{DY}}}{\mll}
  +\etaCIADD{X}\mathcal{I}(\mll)
  +\etaCIADD{X}^{2}\mathcal{S}(\mll)\text{,}
  \label{eq:dilepxs}
\end{equation}
\end{linenomath}
where $\text{d}\sigma_{\text{DY}}/\text{d}\mll$ is the SM DY differential cross section, \etaCIADD{X} is a model-specific parameter, and the signal contribution terms are separated into an interference term ($\mathcal{I}$) and a pure signal term ($\mathcal{S}$).
Interference between the signal process and the SM DY background is possible when the new process acts on the same initial state and yields the same final state, and can be either constructive or destructive, depending on the sign of \etaCIADD{X}. 

In the ADD model, which is an attempt to describe quantum gravity with an effective field theory, spacetime is extended by $n$ additional compactified spatial dimensions of size $L$. SM particles are confined to the four-dimensional subspace (the brane), while gravity can propagate to all $D=n+4$ dimensions (the bulk).
If $L$ is sufficiently large, the $D$-dimensional fundamental Planck mass \massADD{D}, which is related to the effective Planck mass \massADD{Pl} in three dimensions by
\begin{linenomath}
\begin{equation}
  \massADD{D}^{2+n} \sim \massADD{Pl}^2/L^n\text{,}
  \label{eq:led-planck}
\end{equation}
\end{linenomath}
can then be probed at the \TeVns scale.
In contrast to the RS model, the ADD model predicts the presence of a quasi-continuous spectrum of KK graviton modes as a result of the compactification of the extra dimensions with a large compactification radius. The number of excited modes increases with the interaction scale, resulting in a nonresonant excess at high dilepton masses from the decay of the virtual gravitons, which cannot be individually resolved with the LHC detectors. 

These processes can be characterized by the single energy cutoff scale \lambdaCIADD{T} in the Giudice--Rattazzi--Wells (GRW) convention \cite{GRW99:extradim}, the string scale \massADD{S} in the Hewett convention~\cite{Hewett:1998sn}, or the number of additional dimensions $n$ in conjunction with \massADD{S} in the Han--Lykken--Zhang (HLZ) convention~\cite{han99:hlz}.
The generic form factor \etaCIADD{X} in Eq.~(\ref{eq:dilepxs}) is replaced by \etaCIADD{G}, which depends on the chosen convention:
\begin{linenomath}
\begin{align}
  \label{eq:grw}
  & \text{GRW:}    & \etaCIADD{G} &= \frac{1}{\lambdaCIADD{T}^4}; \\
  \label{eq:hew}
  & \text{Hewett:} & \etaCIADD{G} &= \frac{2}{\pi} \frac{\lambda}{\massADD{S}^4} \quad\text{with}\,\lambda=\pm 1; \\
  \label{eq:hlz}
  & \text{HLZ:}    & \etaCIADD{G} &=
  \begin{cases}
    \ln\left({\massADD{S}^2}/{\hat{s}}\right)\frac{1}{\massADD{S}^4} & \text{for}\,n = 2,        \\
    \frac{2}{n-2} \frac{1}{\massADD{S}^4}                            & \text{for}\,n > 2\text{.}
  \end{cases}
\end{align}
\end{linenomath}
Here $\hat{s}$ is the square of the parton center-of-mass energy. In the ADD model, interference with DY is limited, as the production of virtual gravitons is dominated by gluon-induced processes. Positive interference is assumed in this analysis, but the choice has negligible effects on the results. The effective field theory will only yield reliable results up to a certain energy scale, above which a more exact theory of quantum gravity is necessary. Both  \lambdaCIADD{T} and \massADD{S} act as this ultraviolet cutoff parameter in their respective conventions.

In case of the CI model, assuming that quarks and leptons share common constituents, the Lagrangian for the CI process $\PQq\PAQq\to\Pell\Pell$, where $\Pell$ is a charged lepton, can be expressed as
\begin{linenomath}
\begin{equation}
    \begin{aligned}
  \mathcal{L}_{\PQq\Pell}=\frac{g^2_{\text{contact}}}{\Lambda^2} 
        \Big[&\etaCIADD{LL}(\PAQq_\text{L}\gamma^\mu\PQq_\text{L})(\overline{\Pell}_\text{L}\gamma_{\mu}\Pell_\text{L})
      + \etaCIADD{RR}(\PAQq_\text{R}\gamma^\mu\PQq_\text{R})(\overline{\Pell}_\text{R}\gamma_{\mu}\Pell_\text{R}) \\ %
      + &\etaCIADD{LR}(\PAQq_\text{L}\gamma^\mu\PQq_\text{L})(\overline{\Pell}_\text{R}\gamma_{\mu}\Pell_\text{R})
      + \etaCIADD{RL}(\PAQq_\text{R}\gamma^\mu\PQq_\text{R})(\overline{\Pell}_\text{L}\gamma_{\mu}\Pell_\text{L})\Big],
  \label{eq:cil}
    \end{aligned}
\end{equation}
\end{linenomath}
where $\PQq_\text{L}=(\PQqu,\PQqd)_\text{L}$ is a left-handed quark doublet;
$\PQq_\text{R}$ represents a sum over the right-handed quark singlets (\PQqu- and \PQqd-type); the $\etaCIADD{ij}$ are real numbers between $-1$ and 1;
and $\Pell_\text{L}$ and $\Pell_\text{R}$ are the left- and right-handed lepton fields, respectively.
By convention, $g_{\text{contact}}^2/{4\pi}=1$ and the helicity parameters \etaCIADD{ij} are taken to have unit magnitude.
The compositeness scale, represented by $\Lambda$, is potentially different for each of the individual terms in the Lagrangian.
Therefore, the individual helicity currents for ``left-left'' ($\text{LL}$), ``right-right'' ($\text{RR}$), and the combination of ``left-right'' ($\text{LR}$) and ``right-left'' ($\text{RL}$) in Eq.~(\ref{eq:cil}), together with their scales (\lambdaCIADD{LL}, \lambdaCIADD{RR}, \lambdaCIADD{LR}, and \lambdaCIADD{RL}), are considered separately in this search, and in each case all other currents are assumed to be zero. While the $\text{LL}$ and $\text{RR}$ models favor positive values of \costheta, as does the DY background, the distribution is inverted for the $\text{LR}$ and $\text{RL}$ models, which favor negative values.

A given \etaCIADD{ij} can be related to the form factor in the differential cross section in Eq.~(\ref{eq:dilepxs}) by
\begin{linenomath}
\begin{equation}
  \etaCIADD{X} = -\frac{\etaCIADD{ij}}{\lambdaCIADD{ij}^{2}}\text{.}
  \label{eq:ciprefact}
\end{equation}
\end{linenomath}

\section{Simulated event samples}
\label{sec:samples}
Numerous simulated event samples are used to describe both background and signal processes. Dedicated samples are generated for each of the three years of data taking considered in this analysis, reflecting the changing beam and detector conditions from year to year. Different generators are used to generate the individual processes.

The dominant DY background is generated with  \POWHEG~v2~\cite{Nason:2004rx,Frixione:2007vw,Alioli:2010xd,Alioli:2008gx,Frixione:2007nw,Re:2010bp} using next-to-leading order (NLO) matrix elements. In the sample generation, PDFs are evaluated using the LHAPDF library~\cite{Whalley:2005nh,Bourilkov:2006cj,Buckley:2014ana}. For the 2016 samples, the NNPDF3.0~\cite{Ball:2014uwa} PDF set at NLO is used, while for the samples describing the 2017 and 2018 data, the NNPDF3.1 PDF set computed at the next-to-next-to-leading order (NNLO)~\cite{Ball:2017nwa} is used. The \PYTHIA~8.212 (8.230)~\cite{Sjostrand:2014zea} generator is used to simulate parton showering and hadronization in the 2016 (2017 and 2018) samples. For all 2016 samples, the \PYTHIA tune CUETP8M1~\cite{Khachatryan:2015pea} is used. It is replaced for 2017 and 2018 with the CP5~\cite{Sirunyan:2019dfx} tune, which is optimized for the NNPDF3.1 PDF set. 

 The NLO cross sections obtained from \POWHEG are corrected for NNLO effects in perturbative QCD, as well as for missing electroweak effects at NLO, using a correction factor that depends on the dilepton invariant mass, obtained using the \FEWZ~3.1.b2 program~\cite{Li:2012wna}. In these calculations, the LUXqed\_\-plus\_\-PDF4LHC15\_\-nnlo\_\-100~\cite{Manohar:2016nzj} PDF set is used. It combines the QCD PDFs, based on the PDF4LHC recommendations~\cite{Butterworth:2015oua}, with the photon PDFs to account for pure quantum electrodynamics effects. This allows the inclusion of the background arising from a $\PGg\PGg$ initial state via $t$- and $u$-channel processes~\cite{Bourilkov:2016qum,Bourilkov:2016oet} in the correction.
  
The \ttbar, $\PQqt\PW$, and $\PW\PW$ backgrounds are simulated using \POWHEG~v2, with parton showering and hadronization described by \PYTHIA, using the same PDF sets as for the DY samples. The \ttbar background is normalized to the NNLO cross section calculated with \textsc{top++}~\cite{Czakon:2011xx} assuming a top quark mass of 172.5\GeV, 
while the $\PQqt\PW$ simulation is normalized to the cross section computed up to next-to-next-to-leading logarithmic accuracy~\cite{Kidonakis:2010ux}.

In addition to the $\PW\PW$ samples,  $\PW\PZ$ and $\PZ\PZ$ events are considered. As the analysis in the dimuon channel for 2016 is unchanged from Ref.~\cite{Sirunyan:2018exx}, samples simulated at leading order (LO) using the \PYTHIA program along with the NNPDF2.3 PDFs at LO are used. For the analysis of the 2016 data in the electron channel, as well as the 2017 and 2018 data in both channels, samples are generated at NLO with \POWHEG and \MGvATNLO version 2.2.2~\cite{Alwall:2014hca}, using the NNPDF3.1 PDF set.
The process involving $\PGg^{\ast}/\PZ\to \Pgt^{+}\Pgt^{-}$ decays, which is not included in the \POWHEG DY samples described above, as well as the $\PW$+jets process, are simulated at LO with the \MGvATNLO generator.
Cross sections for these processes have been calculated up to NNLO with \MCFM~6.6~\cite{Boughezal:2016wmq,Campbell:2015qma,Campbell:2011bn,Campbell:1999ah}.

 The NNPDF3.1 NNLO PDF set was found to predict unphysical cross sections at very high dilepton invariant masses ($>$4\TeV). The DY and \ttbar events generated with \POWHEG are therefore reweighted so that the generated \mll distribution matches the prediction of the NNPDF3.0 NLO set that was used in Ref.~\cite{Sirunyan:2018exx}, which does not exhibit this undesired behavior. For the DY samples, these weights are applied before they are further corrected to the higher order cross section calculated with \FEWZ described above. Thus, the DY background estimate is consistent among the three years of data taking. Other processes also generated with the NNPDF3.1 PDF set are less affected and have only small contributions to the overall background predictions. No reweighting is applied to them.

As the properties of high-mass events containing a spin-1 resonance can be studied by using the DY background samples, only one sample containing a \ZPSSM~boson with a mass of 5000\GeV is generated for illustrative purposes. The RS \GKK samples with the graviton generated at different mass values from 250 to 4000\GeV have been generated for all three years. 
In all samples, the high-mass resonances decay to electron and muon pairs.
These signal samples are generated using the \PYTHIA~8.205 program with the NNPDF2.3 LO PDF set for 2016, and \PYTHIA~8.230 with the NNPDF3.0 LO PDF set for 2017 and 2018 samples.

For the nonresonant signal models, samples for both CI and ADD are simulated at LO using \PYTHIA~8.205 for 2016, and \PYTHIA~8.230 for 2017 and 2018 samples. The NNPDF2.3 LO PDF set is used for 2016 samples while for 2017 and 2018, NNPDF3.1 NNLO is used. For consistency, the generated mass distribution of the 2017 and 2018 samples is reweighted to match that of the 2016 samples. The departures of these weights from unity are assigned as systematic uncertainties. In both cases, signal and DY background are simulated simultaneously to model interference effects. 

The CI samples, as well as ADD samples for 2017 and 2018, are generated with a lower threshold of 300\GeV on the dilepton invariant mass, while for the ADD samples for 2016, a threshold of 1700\GeV is applied. In case of the CI samples, separate samples for the $\text{LL}$, $\text{LR+RL}$, and $\text{RR}$ helicity states are generated, where the \PYTHIA implementation of this signal model summed the $\text{LR}$ and $\text{RL}$ currents into a single process. The generated events in these samples are therefore reweighted to recover the individual $\text{LR}$ and $\text{RL}$ models.
Dedicated \PYTHIA DY samples are produced with the same generator settings and subtracted from the signal samples to obtain the respective pure signal yields. In the case of ADD samples, NNLO calculations in QCD show that the correction factor to the LO cross section can be as high as 1.6~\cite{Ahmed:2017}, and that it always exceeds 1.3 in the considered dilepton mass range. Furthermore, NLO electroweak corrections are not taken into account.
This motivates applying the minimal correction factor of 1.3 in the analysis, which also allows a direct comparison to previous results~\cite{Khachatryan:2014fba}.

During the 2016 and 2017 data taking, the CMS L1 trigger was affected by a slowly increasing shift of the reconstructed cluster time in the ECAL, predominantly at high $\eta$. This led to a loss of a fraction of events in the trigger, as these clusters could be assigned to the wrong LHC bunch crossing (``trigger prefiring'')~\cite{Sirunyan:2020zal}. As this effect is not present in simulation, simulated events in the dielectron channel are reweighted to account for this inefficiency.

The detector response is simulated using the \GEANTfour~\cite{Agostinelli:2002hh} package. The presence of additional $\Pp\Pp$ interactions in the same or adjacent bunch crossing observed in data (pileup) is incorporated into simulated events by including extra $\Pp\Pp$ interactions that have been generated with \PYTHIA.
In the dielectron channel, simulated samples are reweighted to reproduce the same pileup distribution as measured in data (pileup reweighting). In the dimuon channel, the trigger, reconstruction, and identification efficiencies exhibit no significant pileup dependence, and no such reweighting is performed.

\section{Lepton reconstruction and event selection}
\label{sec:objsel}
The electron and muon reconstruction algorithms remain close to those used in Ref.~\cite{Sirunyan:2018exx}, with a few updates implemented. In the electron channel, the 2016 data are reprocessed using an updated electron calibration that corrected for a drift in the energy response at very high energy. For the 2017 and 2018 data sets, there are some modifications to the criteria for both electron and muon identification and electron isolation, as well as for the overall event selection. These improve the treatment of the higher pileup and the selection efficiency for leptons with very high (several \TeVns) transverse momentum \pt. 

\subsection{Electron reconstruction and selection}
To select dielectron events in the L1 trigger, events are required to contain at least two electron candidates. The \pt thresholds evolved with time, but never exceeded 25\GeV for the electron with the higher \pt and 17\GeV for the electron with the lower \pt. As a safeguard against cases where one of the electrons fails the L1 trigger, events containing at least one electron, jet, or tau lepton candidate passing the higher \pt thresholds at L1 are also considered at HLT level. In the HLT, dielectron events were collected with a trigger requiring two electrons with $\pt > 33\GeV$ and $\abs{\eta} < 2.5$ in 2016 and 2017. These electrons were also required to pass loose identification criteria using the shower shape in the calorimeter and the matching of a pixel detector track with the energy deposit. The \pt threshold was lowered to 25\GeV in 2018, since the electron trigger was improved by making use of the upgraded pixel detector. Secondary trigger paths with progressively higher \pt thresholds and fewer selection requirements are employed to monitor the performance of the main trigger. 

Electron candidates are reconstructed from energy clusters in the ECAL and tracks from the inner tracker.
Typically, electrons lose some of their energy due to bremsstrahlung emission in the tracking detector, which is recovered by adding compatible energy deposits to the ECAL cluster. The final clusters are then associated geometrically with tracks to form electron candidates. As the measurement of the track parameters can be unreliable for high-energy electrons because of bremsstrahlung, the energy of the candidate is taken directly from the ECAL cluster without combination with track information. However, the direction of the electron used in the analysis is still obtained from the combined track and ECAL cluster information. This minor mis-reconstruction has a negligible effect on the dielectron mass measurement.

We select electron candidates with  $\pt>35\GeV$ that satisfy $\abs{\eta_C}<1.44$ (ECAL barrel region) or $1.57<\abs{\eta_C}<2.50$ (ECAL endcap region), where $\eta_C$ is the pseudorapidity of the cluster of ECAL deposits making up the electron with respect to the nominal center of the CMS detector. The transition region $1.44<\abs{\eta_C}<1.57$ is excluded since it leads to lower quality reconstructed clusters, 
owing mainly to inactive material consisting of services and cables exiting between the barrel and endcap calorimeters.

Electrons are required to pass a set of selection criteria optimized for the identification of high-energy electrons~\cite{CMS-ele-paper}. The lateral spread of energy deposits in the ECAL associated with the electron's cluster is required to be consistent with that of a single electron. The electron's track is required to be matched geometrically to the ECAL cluster and to be compatible with originating from the nominal interaction point. Energy deposits in the HCAL in the direction of the electron, corrected for noise and pileup, are required to be less than 5\% of the electron's energy. Misreconstructed electrons, and electrons in jets, are suppressed by requiring that the electron be isolated in a cone of radius $\Delta R = \sqrt{\smash[b]{(\Delta\eta)^2+(\Delta\phi)^2}} = 0.3$ in both the calorimeter and tracker~\cite{Khachatryan:2016zqb}.
Only well-measured tracks that are consistent with originating from the same vertex as the electron are included in the isolation sum.
The efficiency of the trigger to select events with two electrons passing the analysis selection requirements, measured in data as discussed below, is at least 95\% for barrel (endcap) electrons satisfying $\pt > 36 (37)\GeV$ in 2016, $\pt > 38 (40)\GeV$ in 2017, and $\pt > 27 (29)\GeV$ in 2018.

In the endcap region of the detector, there is a large background of events with hadronic jets misreconstructed as electrons, which comes from $\PW\text{+jets}$ events and from events composed uniquely of jets produced through the strong interaction, referred to as QCD multijet events. To reduce this background, we require that at least one of the electrons be located in the barrel region. The subset of events with both electrons in the endcap region is expected to be overwhelmingly dominated by this background, and is used as a control region for the jet background estimate. 

No requirement is made on the sign of electrons because of the increasing probability of charge misidentification for high-energy electrons, which would result in a significant efficiency loss from an opposite-sign selection. The fraction of same-sign events in DY simulation was found to increase from about 1\% at the \PZ boson peak to $\approx$10\% for masses of several TeV. In data, it was found to be consistent with this expectation.

The efficiency to trigger, reconstruct, and select an electron pair with invariant mass $\mee =1\TeV$ within the detector acceptance is 72 (68)\% for 2016 (2017 and 2018) for barrel-barrel events and 67, 60, and 67\% for the three years, for barrel-endcap events. It  remains stable within a few percent for larger masses. The lower efficiency for barrel-barrel events in 2017 and 2018 is caused by changes to the trigger making use of the upgraded pixel detector to control the trigger rate. The reduction for barrel-endcap events in 2017 is mostly due to trigger prefiring. The total efficiency is estimated using simulated DY events, and the efficiencies of the various selection steps are compared in data and simulation using a method similar to that described in Ref.~\cite{Sirunyan:2020ycc}. Using high-\pt \PZ bosons, it is possible to probe the efficiencies up to $\pt = 1$ $(0.7)\TeV$ for electrons in the barrel (endcap) region, with a precision of less than 10\%. Based on those measurements, a correction factor is applied to the trigger efficiency in simulation to model the lower \pt turn-on observed in data. No correction is found to be necessary for the offline selection.

\subsection{Muon reconstruction and selection}
For the selection of dimuon events, in the L1 trigger, events with at least one single muon with $\pt > 22\GeV$ are accepted. At HLT, single-muon triggers requiring $\pt > 50\GeV$ and $\abs{\eta}< 2.4$ and without an isolation requirement were available during the whole data taking period. During 2017 and 2018, an upgraded version of the muon reconstruction in the HLT was deployed, and  as a backup the algorithm used in 2016 was used in an HLT path requiring $\pt > 100\GeV$~\cite{Sirunyan:2021zrd}. The backup trigger increased the overall trigger efficiency for dimuon events with invariant masses above 1\TeV by $\approx$1\%. To collect an event sample unbiased by the trigger turn-on for a data normalization region around the Z boson mass ($60 < \mll < 120\GeV$), a prescaled HLT path with a reduced pT threshold of 27 GeV was used. The average prescale value ranged from about 140 in 2016 to about 460 in 2018.

Muon candidates are reconstructed by combining hits in the inner tracking detector with those in the muon system. For muons with $\pt > 200\GeV$, dedicated reconstruction algorithms take into account radiative energy losses in the detector material~\cite{MUO-10-004-PAS,Sirunyan:2019yvv}. As no calorimeter information is used in the muon reconstruction, data corresponding to an integrated luminosity of about 3\fbinv recorded with ECAL or HCAL in a degraded condition are included in the analysis of the dimuon final state. Muons are required to have $\pt > 53\GeV$, $\abs{\eta} < 2.4$, and to pass a set of selection criteria optimized for high-\pt muons~\cite{Khachatryan:2014fba}. To reject muons produced inside jets, we require that the scalar \pt sum of all tracks identified to come from the same pp collision and within a cone of $\Delta R = 0.3$ around the muon candidate, excluding the muon candidate itself, does not exceed 10\% of the \pt of the muon. 

Dimuon candidates are formed from two muons with invariant mass $\muu >150\GeV$. While the searched-for neutral dilepton final states would decay to truly opposite-sign dileptons, in principle leptons with misreconstructed charge are usable (and are in fact used in the case of electrons, as described above). For muons, misreconstruction of the charge indicates poor resolution of the track fit and thus an unreliable \pt measurement. Therefore, the two muons are required to have opposite charge. As the charge misidentification probability is of the order of $10^{-4} $ for muon momenta of 2\,TeV~\cite{Sirunyan:2019yvv}, this requirement has a negligible effect on the signal efficiency. The muon tracks are fit to a common vertex, which improves mass resolution at high mass.  A loose requirement on the quality of the vertex fit rejects some residual background and potentially badly measured signal events (less than 1\% in simulation). Backgrounds from cosmic ray muons are reduced to a negligible level by requiring that the three-dimensional  angle between the two muons be less than $\pi - 0.02$. The remaining contribution of this background after the full selection for events containing muons with $\pt > 300\GeV$ is less than 0.2\%.

The trigger efficiency for dimuon events that pass these offline muon selection criteria has been measured in data using $\PZ$ boson events. For all three years of data taking,  it is above 99.3\% for events where both muons are in the barrel region ($\abs{\eta} < 1.2$) of the detector and above 99.0\% if one or more muons are located in the endcaps ($\abs{\eta}>1.2$).
Measured in data using \PZ events,
the overall efficiency to trigger on, reconstruct, and identify muon pairs within the detector acceptance is 93\%, with little dependence on the dimuon invariant mass.

Detailed studies of the muon reconstruction performance revealed a slight inefficiency for muons with $\abs{\eta}> 1.6$, which increases with muon momentum, caused by electromagnetic showers in the detector material~\cite{Sirunyan:2019yvv}. 
This effect is more pronounced in data than in simulation. This difference is included in a mass-dependent systematic uncertainty.

The dimuon mass resolution for events with high-\pt muons has been studied using highly Lorentz-boosted \PZ boson events. It was found that for events where at least one muon is in the endcap, the simulation predicted better mass resolution than observed in the data. To correct for this effect, the reconstructed dimuon mass for simulated events in this category is smeared by 15\%, bringing data and simulation into good agreement.

\begin{figure}
\centering
\includegraphics[width=0.45\textwidth]{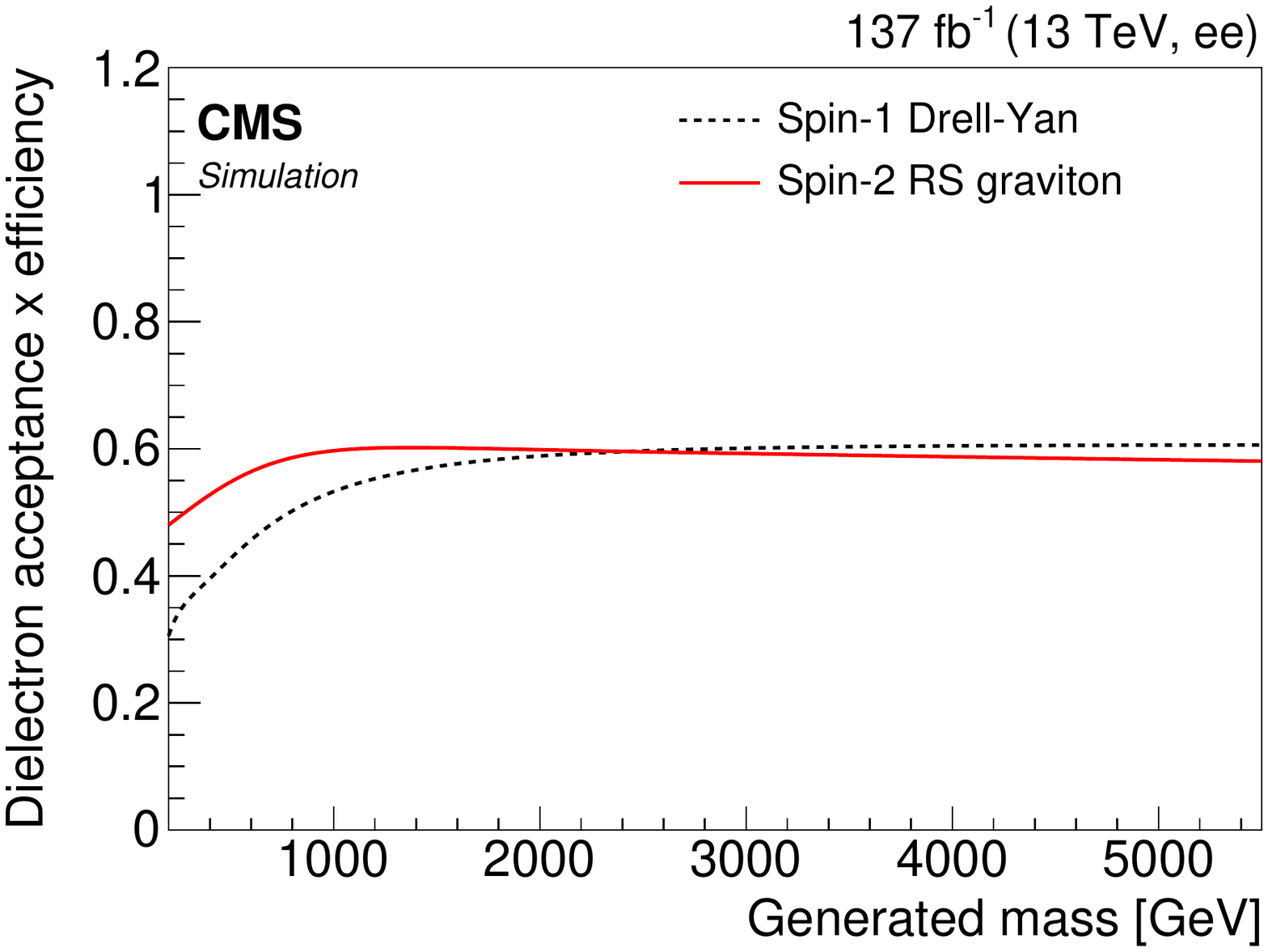}
\includegraphics[width=0.45\textwidth]{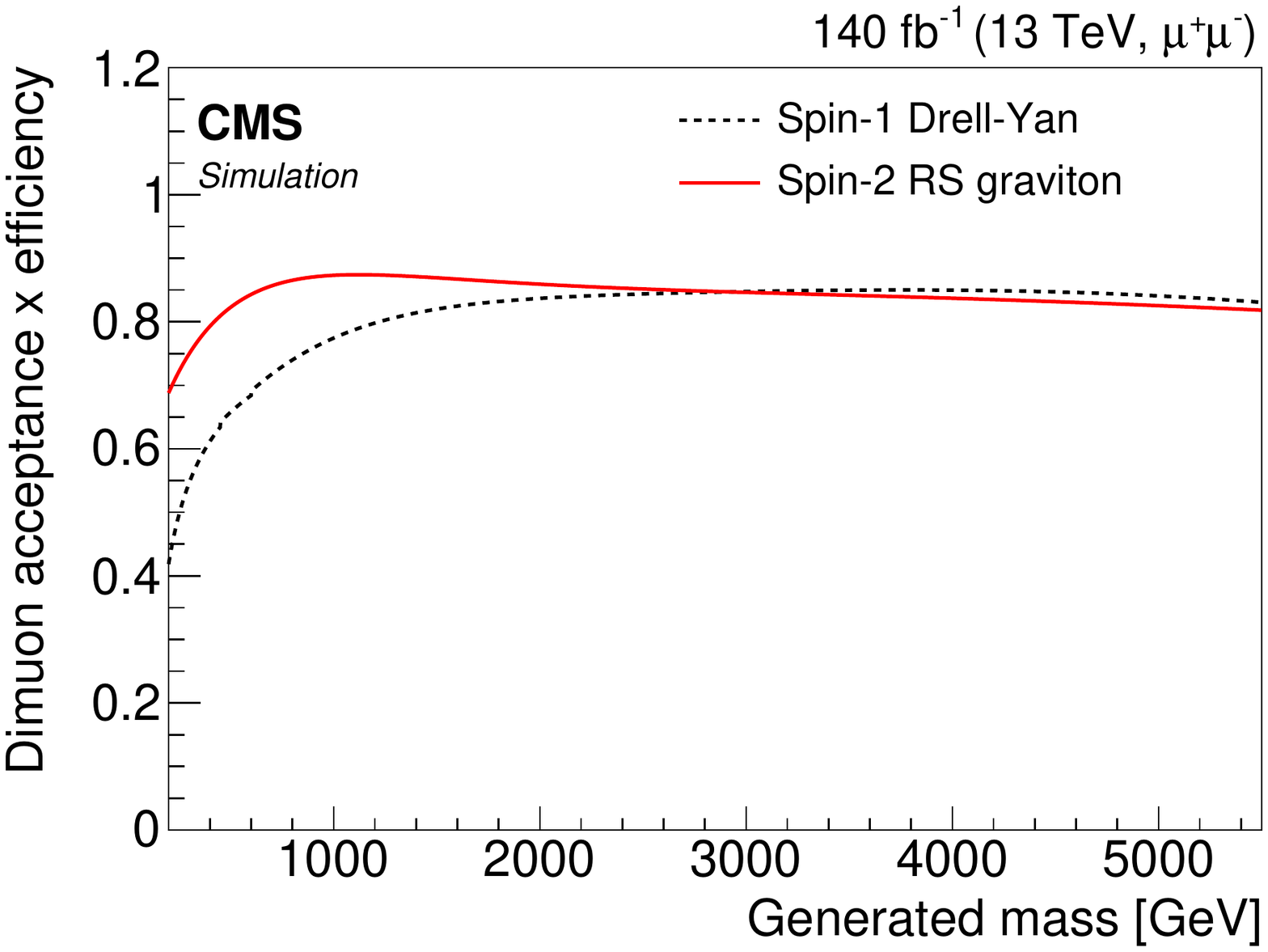}
\caption{Product of the acceptance and the efficiency for (left) dielectron and (right) dimuon pairs as a function of generated mass in simulated events. The DY samples are used to represent spin-1 particles, and RS graviton samples are used for spin-2 particles.}
\label{fig:accEff}
\end{figure}

The combined product of the acceptance and the efficiency as a function of mass is obtained from simulation for both dielectron and dimuon pairs. This product is shown in Fig.~\ref{fig:accEff}, separately for spin-1 and spin-2 particles. The values for the different years of data taking are averaged, weighted by the integrated luminosity of the data sets for each year.  Mixtures of polynomials and exponential functions are used to parameterize the mass dependence of the product of the acceptance and the efficiency. The observed turn-on in the low-mass region is a consequence of the lepton \pt threshold and the \pt-dependent selection efficiency. For $\mll > 200\GeV$, the reconstruction and selection efficiencies do not significantly depend on mass, and changes in the product of acceptance and efficiency are due to the changing acceptance as the angular distribution of the leptons changes. For the spin-2 RS graviton, the slight drop at high mass is related to the different production modes (gg and $\PQq\PAQq$) that lead to different angular distribution for the leptons. The fraction of the latter increases with mass and results in a slightly larger fraction of leptons emitted outside the detector acceptance.

\subsection{Event disambiguation}

In a small fraction of the events ($<$1\%), originating mostly from $\PW\PZ$ and $\PZ\PZ$ production, more than two leptons are present. If several dilepton pairs can be formed in the same event, and if at least one pair has a mass within 20\GeV of the \PZ boson mass, then the pair closest in mass to  the \PZ boson mass is chosen. Otherwise, the pair with the two highest-\pt leptons is selected.
For dimuon events, the analysis has not been changed from the published result based on the 2016 data, and the pair with the two highest \pt muons is always selected~\cite{Sirunyan:2018exx}. Events with both a dimuon and dielectron pair that pass the event selection are very rare and both pairs are considered in the analysis.

For both the CI and ADD models, the sensitivity of the analysis can be improved by studying the angle of the outgoing negatively 
charged lepton with respect to the $z$ axis in the Collins--Soper frame, given by
\begin{linenomath}
\begin{equation}
 \label{eq:ci-CS}
 \costheta = \frac{p_z(\Pell^+\Pell^-)}{\abs{p_z(\Pell^+\Pell^-)}}\frac{2(p_1^+p_2^--p_1^-p_2^+)}{m(\Pell^+\Pell^-)\sqrt{m(\Pell^+\Pell^-)^2+\pt(\Pell^+\Pell^-)^2}}.
\end{equation}
\end{linenomath} 
Here $p^{\pm}$ represents $\frac{1}{\sqrt{2}}(E\pm{p_z})$, and the indices 1 and 2
correspond to the negatively and positively charged leptons,
respectively. Events with same-sign electrons are included in the selected event sample, which introduces an ambiguity in the calculation of \costheta. To resolve the ambiguity, same-sign events in the dielectron channel are assumed to originate from charge misidentification. The measured charge of the electron with the lower value of $\eta_C$ is assumed to be correct and the charge of the other electron is taken to be mismeasured and assigned to be the opposite. This results in the correct lepton being chosen in 76\% of same-sign events, according to the simulation.

\section{Background estimation}
\label{sec:backgrounds}
The dominant (and irreducible) source of background in the signal selection is the DY process. Further sources of high-mass dilepton events are \ttbar and $\PQqt\PW$ production and diboson production (especially $\PW\PW$ events).  In addition, both QCD multijet events and $\PW\text{+jets}$ events can be reconstructed as dilepton events if jets or nonisolated leptons are identified as isolated leptons; this category is called ``jet misidentification'' below.

The DY background shape is estimated directly from the \POWHEG samples described above, including the mass-dependent higher-order corrections to the cross section. For the DY $\to\Pgt\Pgt$ process, the \MGvATNLO sample is used. Given the small size of this background, no mass-dependent corrections are applied. The other backgrounds, with the exception of the jet misidentification backgrounds, are also estimated from simulation. The second-largest background contribution originates from \ttbar events and is estimated from the \POWHEG samples. The simulation of this process is validated in a data control region containing an $\Pe\PGm$ pair passing the same lepton requirements as discussed above and is found to describe both the shape of the mass distribution and the event yield observed in data, within uncertainties.

Backgrounds arising from jets misidentified as electrons are estimated from data control regions enriched in QCD multijet events with the methods described in Ref.~\cite{Khachatryan:2014fba}. Similar methods are used to evaluate the background contributions from jets or nonisolated muons to the sample of selected muons. Even though the uncertainty in these estimates is as high as 50\% for the entire mass range, this background has a negligible effect on the results of the analysis, as its contribution to the overall event yield is about 1--3\%. The jet background estimate is found to be independent of \costheta and is therefore scaled by 0.5 when the event sample is split into $\costheta < 0$ and $\geq$0 in the nonresonant interpretation.

Finally, the combined background shape is normalized to the data in the control region around the \PZ boson mass ($60 < \mll < 120\GeV$). For the 2017 and 2018 data, the DY simulation generated with the NNPDF3.1 NNLO PDF set does not accurately describe the \PZ boson \pt spectrum at low \pt. The simulated events are therefore reweighted to match the prediction obtained with the NNPDF3.0 NLO PDF, as used for the 2016 samples. For the dimuon channel, a prescaled trigger with a muon \pt threshold of 27\GeV was used to collect events in this region. The corresponding offline threshold on the muon \pt is 30\GeV. To calculate the normalization factors, the numbers of events in that region are multiplied by the corresponding trigger prescale factor. In the electron channel the normalization factor is 0.95, independent of the year of data taking. In the dimuon channel, these factors are 0.97 for 2016, 1.03 for 2017, and 1.00 for 2018 data.

\section{Systematic uncertainties}
Various effects can impact the shape or the normalization of the dilepton invariant mass distribution and lead to systematic uncertainties in the signal and background estimates. 

One source of uncertainty describes possible residual systematic effects leading to mismodeling in the simulation of the lepton selection efficiency and momentum measurements at high mass, which are not absorbed in the normalization at the \PZ boson peak. These uncertainties are estimated from measurements performed with on-shell \PZ bosons with large transverse boost and measured up to 1\TeV in electron \pt and above 2\TeV in muon momentum. High-energy cosmic ray muons are also used to study the muon energy scale~\cite{Sirunyan:2019yvv}. 

Based on these studies, the overall event selection efficiency including trigger, reconstruction, and identification is assigned a relative uncertainty of 6\% for barrel-barrel and 8\% for barrel-endcap events in the dielectron channel, while an uncertainty of 1--2\% is assigned everywhere for the dimuon channel. As discussed in Section~\ref{sec:objsel}, a loss of efficiency for very energetic muons caused by electromagnetic
showers in the detector material was observed. As this effect is significantly stronger in data compared to simulations, an additional mass-dependent, one-sided uncertainty is assigned. Its magnitude is the largest in 2016 data where it reaches $-1.0$ ($-6.5$)\% at $\muu = 4\TeV$ for barrel-barrel (barrel-endcap plus endcap-endcap) events. It is $-2$\% or smaller in other years~\cite{Sirunyan:2019yvv}. The difference in the uncertainty between the years is due to the limited size of the data control samples.
 
The uncertainty in the mass scale is 2 (1)\% for dielectron pairs in the barrel-barrel (barrel-endcap) category. For dimuon events, potential biases in the mass scale are studied using the generalized endpoint method~\cite{Sirunyan:2019yvv}. The associated uncertainty is 1 (3)\% in the barrel-barrel (barrel-endcap plus endcap-endcap) category for 2016, independent of mass. In 2017 and 2018, a mass-dependent uncertainty is assigned that reaches 1--2\% at $\muu = 5\TeV$.

For dimuon events, an uncertainty in the mass resolution of 15\% is assigned for 2016 and 8.5\% for 2017 and 2018, based on the studies discussed in Section~\ref{sec:objsel}. The difference in the uncertainty between the years is the result of different parameterizations of the mass resolution being chosen. The uncertainty in the dielectron mass resolution is negligible.

The uncertainty in the pileup reweighting procedure in the dielectron channel is evaluated by recalculating the weights, shifting the assumed total inelastic cross section by 4.6\% around the nominal value~\cite{Sirunyan:2018nqx}. The resulting changes in the event yield estimates are below 0.6\%. 

The uncertainty in the trigger prefiring rate in the forward region of the ECAL endcap in 2016 and 2017 is 20\%, resulting in a 2--3\% uncertainty in the acceptance for dielectron events in which one electron is in the endcap for 2017. Averaged over the full data set, the uncertainty is well below 1\%.

Theoretical uncertainties in the DY cross section at high mass related to higher-order corrections and PDF uncertainties grow with mass and reach 20\% at $\mll = 6\TeV$. Here, the PDF uncertainty is determined with the PDF4LHC procedure~\cite{Butterworth:2015oua} using replicas of the NNPDF3.0 PDF set. The contribution from other backgrounds estimated from simulations is mostly relevant for masses below 1--2\TeV, and a 7\% uncertainty is applied based on the uncertainties in the cross section calculations of the individual processes. 

The estimate for the background from jet misidentification ($\PW\text{+jets}$, QCD), obtained from data, is assigned a 50\% uncertainty in both dimuon and dielectron channels for all years. These estimates are based on the validation of the method in a control region enriched in events with one genuine electron and one misidentified electron in the dielectron channel, and on uncertainties in the measurement of the misidentification rate at high mass in the dimuon channel. The stability of the analysis results against potential further uncertainties in the extrapolation of this background estimate to the highest mass values has been tested and the effect on the results was found to be negligible.

The uncertainty in the \PZ peak normalization factor in the dielectron channel is 1\% in 2016 and 2 (4)\% for barrel-barrel (barrel-endcap) events for 2017 and 2018, driven by the typical electron \pt in this region being close to the trigger threshold. It is 5\% in the dimuon channel for all three years of data taking, arising from the use of a trigger, whose prescale changed with instantaneous luminosity, to collect the \PZ boson sample. 

The uncertainties arising from the limited sizes of the simulated background and signal samples affect the results differently for the resonant and nonresonant signals. In the resonant case they are accounted for in the uncertainty in the parameters in the background parametrization discussed below. For nonresonant signals they are included as a systematic uncertainty in the background and signal yield estimates. 

For nonresonant signals, an additional uncertainty is assigned to the reweighting of the 2017 and 2018 samples to match the 2016 PDF set. The size of this uncertainty is mass dependent and reaches 30\% at a mass of 5\TeV. The impact of this uncertainty on the limits on the model parameters $\Lambda$ and $\Lambda_{\mathrm{T}}$ in these signal models is  as large as 5\%. As no signal simulation is used in the resonant case, it is unaffected by this uncertainty.

In the statistical interpretation for resonant signals, the signal is normalized to the data at the \PZ boson peak and the backgrounds are normalized to the data in the mass windows around the resonance mass. Uncertainties in the background normalization are taken into account with an overall statistical uncertainty and any remaining uncertainties in the background shape are covered by it. Therefore, only uncertainties in the signal modeling whose impact on the analysis depend on the dilepton mass have to be considered in this case. PDF uncertainties in the signal cross section are considered to be theoretical uncertainties and are not included here. These uncertainties are summarized in Table~\ref{tab:syst}.

\begin{table}[!htb]
\centering
\topcaption{
Sources of systematic uncertainties considered in the search for resonant signals and their relative magnitude.}
\centering
\begin{tabular}{lc}
\hline
Source & Uncertainty \\\hline 
Electron selection efficiency & 6--8\%\\
Muon selection efficiency &1--2\% (two-sided), 0--6.5\% (one-sided)\\
Mass scale uncertainty & 0--3\%\\
Dimuon mass resolution uncertainty & 8.5--15\%\\
\hline
\end{tabular}
\label{tab:syst}
\end{table}

For the limits on nonresonant signals and the measurement of \Rmumuee, the full set of uncertainties is taken into account. The full set of uncertainties is also taken into account in the uncertainty bands for data to simulation comparisons. The impact of these uncertainties on the background estimate for different mass thresholds is shown in Table~\ref{tab:systNonRes}.

\begin{table}[!hbtp]
\centering
\topcaption{Systematic uncertainties considered in the search for nonresonant signals. The relative impact of the uncertainties on the background yield estimates is shown for two dilepton invariant mass thresholds, 1 and 3\TeV. The uncertainty in the jet misidentification background has a negligible effect on the overall background estimate and is not listed.}
\begin{tabular}{lcccc} 
\hline
\multirow{3}{*}{Uncertainty source} & \multicolumn{4}{c}{Impact on background [\%]}\\
 & \multicolumn{2}{c}{$\mll > 1\TeV$}  & \multicolumn{2}{c}{$\mll > 3\TeV$} \\
 & $\Pe\Pe$  & $\PGm\PGm$ & $\Pe\Pe$ & $\PGm\PGm$ \\ \hline
  Lepton selection efficiency & 6.8 & 0.8 & 6.4 & 1.3 \\  
   Muon trigger efficiency & \NA & 0.9 & \NA & 0.9 \\   
  Mass scale & 7.0 & 2.7 & 15.4 & 2.4 \\ 
  Dimuon mass resolution & \NA & 0.1 & \NA & 0.6 \\ 
  Pileup reweighting & 0.3 & \NA & 0.5 & \NA \\ 
  Trigger prefiring & 0.5 & \NA & 0.2 & \NA \\ 
  PDF & 3.7 & 3.0 & 9.4 & 10.2 \\ 
  Cross section for other simulated backgrounds & 0.6 & 0.8 & 0.2 & 0.4 \\ 
  \PZ peak normalization & 2.3 & 5.0 & 2.0 & 5.0 \\ 
  Simulated sample size & 0.4 & 0.4 & 1.3 & 1.6 \\  
\hline
\end{tabular}
\label{tab:systNonRes}
\end{table}

\label{sec:systematics}

\section{Results}

\begin{figure}[!hbtp]
\centering
\includegraphics[width=.48\textwidth]{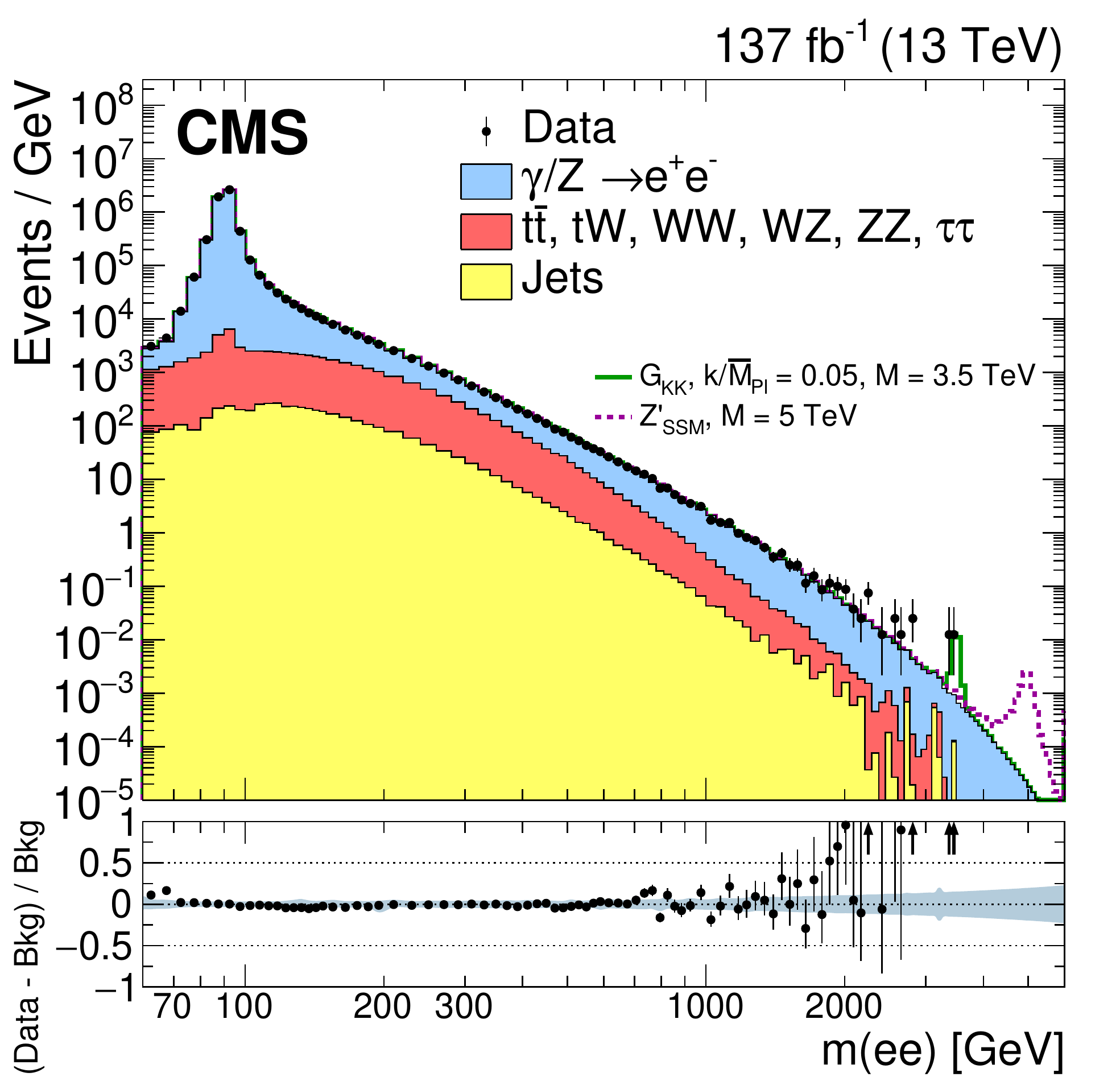}
\includegraphics[width=.48\textwidth]{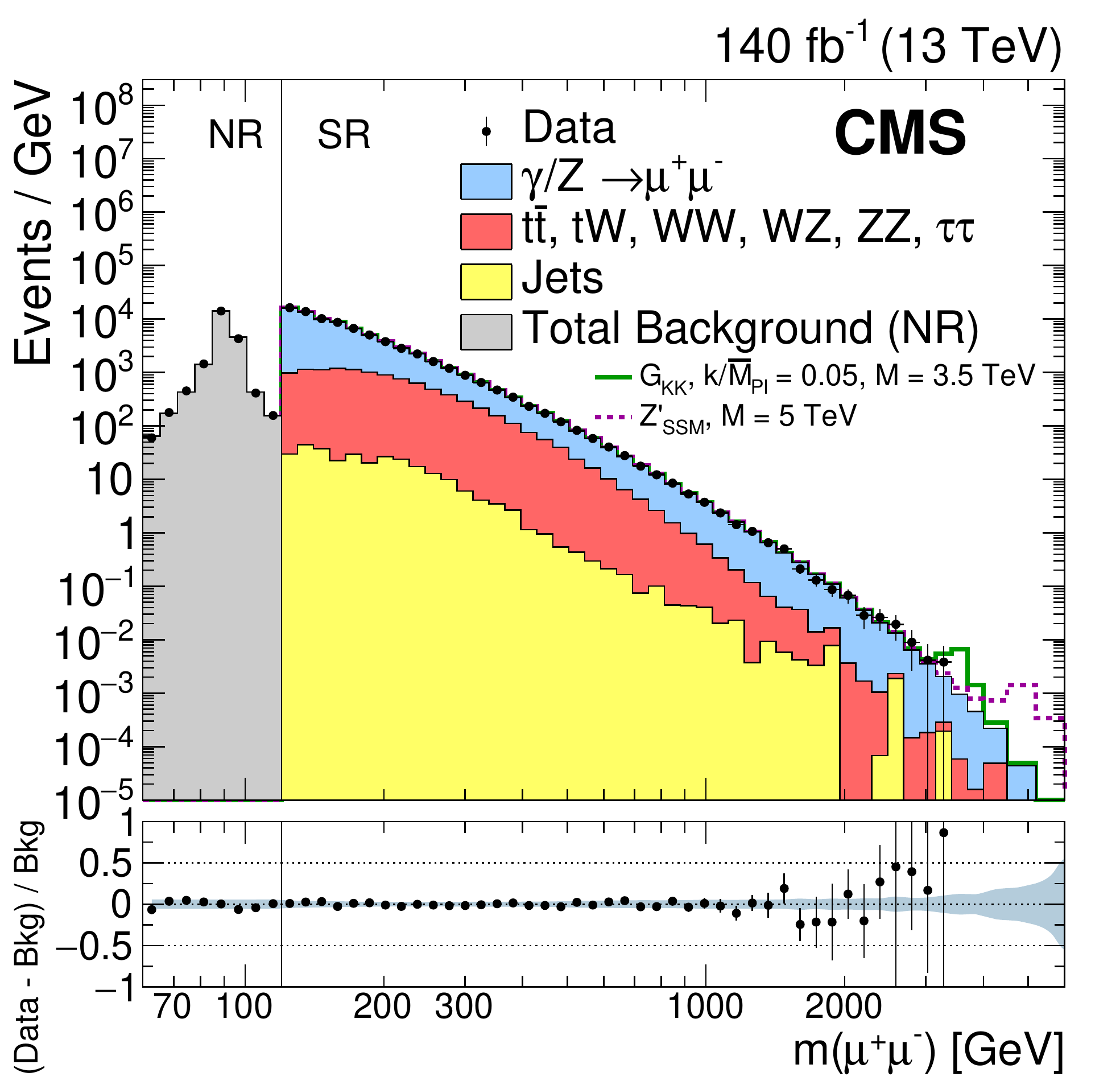}
\caption{The invariant mass distribution of pairs of (left) electrons and (right) muons observed in data (black dots with statistical uncertainties) and expected from the SM processes (stacked histograms). 
For the dimuon channel, a prescaled trigger with a \pt threshold of 27\GeV was used to collect events in the normalization region (NR) with $\muu < 120\GeV$. The corresponding offline threshold is 30\GeV. 
Events in the signal region (SR) corresponding to masses above 120\GeV are collected using an unprescaled single-muon trigger.
The bin width gradually increases with mass. The ratios of the data yields after background subtraction to the expected background yields are shown in the lower plots. The blue shaded band represents the combined statistical and systematic uncertainties in the background. Signal contributions expected from simulated \GKK and \ZPSSM resonances with masses of 3.5 and 5\TeV, respectively, are shown.
}
\label{fig:invmass}
\end{figure}

The invariant mass distributions of electron and muon pairs are shown in Fig.~\ref{fig:invmass}, combining the 2016--2018 data sets. For illustration, simulated \GKK and \ZPSSM signals with masses close to the exclusion sensitivity of the analysis are shown. 

Agreement is observed between the data yields and the expected background, which is quantified in Section~\ref{sec:interpretation}, although a slight excess of events is seen in the dielectron channel for masses above 1.8\TeV. Careful inspection of those events did not reveal any pathologies. 
Four events are observed with mass above 3\TeV, two each in the dielectron and dimuon channels. The measured masses are 3.35 and 3.47\TeV in the dielectron channel and 3.07 and 3.34\TeV in the dimuon channel. 

Table~\ref{tab:datamcyieldseemumu} presents the observed and expected number of events for various mass ranges for the dielectron and dimuon pairs. The uncertainty in the total background is calculated taking into account correlations in the uncertainties among the different background sources. No jet background estimate is available in the dimuon channel for $60 < \mll < 120\GeV$, since the data in this control region are collected with the prescaled trigger. However, this background is negligible in this mass range given the large cross section for the \PZ boson resonance.

\begin{figure}[!hbtp]
\centering
\includegraphics[width=.48\textwidth]{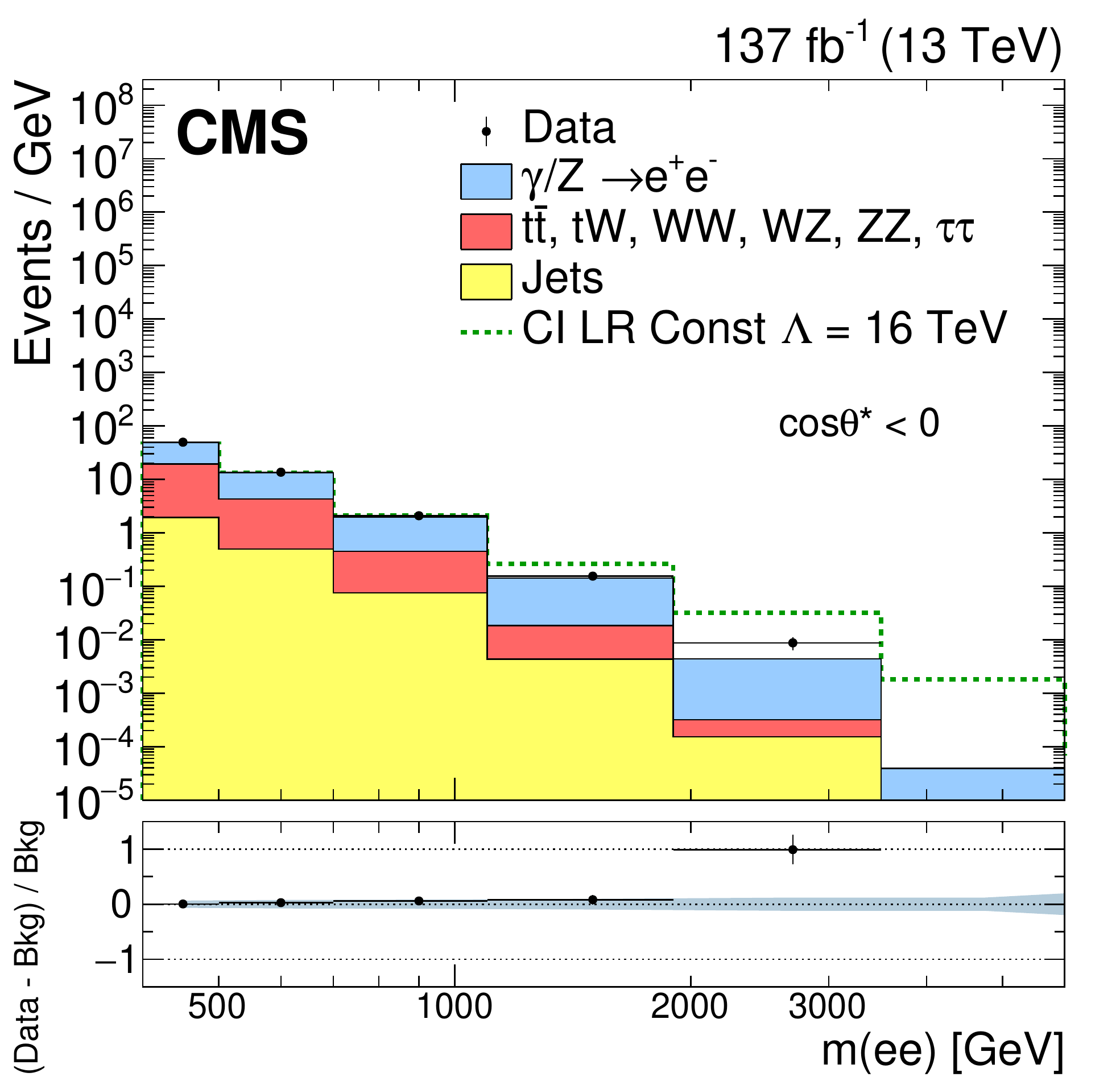}
\includegraphics[width=.48\textwidth]{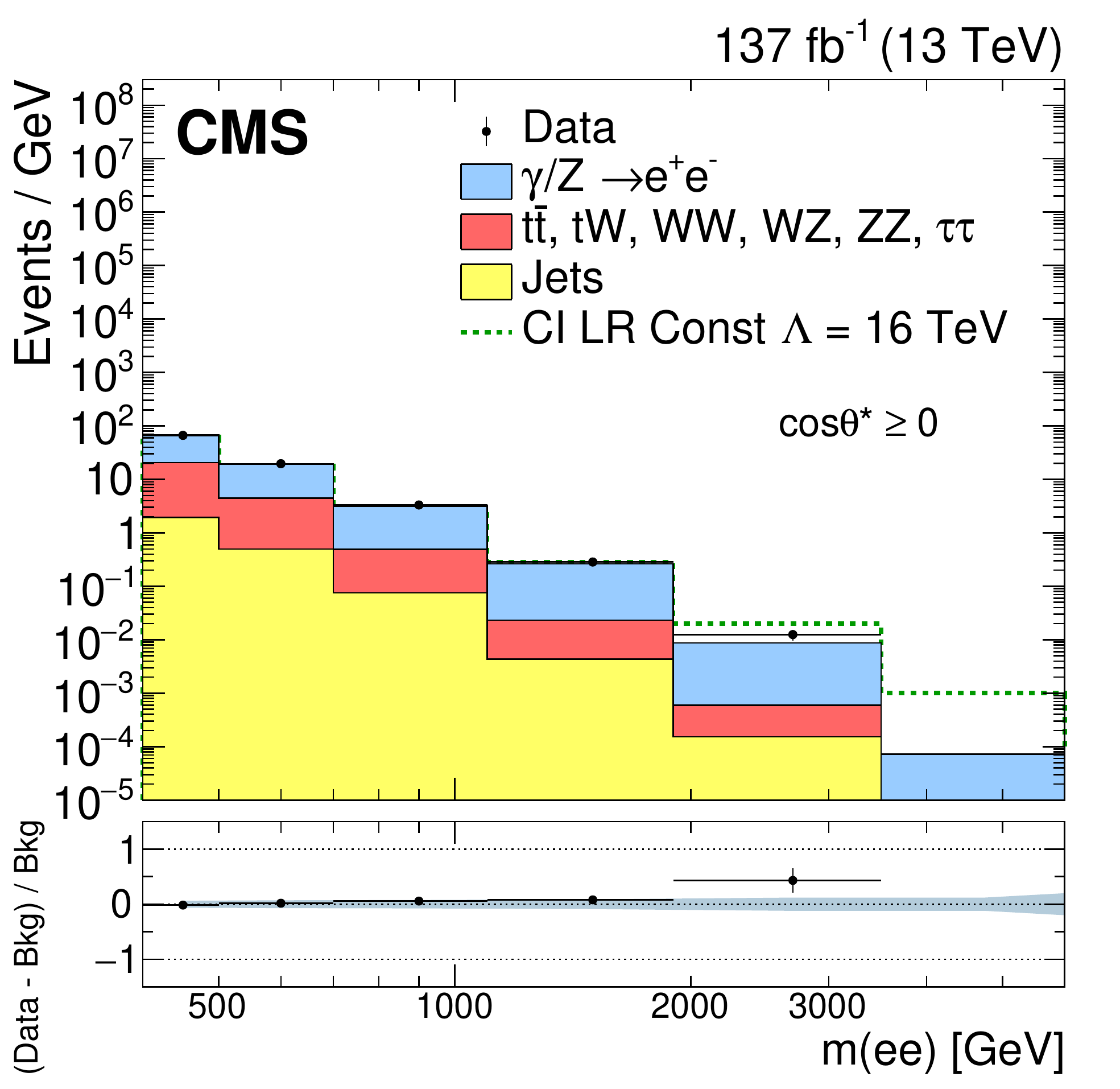}
\includegraphics[width=.48\textwidth]{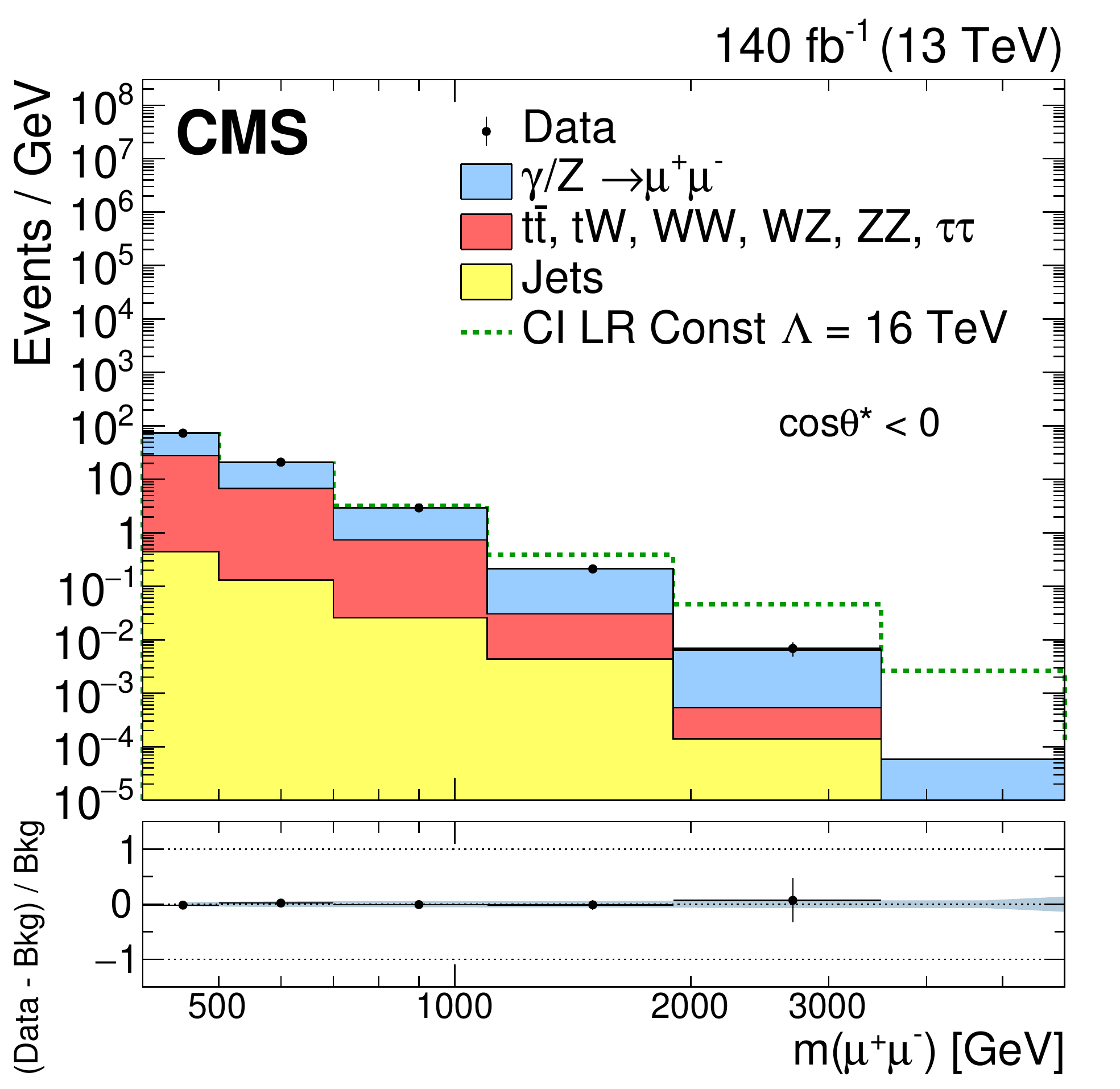}
\includegraphics[width=.48\textwidth]{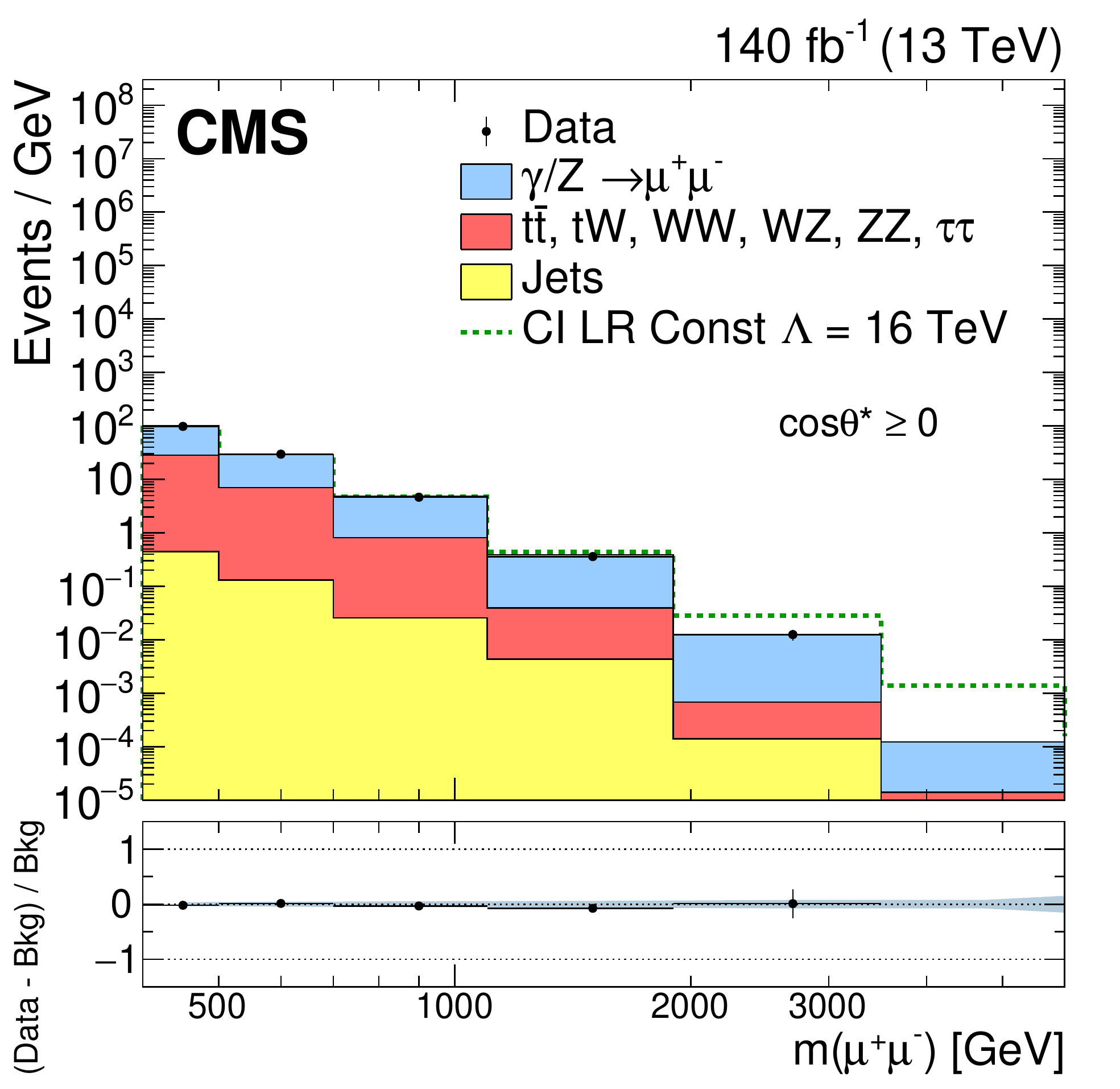}
\caption{The invariant mass distribution of pairs of (upper) electrons and (lower) muons observed in data (black dots with statistical uncertainties) and expected from simulated SM processes (stacked histograms) for (left) $ \costheta < 0$ and (right) $ \costheta \geq 0$. 
The bin width gradually increases with mass. The ratios of the data yields after background subtraction to the expected background yields are shown in the lower plots. The blue band represents the combined statistical and systematic uncertainties in the background. The signal contributions expected from the constructive LL CI model with $\Lambda = 16\TeV$ are shown as dashed green lines.
}
\label{fig:invmassCS}
\end{figure}

The mass distributions are shown again in Fig.~\ref{fig:invmassCS}, split into the two bins in \costheta used in the search for nonresonant signals, and using the same mass bins as those used in the statistical interpretation for the CI signal, as described in Section~\ref{sec:nonreslimits}. A CI signal from the LR model is shown to illustrate the improved signal-to-background ratio in the  $\costheta < 0$ bin for these types of signals. 

\begin{table}[htb] 
\centering
\topcaption{Observed and expected background yields for different mass ranges in the (upper) dielectron channel and (lower) dimuon channel. The sum of all background contributions is shown as well as a breakdown into the three main categories. The quoted uncertainties include both the statistical and the systematic components.}

\resizebox{\linewidth}{!}{
  \begin{tabular}{lccccc}
  \hline
$\mee$ range    & Observed      & \multicolumn{1}{c}{Total}                 & DY     & Other prompt  & Jet mis-\\
\big[GeV\big] & yield & \multicolumn{1}{c}{background} & & lepton backgrounds & identification \\\hline  
 60--120   & 28194452                    & 28200000 $\pm$ 710000~~  & 28000000 $\pm$ 710000~~        & 153000  $\pm$ 8000~~          & 11300 $\pm$ 5700~~ \\  
120--400   & 912504                      & 942000  $\pm$ 37000   & 744000  $\pm$ 31000         & 179000  $\pm$ 11000         & 18900   $\pm$ 9500~~~\\
400--600   & 16192                       & 16400   $\pm$ 770~~     & 10900   $\pm$ 477~~           & 4910    $\pm$ 340           & 534     $\pm$ 267\\
600--900   & 3756                        & 3660    $\pm$ 190     & 2800    $\pm$ 150           & 757     $\pm$ 52            & ~103     $\pm$ 51.4\\
900--1300  & 704                         & 696     $\pm$ 47      & 590     $\pm$ 42            & 89.8      $\pm$ 6.8             & 16.0      $\pm$ 8.0~~~\\
1300--1800  & 135                         & 131     $\pm$ 12      & 118     $\pm$ 11            & 11.0      $\pm$ 1.0             & 2.82       $\pm$ 1.41 \\
$>$1800  & 44                          & 29.2      $\pm$ 3.6       & 26.8      $\pm$ 3.5             & ~~1.60       $\pm$ 0.22             & 0.82       $\pm$ 0.41 \\
\end{tabular}} 
    \resizebox{\linewidth}{!}{
      \begin{tabular}{lccccc}
\hline
        $\muu$ range    & Observed      & \multicolumn{1}{c}{Total}                 & DY     & Other prompt  & Jet mis-\\
\big[GeV\big] & yield & \multicolumn{1}{c}{background} & & lepton backgrounds & identification \\\hline  
60--120 & 164075 & 166000  $\pm$ 9360~~ & 165000 $\pm$ 9300~~ & 994 $\pm$ 89 & \NA  \\
120--400 & 977714 & 1050000 $\pm$ 60400~~ & 836000 $\pm$ 47000 & 210000 $\pm$ 19000 & 3070 $\pm$ 1540 \\
400--600 & 24041 & 26100 $\pm$ 1580 & 16700 $\pm$ 970~~ & 9120 $\pm$ 820 & 212 $\pm$ 106 \\
600--900 & 5501 & 5610 $\pm$ 337 & 4170 $\pm$ 250 & 1370 $\pm$ 120 & 74.0 $\pm$ 37.0 \\
900--1300 & 996 & 1050 $\pm$ 65~~ & 863 $\pm$ 52 & 169 $\pm$ 15 & 19.9 $\pm$ 10.0 \\
1300--1800 & 183 & 195 $\pm$ 13 & 169 $\pm$ 10 & 19.9 $\pm$ 1.8 & 6.7 $\pm$ 3.4 \\
$>$1800 & 42 & 44.3 $\pm$ 3.4 & 38.7 $\pm$ 2.5 & ~~3.3 $\pm$ 0.3 & 2.2 $\pm$ 1.1 \\
\hline
\end{tabular}}
\label{tab:datamcyieldseemumu}
\end{table}

\label{sec:results}

\section{Statistical interpretation}
\label{sec:interpretation}
Limits are calculated at 95\% confidence level (\CL) with Bayesian techniques known to have good frequentist coverage properties~\cite{PDG2020}, using the framework developed for statistically combining Higgs boson searches~\cite{CMS-NOTE-2011-005}, which is based on the \textsc{RooStats} package~\cite{Moneta:2010pm}. For the signal cross section, we use a positive uniform prior, while the nuisance parameters for the uncertainties in dilepton efficiencies, resolution, and scale are modeled with log-normal priors. Two different approaches are followed for the statistical interpretation of the results. To search for resonances in the dilepton invariant mass spectrum, an unbinned maximum likelihood fit is performed to the \mll spectrum in data, while for nonresonant signals a binned likelihood in \mll is constructed. In both cases the likelihood fit is done simultaneously for dielectron and dimuon events, years of data taking, and the $\abs{\eta}$ categories, within the two channels. To further increase the sensitivity to some of the models, the event sample in each of these categories is split into two bins, $\costheta < 0$ and $\costheta \geq 0$, in the nonresonant case.  The likelihoods for all subcategories are then combined to obtain the results.  When the electron and muon channels are combined, we assume that branching fractions of these two signals are the same.

\subsection{Search for resonant signals}
The signal is modeled with the convolution of a Breit--Wigner function to model the intrinsic decay width of the resonance and a double-sided Crystal Ball function~\cite{Oreglia:1980cs} to model the mass-dependent mass resolution, except for 2016 data in the muon channel, where a Gaussian function with exponential tails to either side is used~\cite{delAmoSanchez:2010ae}. Selection efficiency, mass resolution, and mass scale are taken into account as nuisance parameters in the signal description. The parameters are treated as uncorrelated between the different channels. The relative mass scale between them is the only uncertainty with a noticeable impact.
The background is modeled with two different functions for the two dilepton channels. In the dielectron channel it is given by:
\begin{linenomath} 
\begin{equation}
\label{eq:shapebckgELE}
\begin{split}
m^{\kappa}\exp\Biggl(\sum\limits_{i=0}^3 \alpha_{i} m^{i}\Biggr),\qquad \text{if $m \le m_{\text{threshold}}$}, \\
m^{\lambda}\exp\Biggl(\sum\limits_{i=0}^3 \beta_{i} m^{i}\Biggr),\qquad \text{if $m > m_{\text{threshold}} $},
\end{split}
\end{equation}
\end{linenomath} 
while for the dimuon channel the following functional form is used:
\begin{linenomath} 
\begin{equation}
\label{eq:shapebckgMUO}
\begin{split}
m^{\mu}\exp\Biggl(\sum\limits_{i=0}^2 \gamma_{i} m^{i}\Biggr),\qquad \text{if $m \le m_{\text{threshold}}$}, \\
m^{\nu}\exp\Biggl(\sum\limits_{i=0}^3 \delta_{i} m^{i}\Biggr),\qquad \text{if $m > m_{\text{threshold}}$}.
\end{split}
\end{equation}
\end{linenomath}
Here $m$ is the dilepton invariant mass; the other parameters do not represent physical quantities.
In the dimuon channel, these functions are required to be continuously differentiable at the transition point $m_{\mathrm{threshold}}$ between the low- and high-mass parameterizations, which is left free in the fits and ranges from 350 to 750\GeV. In the electron channel, $m_{\mathrm{threshold}}$ is set to 600\GeV.
They are first fit to the sum of the background estimates, the vast majority of which comes from simulation, as described in Section~\ref{sec:backgrounds}, for masses above 150\GeV. In the limit calculation, the resulting background shapes are then normalized to data within mass windows. The function parameters are treated as nuisance parameters and left floating in the limit calculation. Their initial values are set to those obtained from the fit to the background estimates and they are constrained by log-normal priors whose widths are set to the uncertainties obtained from the same fits. Correlations between the parameters are not considered by these constraints.    

The limits are calculated in a mass window of $\pm$4 times the signal width, defined as the sum of intrinsic width and mass resolution, with this window being symmetrically enlarged until there is a minimum of 100 data events in it. This sets an upper limit of 10\% on the statistical uncertainty in the local background estimate in the mass window. It is chosen to be large enough to  dominate the expected systematic uncertainties in the background shape at high mass, which do not have to be considered explicitly for this reason. To allow the background yield to be constrained by its statistical uncertainties, a log-normal prior with a width of three times these uncertainties is used. 

The results have been found to be largely robust against the impact of any remaining uncertainty in the background shape, as well as the choice of background shape parameterization for the full range of considered resonance masses, in the case of a narrow resonance. However, for the largest signal width hypothesis considered, 10\%, a significant bias in the limit values for low resonance masses from the choice of the parameterization can occur. Therefore, no results below a resonance mass of 700\GeV are shown for this width.

The parameter of interest is chosen to be the ratio of the cross section for dilepton production via a \PZpr boson to the observed cross section for $\PZ \to \Pell\Pell$ in the data control region of $60 < \mll < 120\GeV$, $R_{\sigma}$:
\begin{linenomath} 
\begin{equation}
\label{eq:rsigma}
R_\sigma = \frac{\sigma(\Pp\Pp\to \cPZpr+X\to\Pell\Pell+X)}
                {\sigma(\Pp\Pp\to \cPZ+X  \to\Pell\Pell+X)}.
\end{equation}
\end{linenomath}                 
This variable has reduced dependence on the theoretical predictions for the \PZpr cross section and on systematic uncertainties that are correlated between high and low \mll, such as trigger and selection efficiencies, for which only uncertainties in their stability at high mass remain. To ease comparison with the predicted cross section for various signal models, the limits on $R_\sigma$ are presented multiplied by the theoretical prediction for $\sigma(\Pp\Pp\to \cPZ+X\to\Pell\Pell+X)$ of 1928\unit{pb}, calculated at NNLO in QCD and NLO in EWK theory with the \FEWZ~3.1 program~\cite{Li:2012wna}.

\begin{figure}[htb]
\centering
\includegraphics[width=0.49\textwidth]{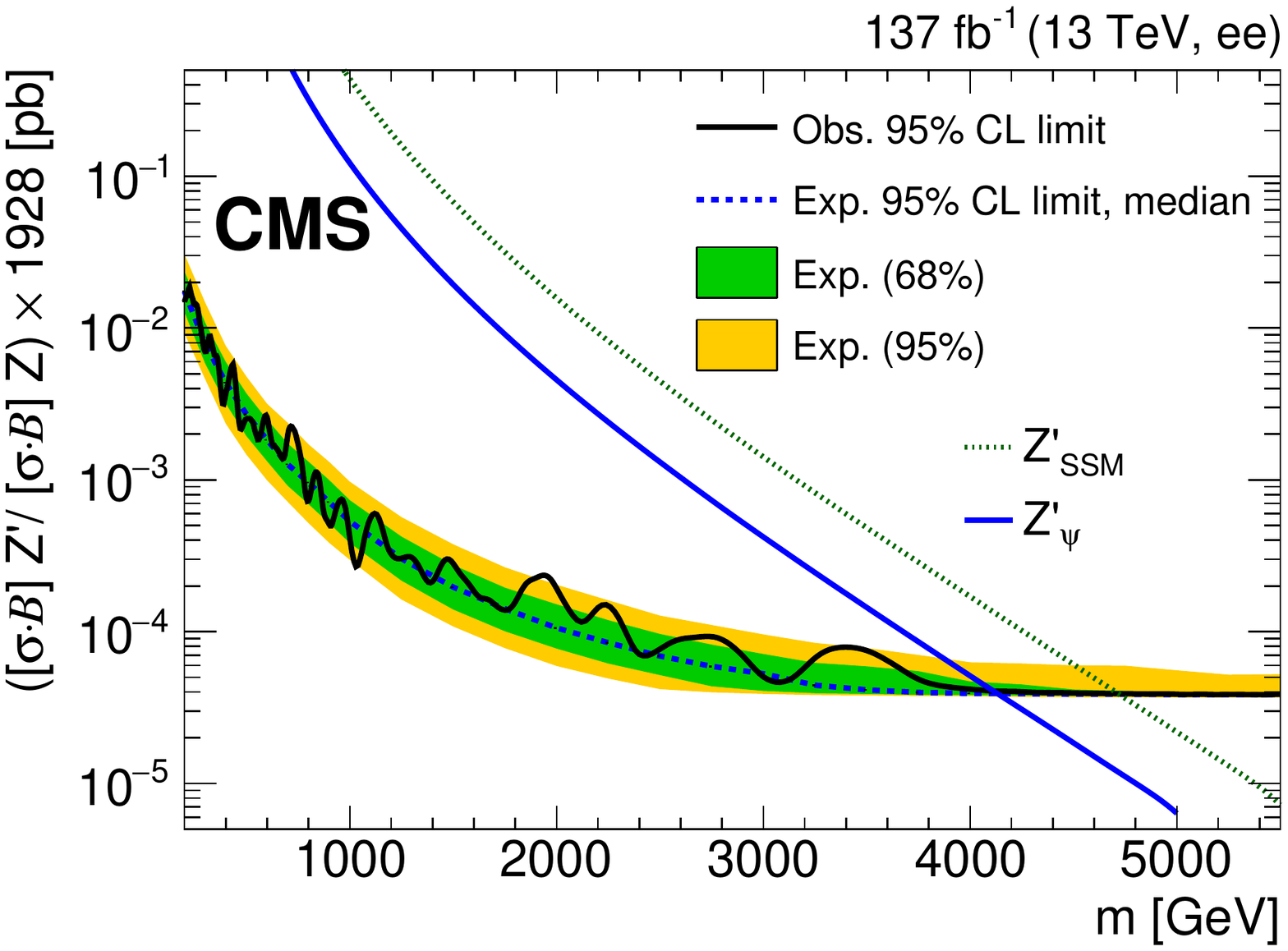}
\includegraphics[width=0.49\textwidth]{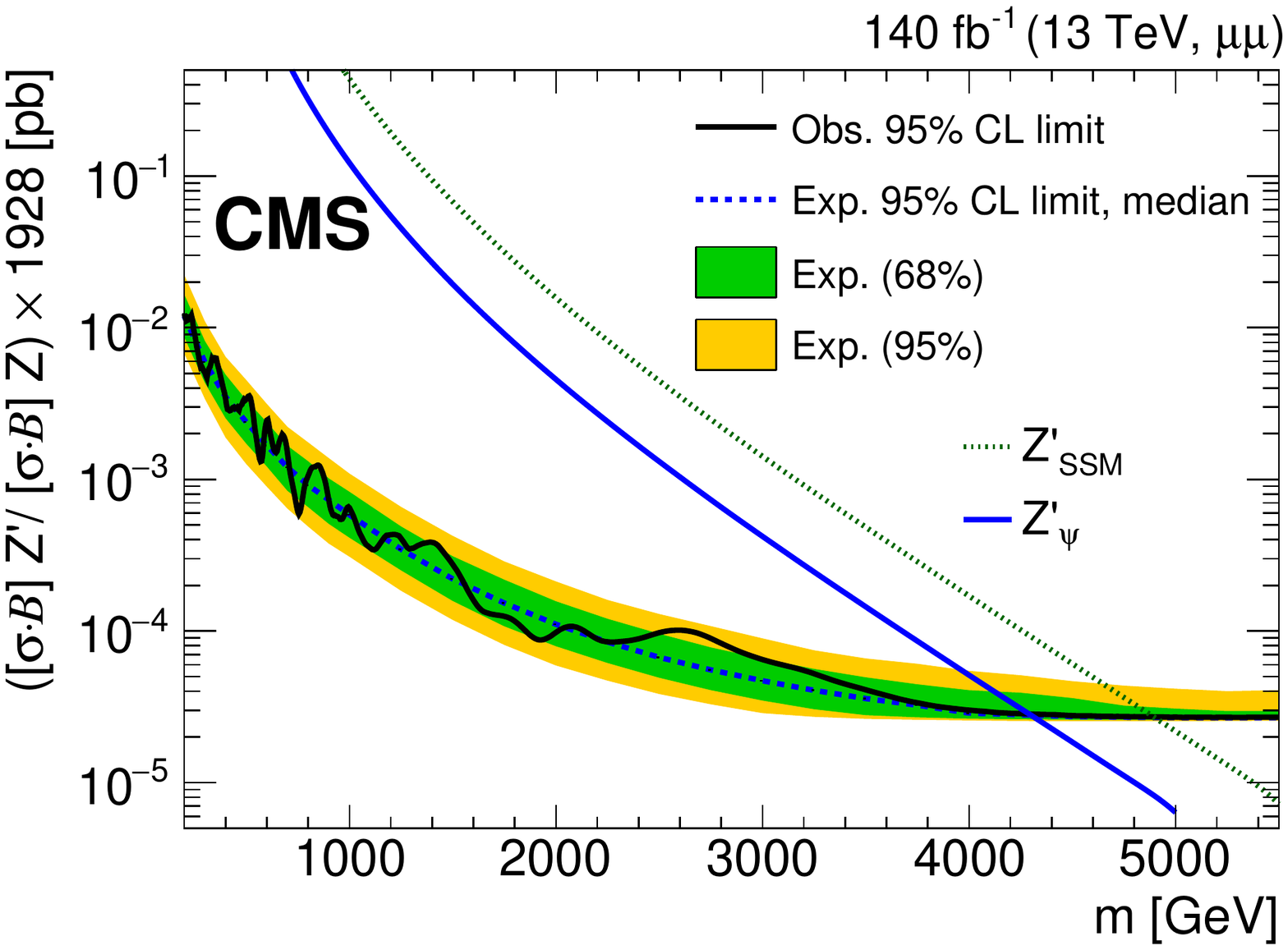}
\includegraphics[width=0.49\textwidth]{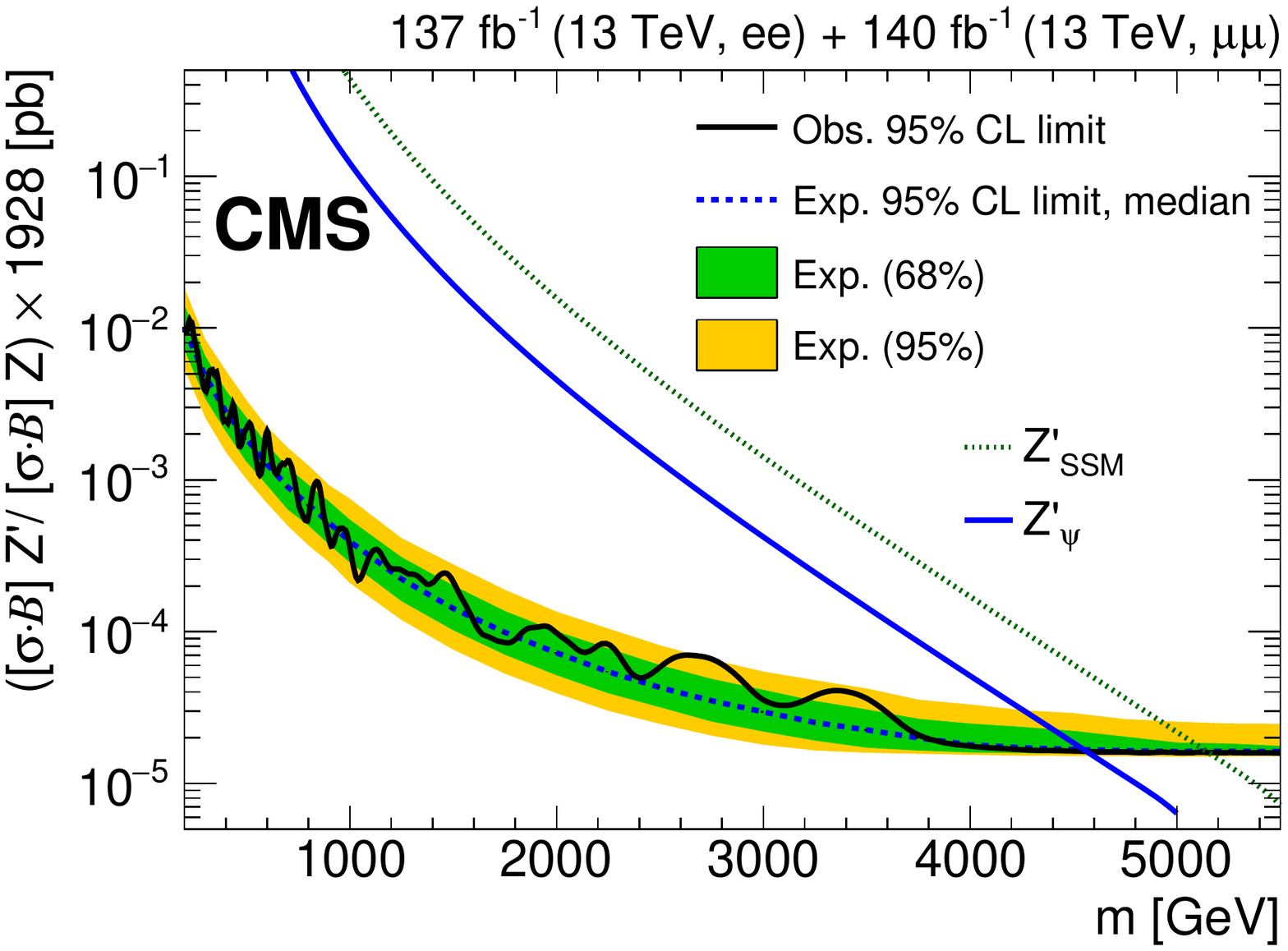}
\caption{
The upper limits at 95\% \CL on the product of the production cross section and the branching fraction for a spin-1 resonance with a width equal to 0.6\% of the resonance mass, relative to the product of the production cross section and the branching fraction of a \PZ boson, multiplied by the theoretical value of $\sigma(\Pp\Pp\to \cPZ+X\to\Pell\Pell+X)$ of 1928\unit{pb},  for (top left) the dielectron channel, (top right) the dimuon channel, and (bottom) their combination.
The shaded bands correspond to the 68 and 95\% quantiles for the expected limits.
Simulated predictions for the spin-1 \ZPSSM and \ZPPSI resonances are shown for comparison.
}
\label{fig:limits2}
\end{figure}

The resulting limits for a narrow resonance with an intrinsic width of 0.6\% are shown in Fig.~\ref{fig:limits2}. At high mass, where the background estimate approaches zero, the limits are independent of the resonance width. The limits are therefore applicable for both \ZPSSM and \ZPPSI. The observed limits are consistent with the expectation from SM backgrounds. The expected and observed mass limits in the \ZPSSM and \ZPPSI models are shown in Table~\ref{tab:massLimitsSpin1}. For the combination of the dielectron and dimuon channels, the observed limits are 5.15\TeV for the \ZPSSM and 4.56\TeV for the \ZPPSI, well above the highest mass event observed in the data. For these models, the ATLAS Collaboration has recently set similar limits~\cite{Aad:2019fac}.

\begin{table}[!hbt]
\centering
\topcaption{
The observed (Obs.) and expected (Exp.) 95\% \CL lower limits on the masses of spin-1 \ZPSSM and \ZPPSI bosons, assuming a signal width of 0.6 (3.0)\% of the resonance mass for \ZPPSI (\ZPSSM).
}
\begin{tabular}{lcccc}
\hline
\multirow{2}{*}{Channel}  & \multicolumn{2}{c}{\ZPSSM} & \multicolumn{2}{c}{\ZPPSI}  \\
                          & Obs. [\TeVns{}] & Exp. [\TeVns{}]      & Obs. [\TeVns{}]  & Exp. [\TeVns{}]      \\\hline
$\Pe\Pe$                    &  4.72      & 4.72            & 4.11        & 4.13            \\
$\PGm\PGm$              &  4.89      & 4.90            & 4.29        & 4.30            \\
$\Pe\Pe$ ~+ $\PGm\PGm$    &  5.15     & 5.14            & 4.56        & 4.55            \\
\hline
\end{tabular}
\label{tab:massLimitsSpin1}
\end{table}

\begin{figure}[htb]
\centering
\includegraphics[width=0.49\textwidth]{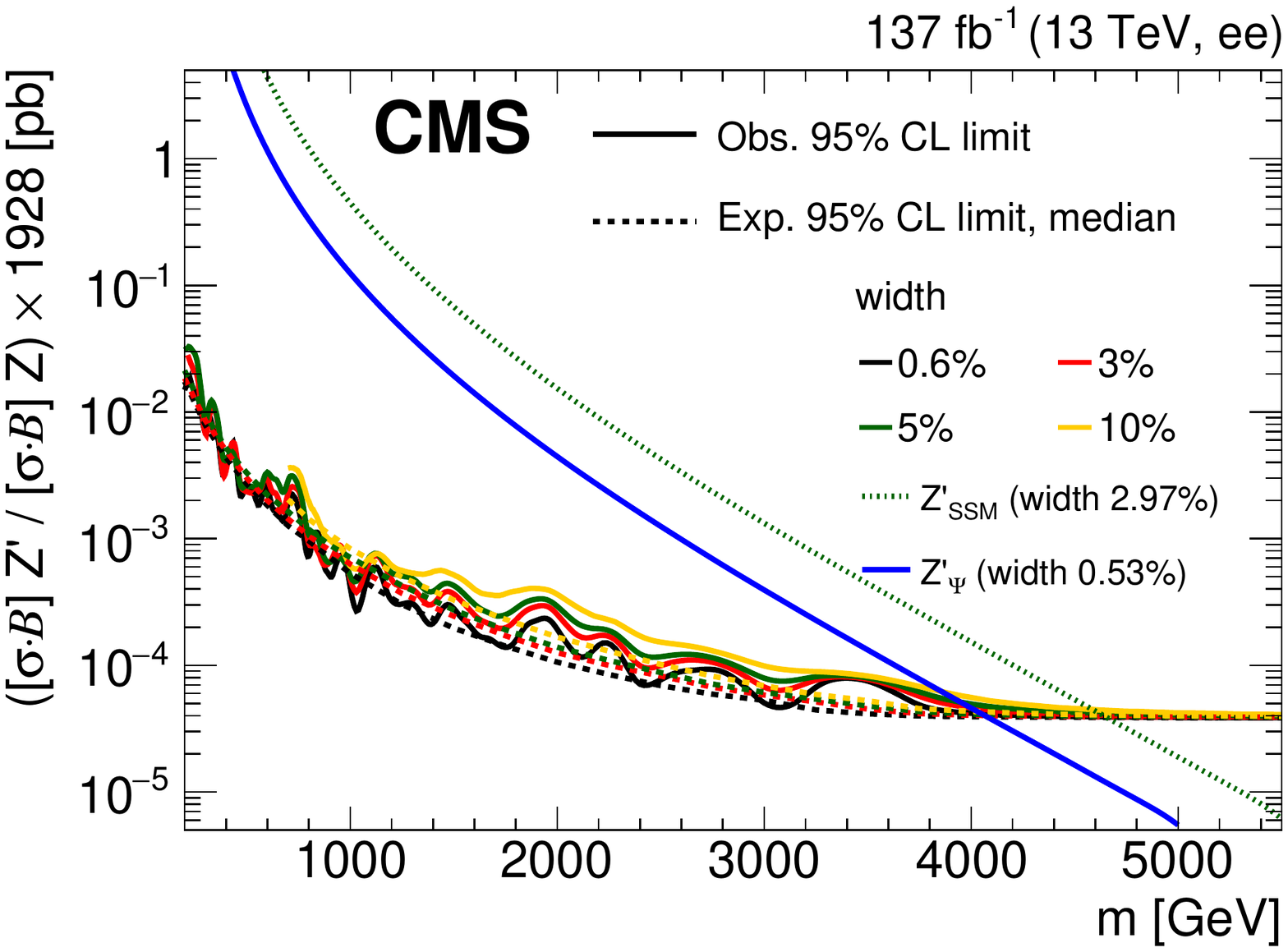}
\includegraphics[width=0.49\textwidth]{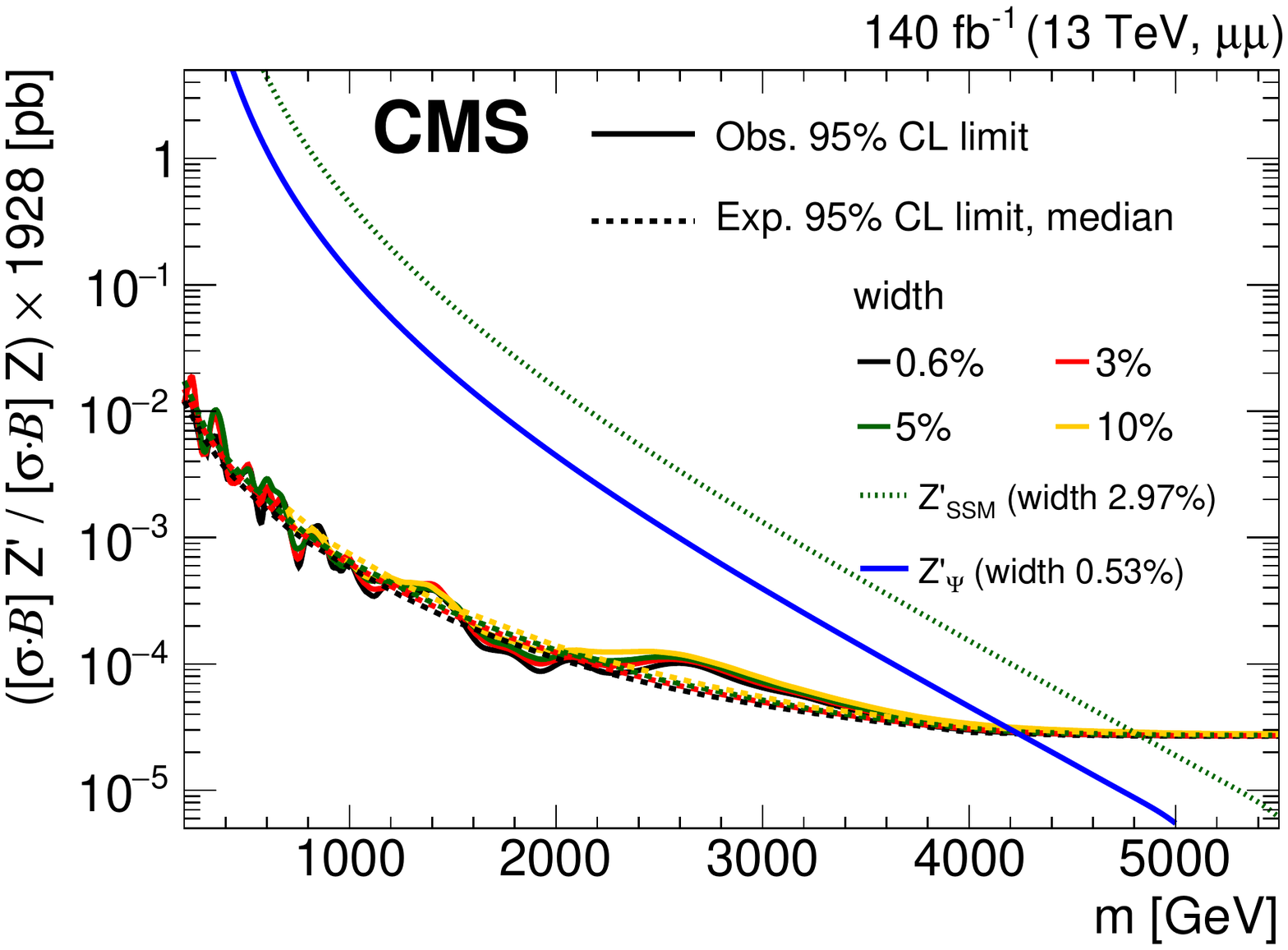}
\includegraphics[width=0.49\textwidth]{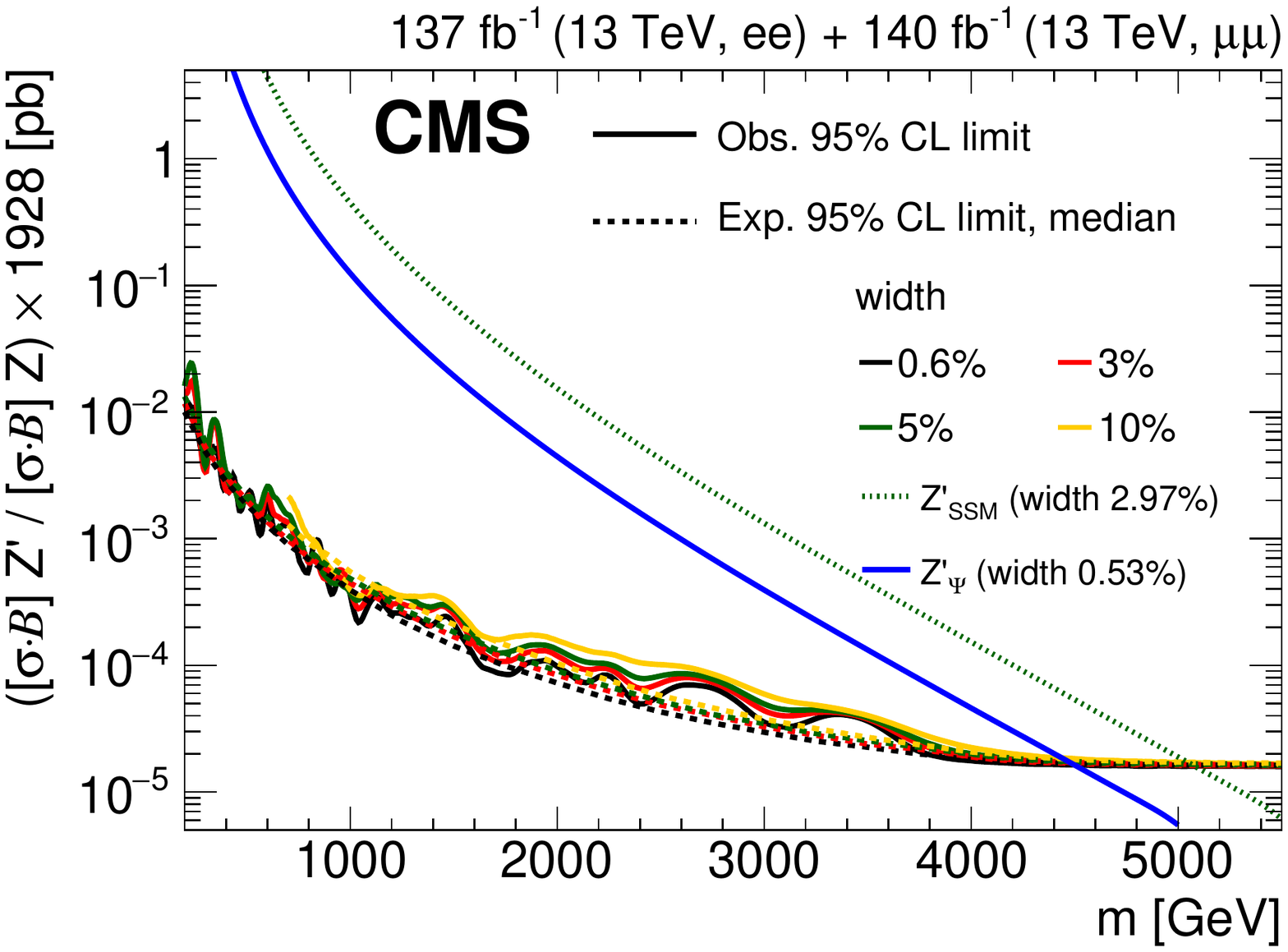}
\caption{
The upper limits at 95\% \CL on the product of the production cross section and the branching fraction for a spin-1 resonance, for widths equal to 0.6, 3, 5, and 10\% of the resonance mass, relative to the product of the production cross section and the branching fraction for a \PZ boson, multiplied by the theoretical value of $\sigma(\Pp\Pp\to \cPZ+X\to\Pell\Pell+X)$ of 1928\unit{pb}, for (upper left) the dielectron channel, (upper right) the dimuon channel, and (lower) their combination.
Theoretical predictions for the spin-1 \ZPSSM and \ZPPSI resonances are also shown.
}
\label{fig:limits3}
\end{figure}

Limits are computed for several hypotheses for the resonance width. In Fig.~\ref{fig:limits3}, expected and observed limits are shown for 0.6\% (identical to curves in Fig.~\ref{fig:limits2}), 3, 5, and 10\%. As discussed above, lower thresholds on the resonance mass are chosen to avoid biasing the limits due to the choice of background shape for the wider resonances. The thresholds are 200\GeV for 0.6, 3, and 5\%, and, as mentioned earlier, 700\GeV for 10\% width. The limit curves for the different widths converge for high resonance masses, where the background expectation approaches zero. For lower masses, the limits become weaker with increasing width as wider resonances are not as easily distinguished from the background. The features in the observed limit curves corresponding to statistical fluctuations in the observed data are smoothed out for wider resonances. 

\begin{figure}[htb]
\centering
\includegraphics[width=0.65\textwidth]{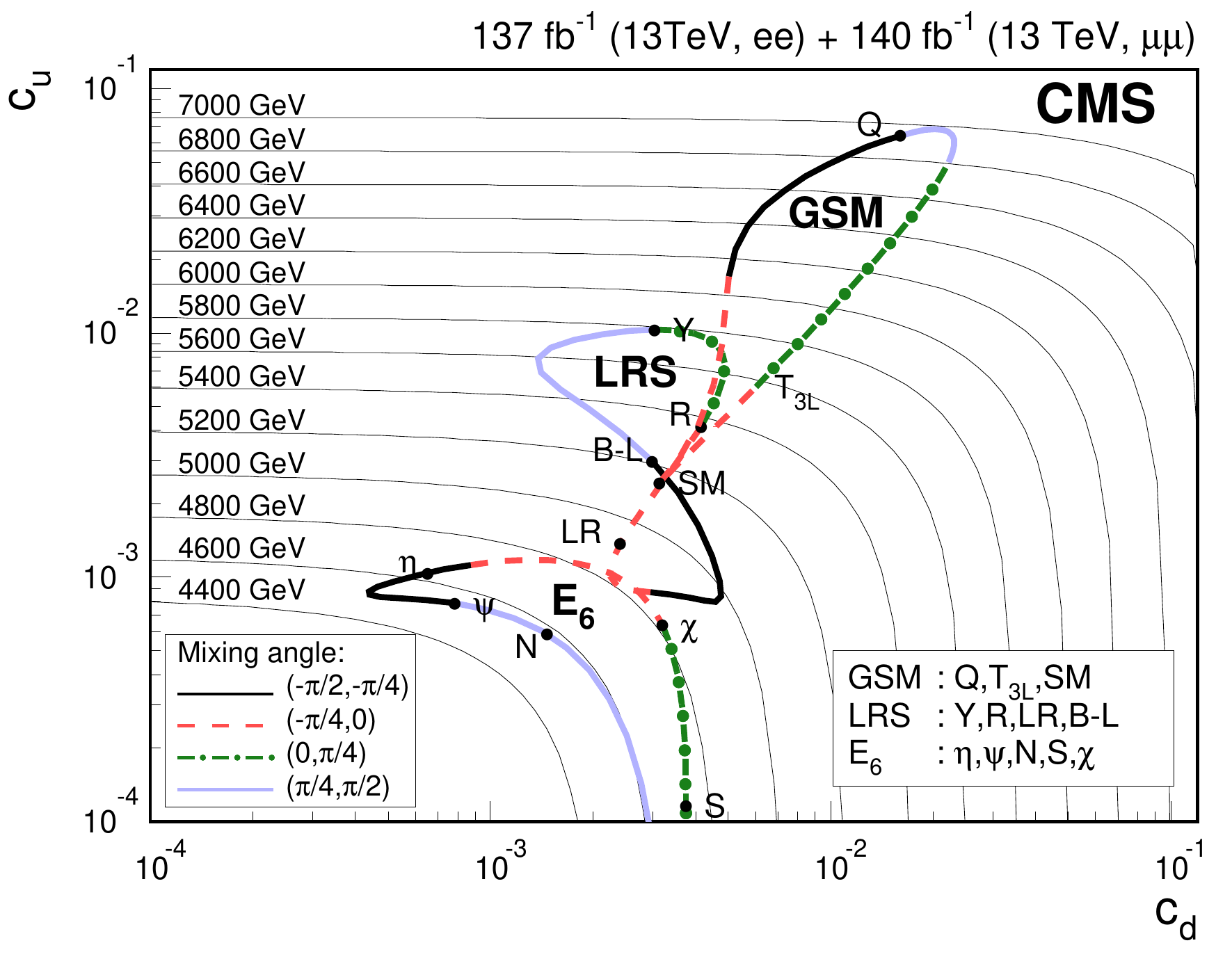}
\caption{
Lower limits in the $(c_{\PQd},c_{\PQu})$ plane obtained by recasting the combined limit at 95\% \CL on the \PZpr boson cross section for narrow resonances from dielectron and dimuon channels.
For a given \PZpr boson mass, the cross section limit results in a solid thin black line.
These lines are labeled with the relevant \PZpr boson masses.
The closed contours representing the GSM, LRS, and E$_6$ model classes are composed of thick curve segments.
Each point on a segment corresponds to a particular model, and the location of the point gives the mass limit on the relevant \PZpr boson.
As indicated in the lower left legend, the curve segment styles correspond to ranges of the particular mixing angle, for each considered model.
The lower right legend indicates constituents of each model class.
}
\label{fig:CuCd}
\end{figure}

The limits in the narrow width approximation are interpreted as limits on the generalized couplings in the  ($c_\PQd, c_\PQu$) plane as shown in Fig.~\ref{fig:CuCd} for the combination of the dielectron and dimuon channels. The closed colored curves show the observed lower limits in the three classes of models (GSM, LRS, and $E_6$), as functions of the mixing angle within each class. For a given point in the plane, a lower mass limit can be obtained with the help of the thin black lines, which are obtained from the observed limit and the ratio of parton luminosities~\cite{Accomando:2010fz}. The observed limit exceeds 4.5\TeV for all models considered and reaches close to 7\TeV in the GSM class of models.

To quantify any discrepancies between the observed data and the background estimates, the $p$-value for the background-only hypothesis is calculated, as a function of resonance mass hypothesis, following the methodology described in Ref.~\cite{Chatrchyan:2012xdj}. The same mass windows as for the limit calculation are used and the resulting $p$-values therefore depend on the assumed resonance width. Similar to the exclusion limits computed for different width hypotheses, lower mass thresholds are determined based on the potential bias in the fitted number of signal events introduced by the choice of the background parameterization. As this extraction of signal yields is found to be more susceptible to these biases, the resulting thresholds are higher than for the limit setting. The threshold is 200\GeV for a width of 0.6\%, 600\GeV for a width of 3\%, and 1\TeV for widths of 5 and 10\%. 

\begin{figure*}[htb]
\centering
\includegraphics[width=0.49\textwidth]{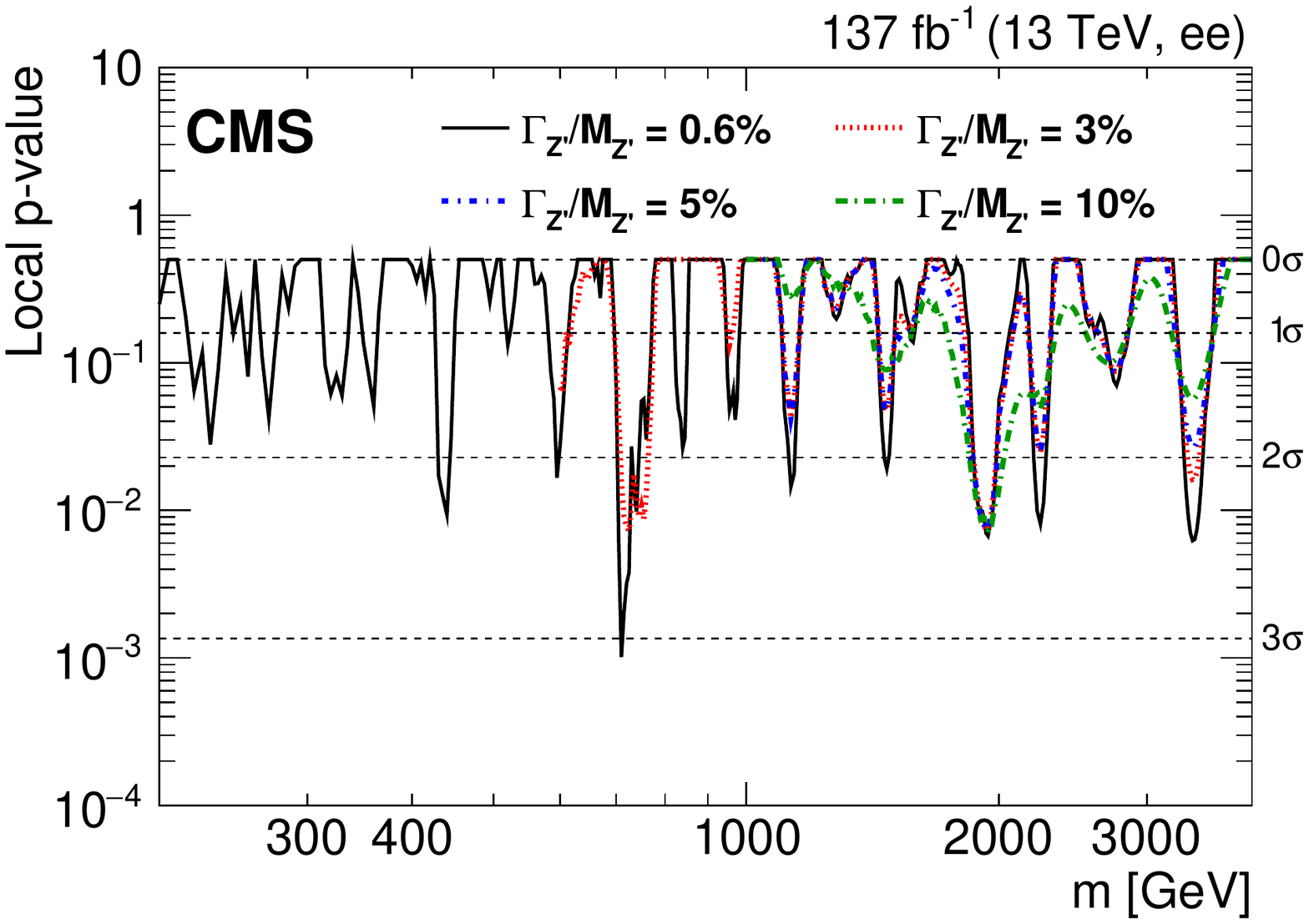} \hfill
\includegraphics[width=0.49\textwidth]{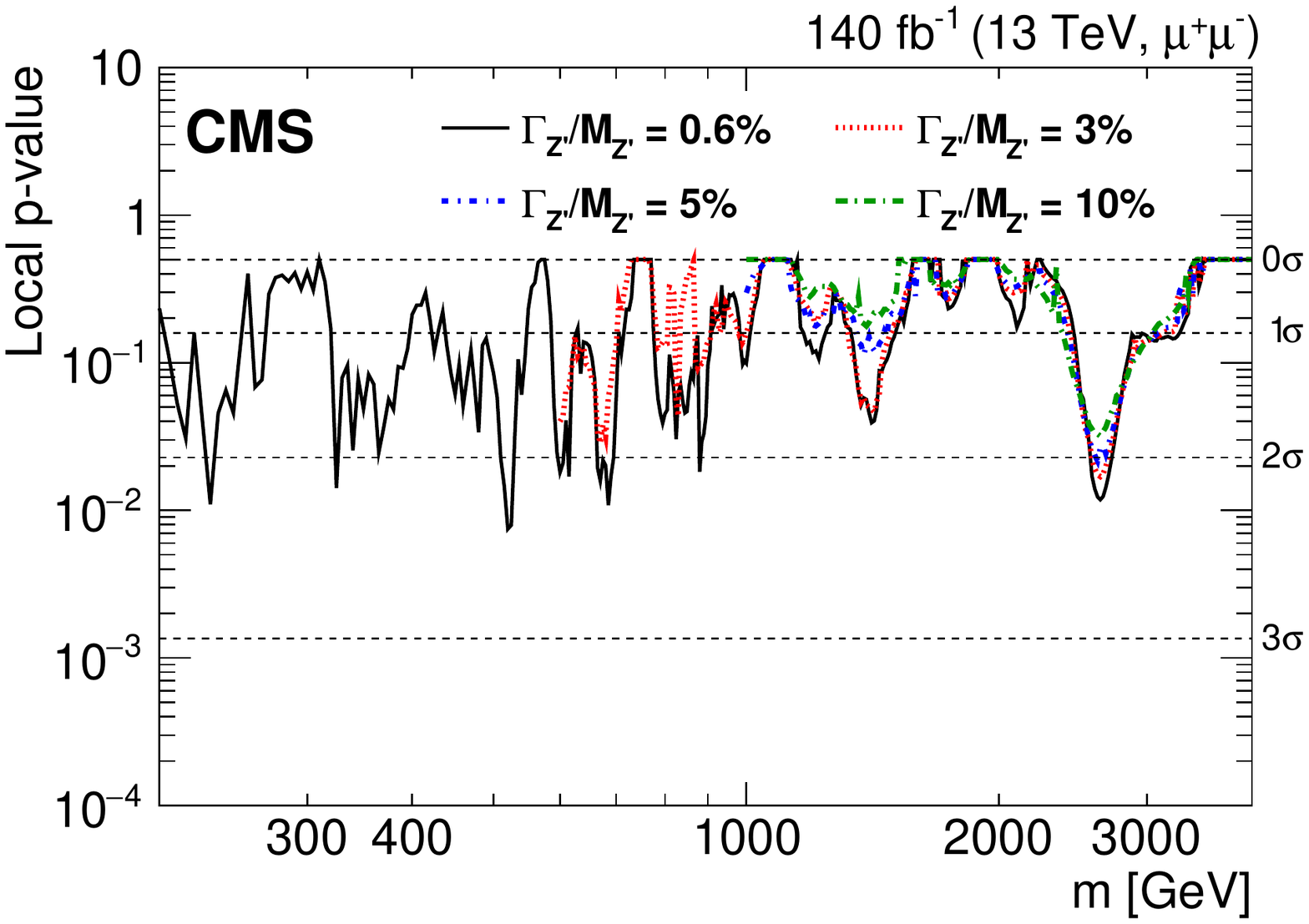}
\includegraphics[width=0.49\textwidth]{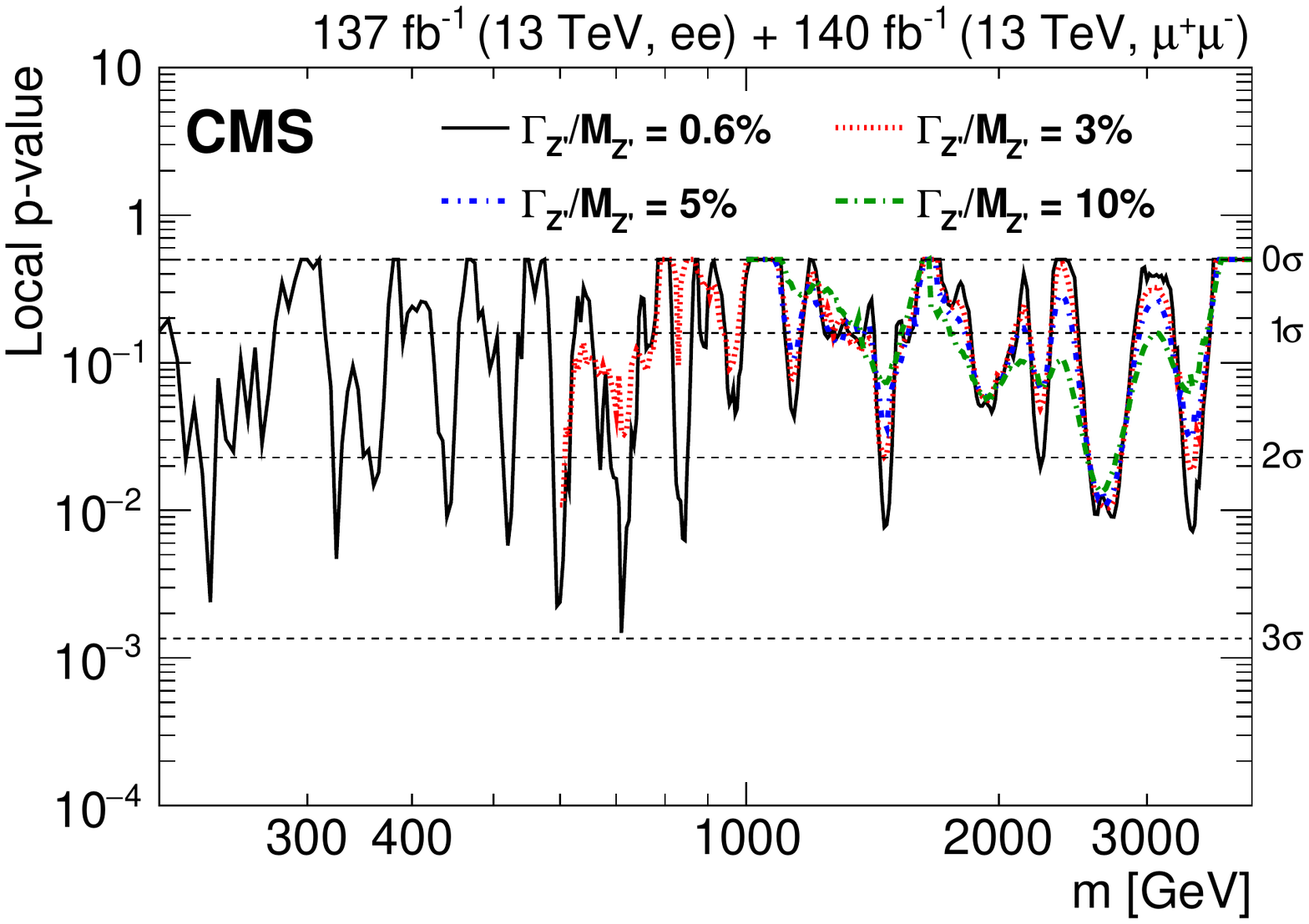}
\caption{
The observed local $p$-value for a given resonance mass hypothesis for (upper left) the dielectron channel, (upper right) the dimuon channel, and (lower) their combination, as a function of the dilepton invariant mass. The four different lines correspond to different signal width hypotheses.
}
\label{fig:pvalue}
\end{figure*}

The results are shown in Fig.~\ref{fig:pvalue}. In the narrow width case, the most significant deviations in the dimuon channel are observed at 520\GeV with a local $p$-value of 0.0074, corresponding to a local significance of 2.4 standard deviations. The local $p$-value corresponds to the hypothesis in which a peak at the observed location (520\GeV in this case) is predicted. In both the dielectron channel and the combination, the largest deviation is located at 710\GeV, with local $p$-values of 0.0010 and 0.0014, respectively, corresponding to local significances of 3.1 and 3.0 standard deviations. By repeating the $p$-value scan on an ensemble of toy data sets generated from the background parameterizations, a global $p$-value can also be computed, giving the probability of finding a local $p$-value as great or greater than that observed, anywhere in a specified mass range.  The global $p$-value for the combination of the two channels is 0.76 in the mass range 500 to 5500\GeV. For such one-tailed tests (for which a $p$-value of 0.5 corresponds to 0 standard deviations) this formally corresponds to ${-}$1.4 standard deviations. Focusing on the vicinity of the observed fluctuation, the global $p$-value in the mass range 700 to 800\GeV is 0.19, corresponding to 0.9 standard deviations.

\begin{figure}[!htb]
\centering
\includegraphics[width=0.49\textwidth]{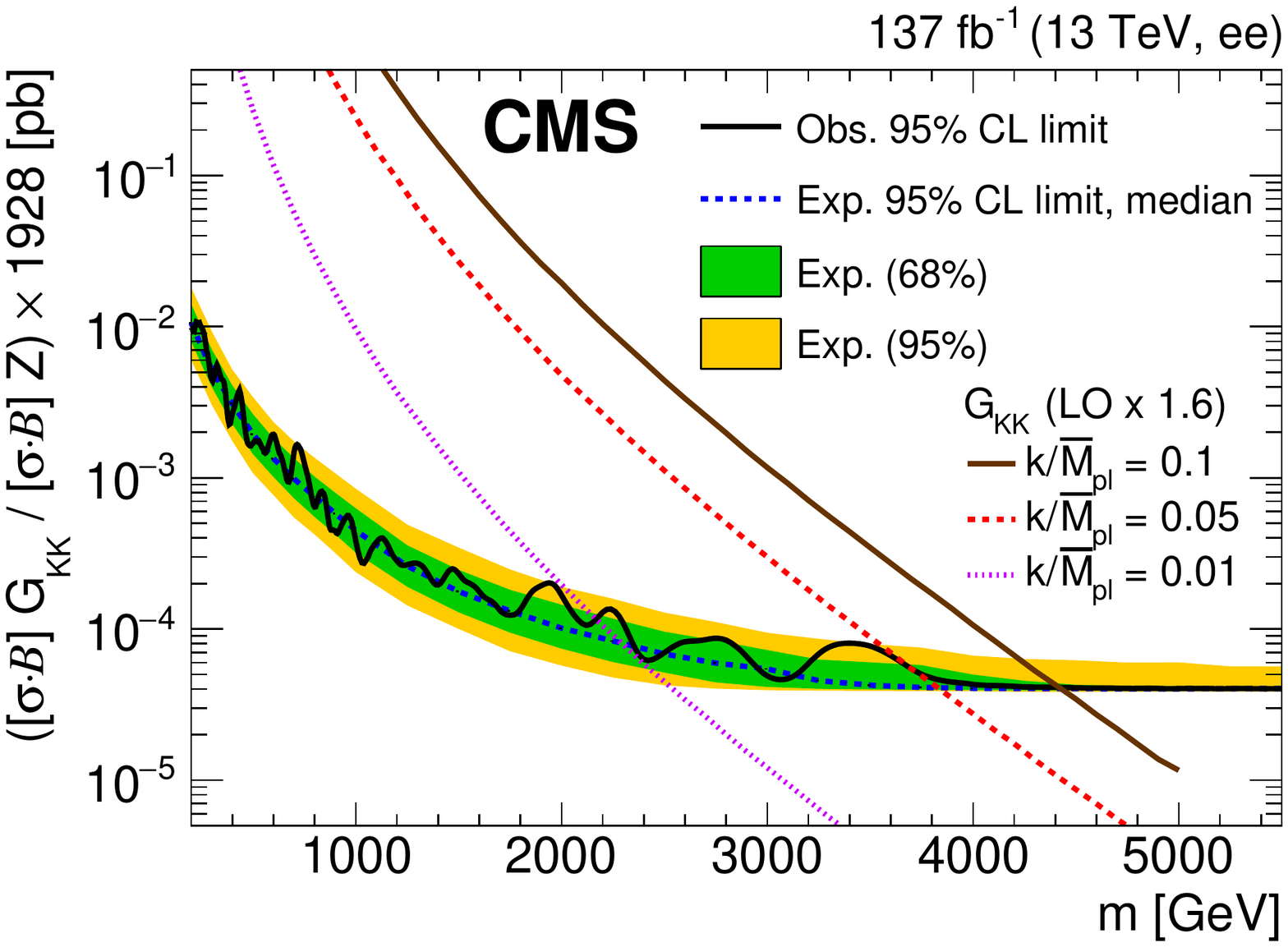}
\includegraphics[width=0.49\textwidth]{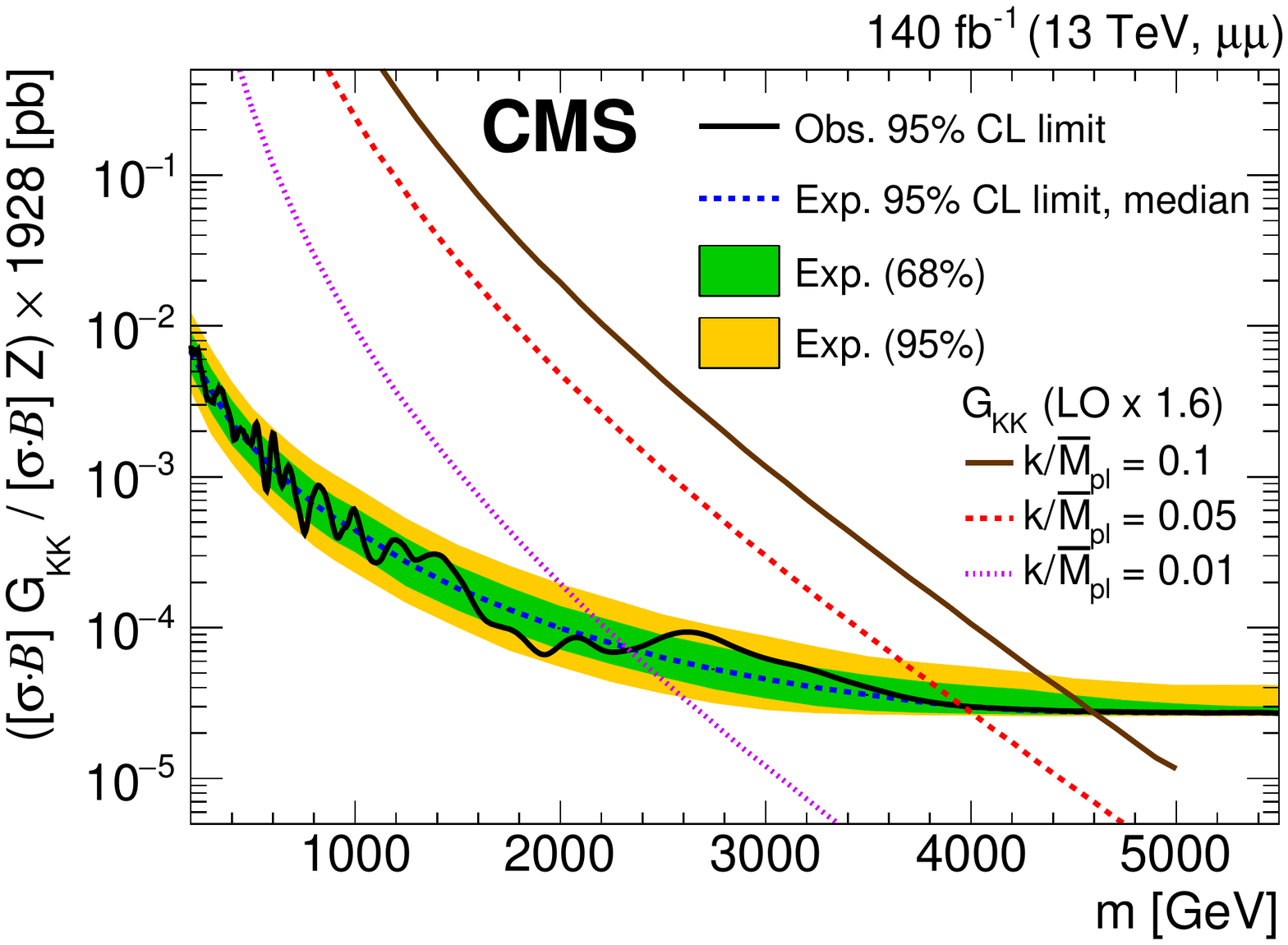}
\includegraphics[width=0.49\textwidth]{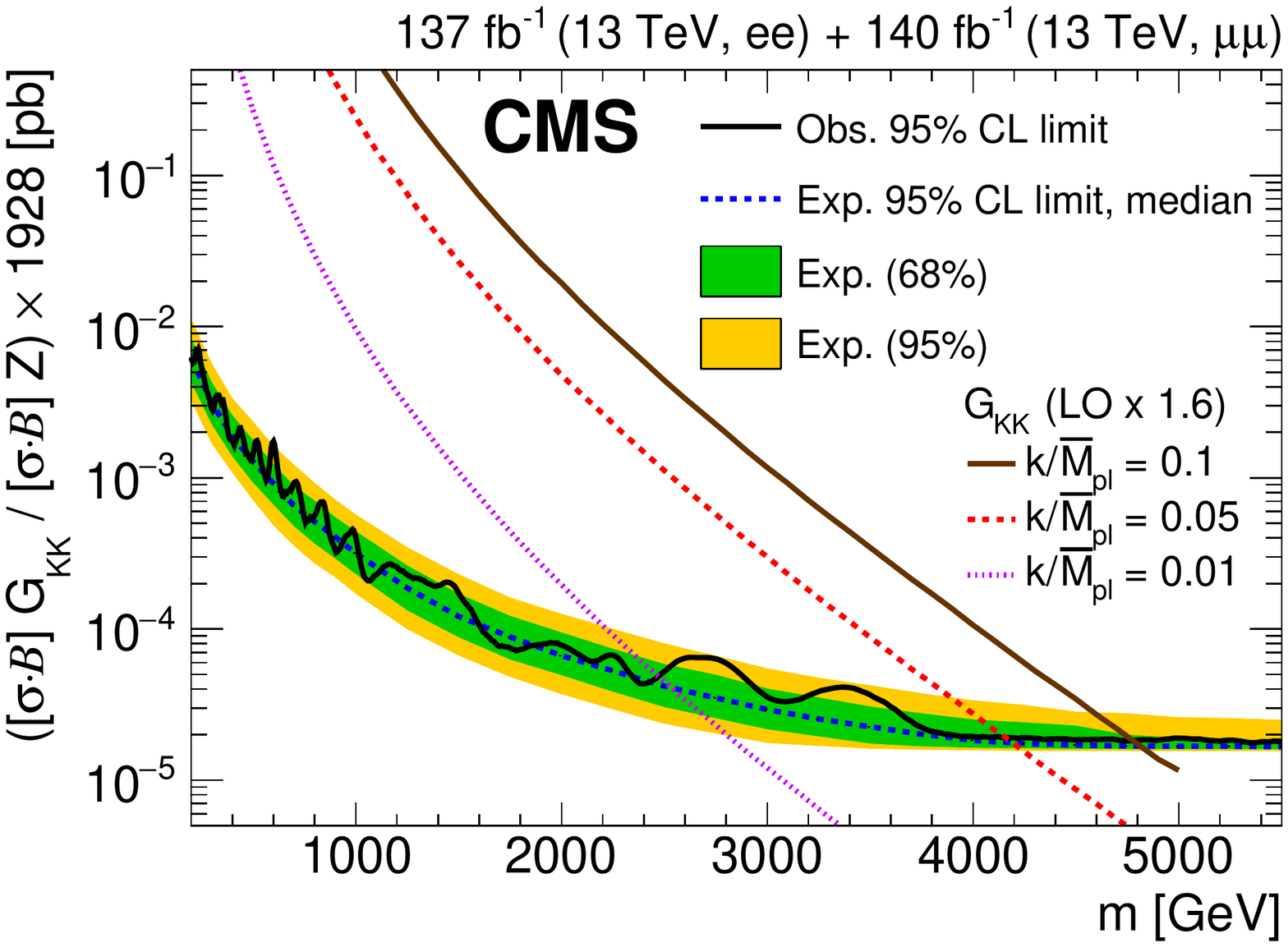}
\caption{
The upper limits at 95\% \CL on the product of the production cross section and the branching fraction for a spin-2 resonance, relative to the product of the production cross section and the branching fraction of a \PZ boson, multiplied by the theoretical value of $\sigma(\Pp\Pp\to \cPZ+X\to\Pell\Pell+X)$ of 1928\unit{pb}, for (upper left) the dielectron channel, (upper right) the dimuon channel, and (lower) their combination.
The shaded bands correspond to the 68 and 95\% quantiles for the expected limits.
Theoretical predictions for the spin-2 resonances for widths equal to 0.01, 0.36, and 1.42\% corresponding to coupling parameters $k/\overline{M}_\mathrm{Pl}$ of 0.01, 0.05, and 0.10, respectively, are shown for comparison.
}
\label{fig:limitsRS}
\end{figure}

In addition to the spin-1 resonances considered thus far, limits are also calculated for spin-2 resonances and the results are shown in Fig.~\ref{fig:limitsRS}. The calculation is performed assuming a narrow width and differs from that for spin-1 resonances only in the detector acceptance for the different spin configurations. The limits in the spin-2 case are more stringent at low mass as the leptons are produced more centrally in the detector, which increases the acceptance, especially in the electron channel where events with two electrons in the endcaps are rejected. Lower mass limits for three example values of the coupling parameter $k/\overline{M}_\mathrm{Pl}$ of 0.01, 0.05, and 0.10, corresponding to intrinsic widths of 0.01, 0.36, and 1.42\%, are shown in Table~\ref{tab:massLimitsSpin2} and range from slightly over 2 to almost 5\TeV. This improves the previous CMS limits by about 370 to 530\GeV.

\begin{table}[!hbt]
\centering
\topcaption{
The observed and expected 95\% \CL lower limits on the masses of spin-2 resonances for widths equal to 0.01, 0.36, and 1.42\%, corresponding to coupling parameters $k/\overline{M}_\mathrm{Pl}$ of 0.01, 0.05, and 0.10.
}
\begin{tabular}{lcccccc}
\hline
\multirow{2}{*}{Channel}  & \multicolumn{2}{c}{$k/\overline{M}_\mathrm{Pl} = 0.01$} & \multicolumn{2}{c}{$k/\overline{M}_\mathrm{Pl} = 0.05$}  & \multicolumn{2}{c}{$k/\overline{M}_\mathrm{Pl} = 0.1$}\\
                          & Obs. [\TeVns{}] & Exp. [\TeVns{}]      & Obs. [\TeVns{}]  & Exp. [\TeVns{}]   & Obs. [\TeVns{}]  & Exp. [\TeVns{}]      \\\hline
$\Pe\Pe$                    &  2.16      & 2.29            & 3.70        & 3.83   & 4.42  & 4.43       \\
$\PGm\PGm$             &  2.34      & 2.32            & 3.96        & 3.96   & 4.59  & 4.59        \\
$\Pe\Pe$ ~+ $\PGm\PGm$    &  2.47     & 2.53            & 4.16        & 4.19   & 4.78  & 4.81        \\
\hline
\end{tabular}
\label{tab:massLimitsSpin2}
\end{table}

The results are also interpreted in the context of the simplified DM models described in Section~\ref{sec:models}. For this purpose, dilepton production cross sections in these models are calculated at NLO in QCD using the \textsc{DMsimp} implementation~\cite{Backovic:2015soa} of the simplified model in \MGvATNLO version 2.6.5~\cite{Alwall:2014hca}. The partial and total mediator decay widths are included using the \textsc{MadWidth} package~\cite{Alwall:2014bza} and are calculated at LO. In case of the axial-vector scenario with coupling $g_\text{DM}=1.0$ and $g_{\PQq}=g_{\Pell}=0.1$, and assuming $m_\text{DM} > m_{\text{med}}/2$, the production cross section for electron or muon pairs within detector acceptance ranges from approximately 100\unit{pb} for $m_{\text{med}}$ around 200\GeV to 0.1\unit{fb} for $m_{\text{med}}$ around 4\TeV.

\begin{figure*} [htb]
\centering
\includegraphics[width=0.49\textwidth]{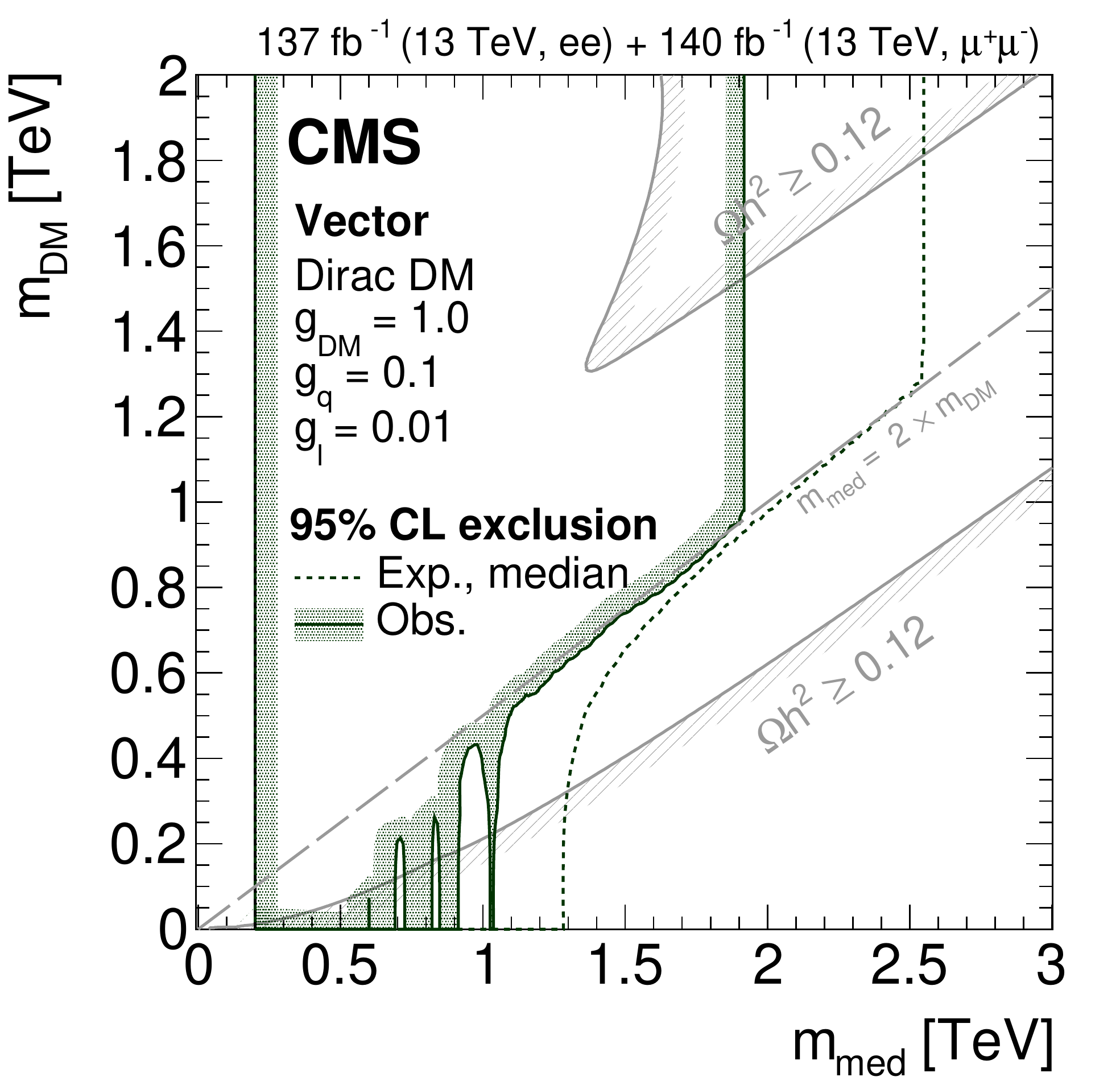} \hfill
\includegraphics[width=0.49\textwidth]{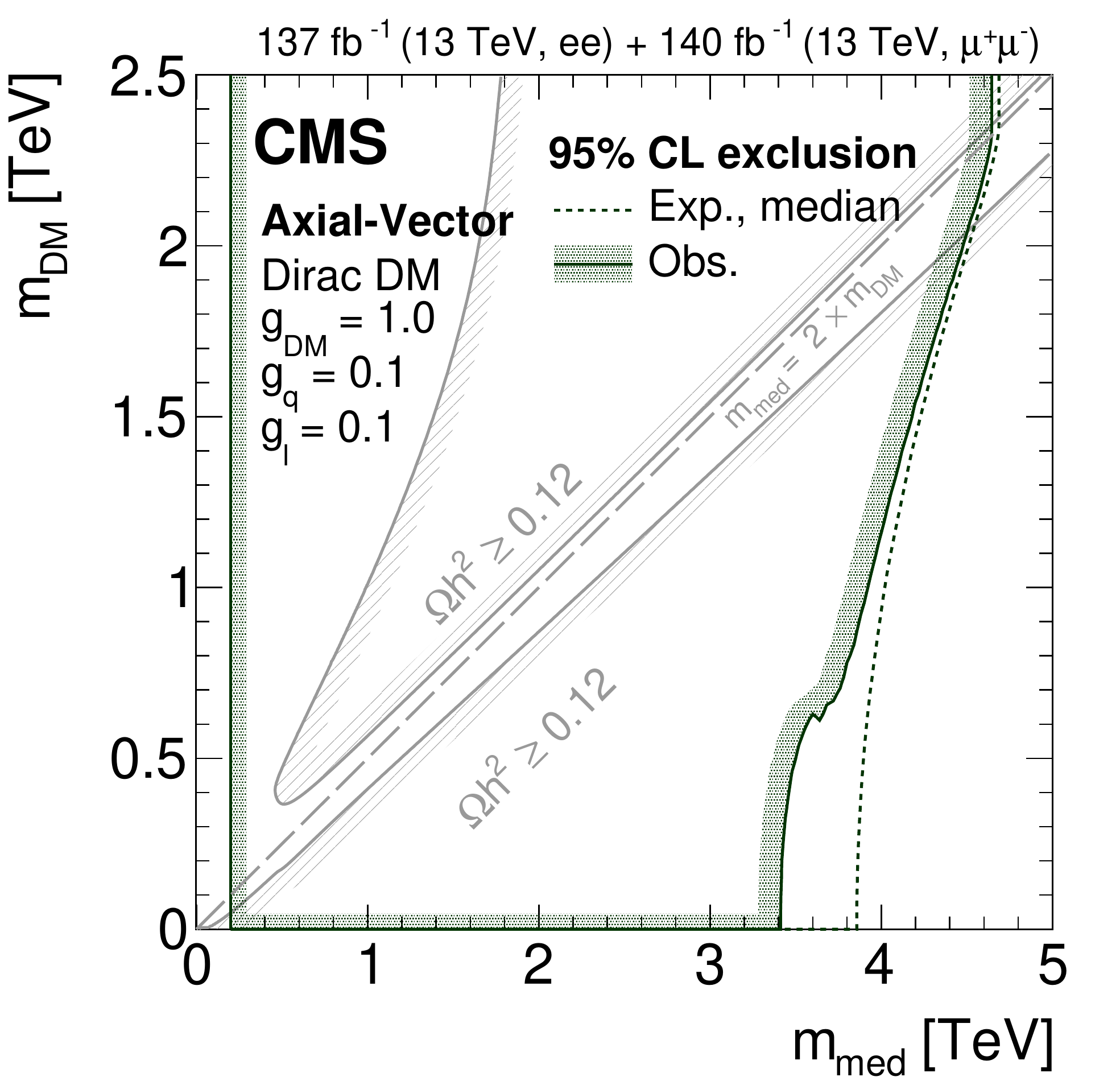}
\caption{
Summary of upper limits at 95\% CL on the masses of the DM particle, which is assumed to be a Dirac fermion, and its associated mediator, in a simplified model of DM production via a (left) vector or (right) axial-vector mediator.
The parameter exclusion regions are obtained by comparing the limits on the product of the production cross section and the branching fraction for decay to a \PZ boson with the values obtained from calculations in the simplified model.
For each combination of the DM particle and mediator mass values, the width of the mediator is taken into account in the limit calculation.
The curves with the hatching represent the excluded regions.
The solid gray curves, marked as ``$\Omega h^{2} \ge 0.12$'', correspond to parameter regions that reproduce the observed DM relic density in the universe~\cite{Backovic:2015tpt,Backovic:2013dpa,Ade:2015xua,Albert:2017onk}, with the hatched area indicating the region where the DM relic abundance exceeds the observed value.
}
\label{fig:DM}
\end{figure*}

In addition to the size of the mediator's coupling to leptons, the sensitivity of this search also depends on $m_\text{DM}$ relative to $m_{\text{med}}$. For small $m_\text{DM}$, the mediator will dominantly decay to DM particles. This decay becomes suppressed for  $m_\text{DM} > m_{\text{med}}/2$, enhancing the decays into leptons and increasing the sensitivity of the dilepton channel. Additionally, the decay width of the mediator is larger if the decay into DM particles is available. Therefore, the width of the mediator changes over the    $m_\text{med}$-$m_{\text{DM}}$ plane. Expected and observed limits are therefore calculated for widths between 0.5 and 3.5\% in steps of 0.25\%. To determine if a point in the mass plane is excluded, the limit obtained with the width closest to the theoretical value for that point is used. The resulting exclusion contours are shown in Fig.~\ref{fig:DM} for both the vector and axial-vector coupling models. The limits are strongest for large values of $m_\text{DM}$ where the decay width of the mediator is small. Here, mediators with masses below 1.92 (4.64)\TeV are excluded in the vector (axial-vector) model. For $m_\text{DM} = 0$, the limit is 3.41\TeV in the axial-vector model. In the vector model, the largest excluded mediator mass for $m_\text{DM} = 0$ is 1.04\TeV. However, due to fluctuations in the observed limit, not all masses below that value are excluded. The ATLAS Collaboration has recently set similar limits~\cite{ATL-PHYS-PUB-2020-021}.

\subsection{Search for nonresonant signals}
\label{sec:nonreslimits}
To set limits on nonresonant signal models, the parameter of interest in the statistical analysis is the
ratio $\mu$ of observed to predicted signal cross section. The dilepton mass spectra are divided into multiple exclusive bins. The bin width is optimized separately for the CI and ADD models based on the expected limits. For CI, bins with lower edges of 400, 500, 700, 1100, 1900, and 3500\GeV are used. For the ADD model,  the most sensitive part of the invariant mass spectrum, $\mll>1.8\TeV$, is subdivided into 400\GeV-wide search regions, with the final region covering all events above 3\TeV.  For the 2016 data, the lowest bin ranges from 1900 to 2200\GeV to ensure that the reconstructed mass distribution is not affected by the lower cutoff on the generated mass of 1700\GeV in the signal samples for that year. 

In each bin the background estimate as described in Section~\ref{sec:backgrounds} and the signal prediction obtained from the samples described in Section~\ref{sec:samples} are compared. For both signal and backgrounds, all systematic uncertainties described in Section~\ref{sec:systematics} are considered as nuisance parameters modeled with log-normal distributions, taking into account correlations between the different categories where applicable. 

The limits for ADD models are obtained using the signal samples produced in the GRW convention. The theoretical description of the model is only valid up to the ultraviolet cutoff parameter $\Lambda_{\mathrm{T}}$. This is not taken into account in the signal simulation with \PYTHIA. However, the signal cross section above the cutoff is small in the $\Lambda_{\mathrm{T}}$ range where limits are set, and has negligible impact on the results of this analysis. 

\begin{figure}[h]
  \centering
  \includegraphics[width=0.48\textwidth]{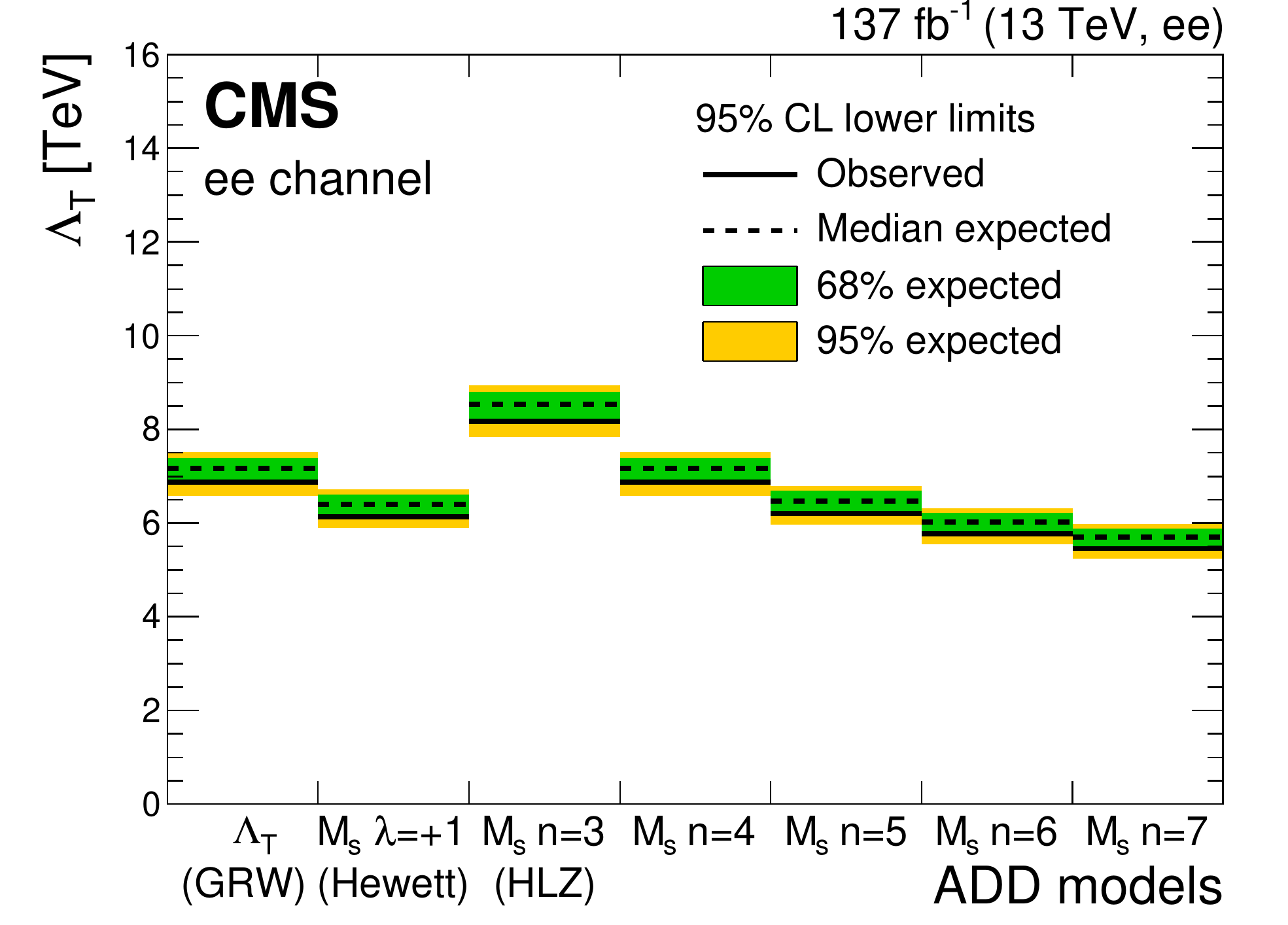}
  \includegraphics[width=0.48\textwidth]{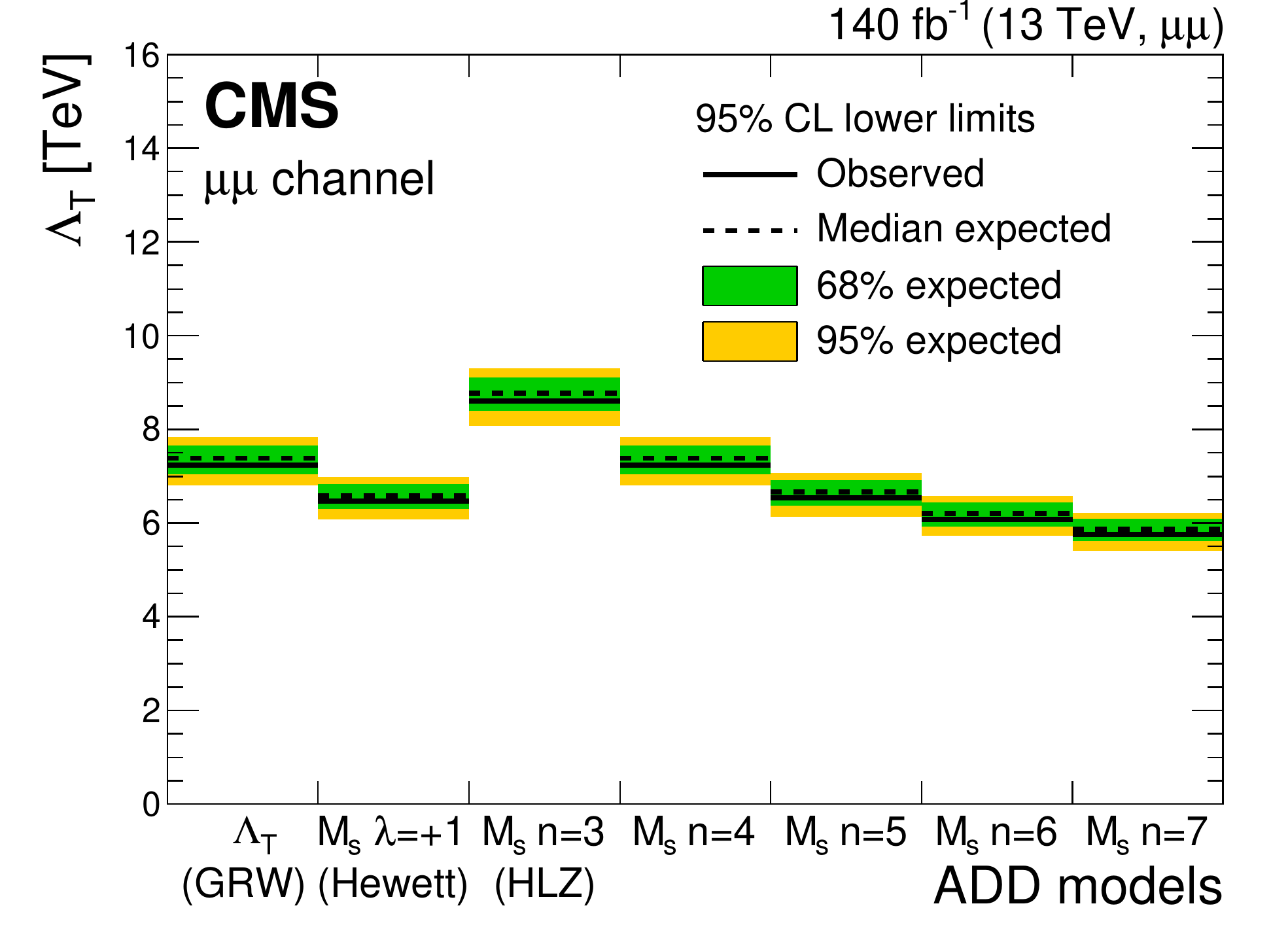}
  \par\smallskip
  \includegraphics[width=0.48\textwidth]{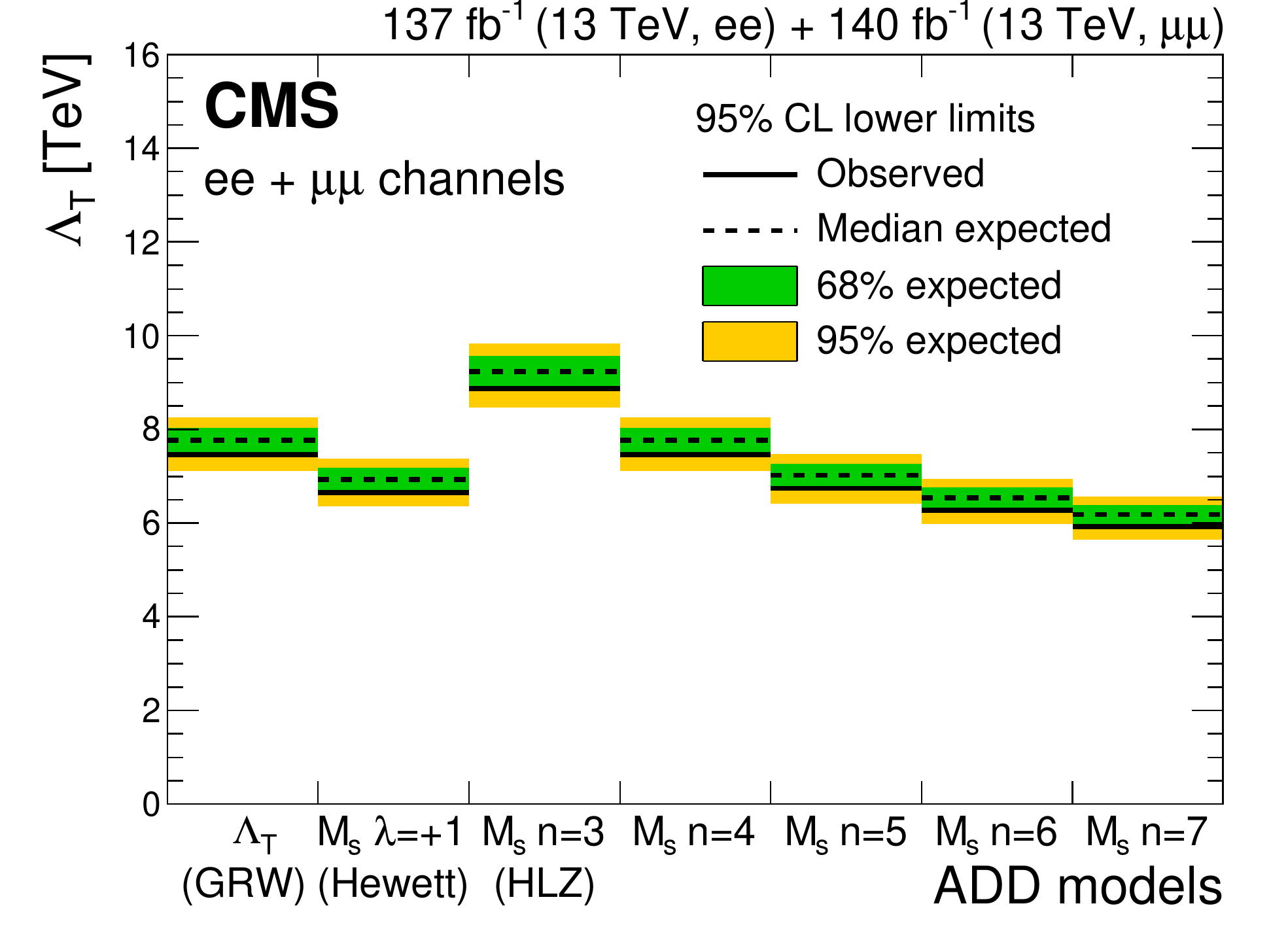}
  \caption{
    Exclusion limits at 95\% \CL on the ultraviolet cutoff for (upper left) the dielectron channel, (upper right) the dimuon channel, and (lower) their combination,  with $\mll>1.8\TeV$ ($\mll>1.9\TeV$ for 2016) in the GRW (first bin), Hewett (second bin), and HLZ conventions (third to seventh bin) for the ADD model.
    Signal model cross sections are calculated up to LO, and an NNLO correction factor of 1.3 is applied.
  }
  \label{fig:led-limit-channels}
\end{figure}

In Fig.~\ref{fig:led-limit-channels}, the limits on the model parameters in the different ADD conventions are shown in the dielectron, dimuon, and combined channels. In the GRW convention, taking into account the NNLO correction factor of 1.3, the observed (expected) limits are 6.9 (7.2)\TeV in the dielectron channel, 7.2 (7.4)\TeV in the dimuon channel, and 7.5 (7.8)\TeV for the combination. The limits are also translated into the Hewett convention and the HLZ convention for different numbers of extra dimensions using Eqs.~(\ref{eq:hew}) and~(\ref{eq:hlz}). All of the limits are shown in Table~\ref{tab:led-limit-summary}.  The case of $n=2$ in the HLZ convention is not considered as there are very strong bounds on $M_{\mathrm{S}}$ from gravitational and astrophysical experiments far in excess of the reach of this analysis~\cite{Torsion,Hannestad:2003yd}. For the considered models, the limits range from 5.9 to 8.9\TeV, depending on the model. These results are the best to date and improve on the previous most stringent limits by 0.5--1.0\TeV~\cite{Sirunyan:2018ipj}. 

\begin{table}[htb]
  \centering
  \topcaption{
    Exclusion limits at 95\% \CL for the electron and muon channels, and their combination, for various parameter conventions of the ADD model.
    Signal model cross sections are calculated up to LO and the NNLO correction factor of 1.3 is applied.
    For each of the model parameters, the observed limit is shown first, followed by the expected limit in parentheses.}

    \begin{tabular}{lccccccc}
        \hline 
      & GRW & Hewett & \multicolumn{5}{c}{HLZ} \\
      \multicolumn{1}{c}{Order} & $\lambdaCIADD{T}$ [{\TeVns}] & $\massADD{S}$ [{\TeVns}] & \multicolumn{5}{c}{$\massADD{S}$ [{\TeVns}]} \\
      &  & $\lambda = +1$ & $n = 3$ & $n = 4$ & $n = 5$ & $n = 6$ & $n = 7$ \\
      \hline
      \multicolumn{8}{c}{{$\Pe\Pe$}} \\
         LO & 6.7 (6.9) & 5.9 (6.2) & 7.9 (8.2) & 6.7 (6.9) & 6.0 (6.3) & 5.6 (5.8) & 5.3 (5.5) \\
         LO $\times 1.3$ & 6.9 (7.2) & 6.1 (6.4) & 8.2 (8.5) & 6.9 (7.2) & 6.2 (6.5) & 5.8 (6.0) & 5.5 (5.7) \\   [\cmsTabSkip]
      \multicolumn{8}{c}{{$\Pgm\Pgm$}} \\
         LO & 7.0 (7.1) & 6.2 (6.4) & 8.3 (8.5) & 7.0 (7.1) & 6.3 (6.4) & 5.9 (6.0) & 5.6 (5.7) \\
         LO $\times 1.3$ & 7.2 (7.4) & 6.5 (6.6) & 8.6 (8.8) & 7.2 (7.4) & 6.5 (6.7) & 6.1 (6.2) & 5.8 (5.9) \\   [\cmsTabSkip]
      \multicolumn{8}{c}{Combined {$\Pe\Pe$} and {$\Pgm\Pgm$}} \\
         LO & 7.3 (7.5) & 6.5 (6.7) & 8.6 (8.9) & 7.3 (7.5) & 6.6 (6.8) & 6.1 (6.3) & 5.8 (6.0) \\
         LO $\times 1.3$ & 7.5 (7.8) & 6.7 (6.9) & 8.9 (9.2) & 7.5 (7.8) & 6.7 (7.0) & 6.3 (6.5) & 5.9 (6.2) \\     [\cmsTabSkip]   
      \hline
    \end{tabular}

  \label{tab:led-limit-summary}
\end{table} 

For the CI model, there can be significant negative interference between the signal and the SM DY process. This can lead to a negative effective signal prediction in some of the mass bins or even an overall negative signal contribution for high values of $\Lambda$. The interference term is therefore taken into account explicitly in the limit calculation. For the combination of the electron and muon channels, a universal CI is assumed.

\begin{figure}[h]
  \centering
  \includegraphics[width=0.48\textwidth]{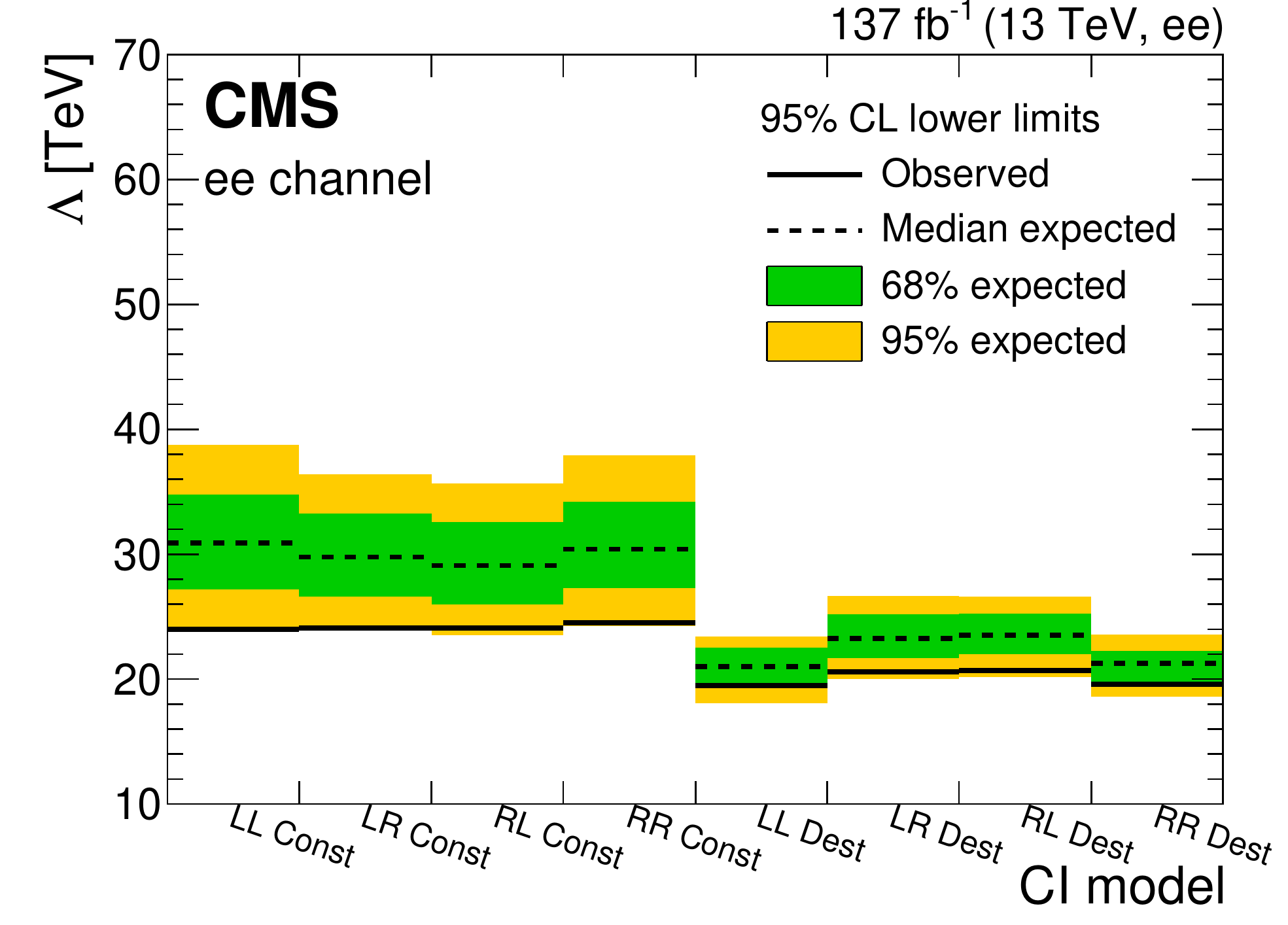}
  \includegraphics[width=0.48\textwidth]{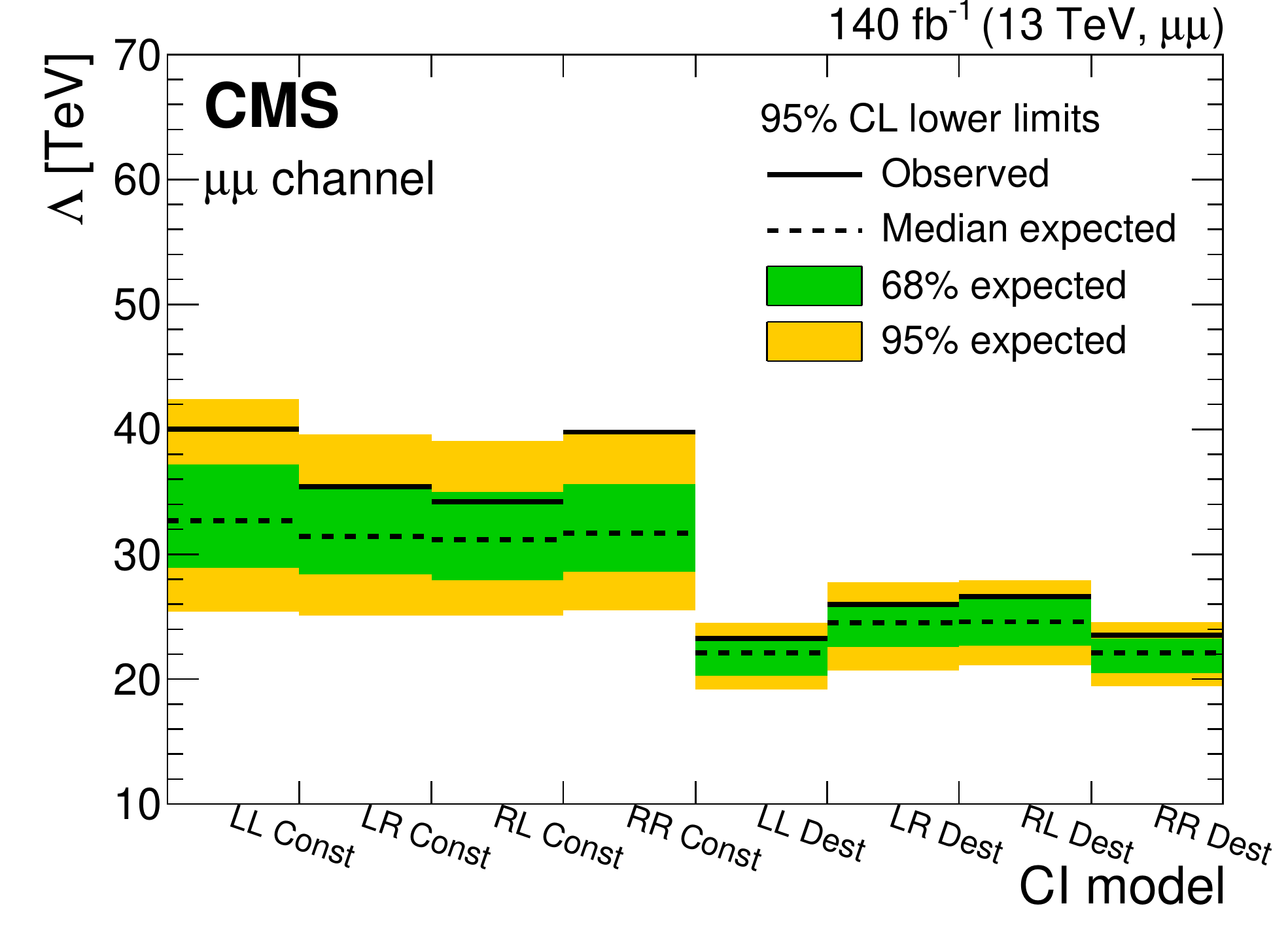}
  \par\smallskip
  \includegraphics[width=0.48\textwidth]{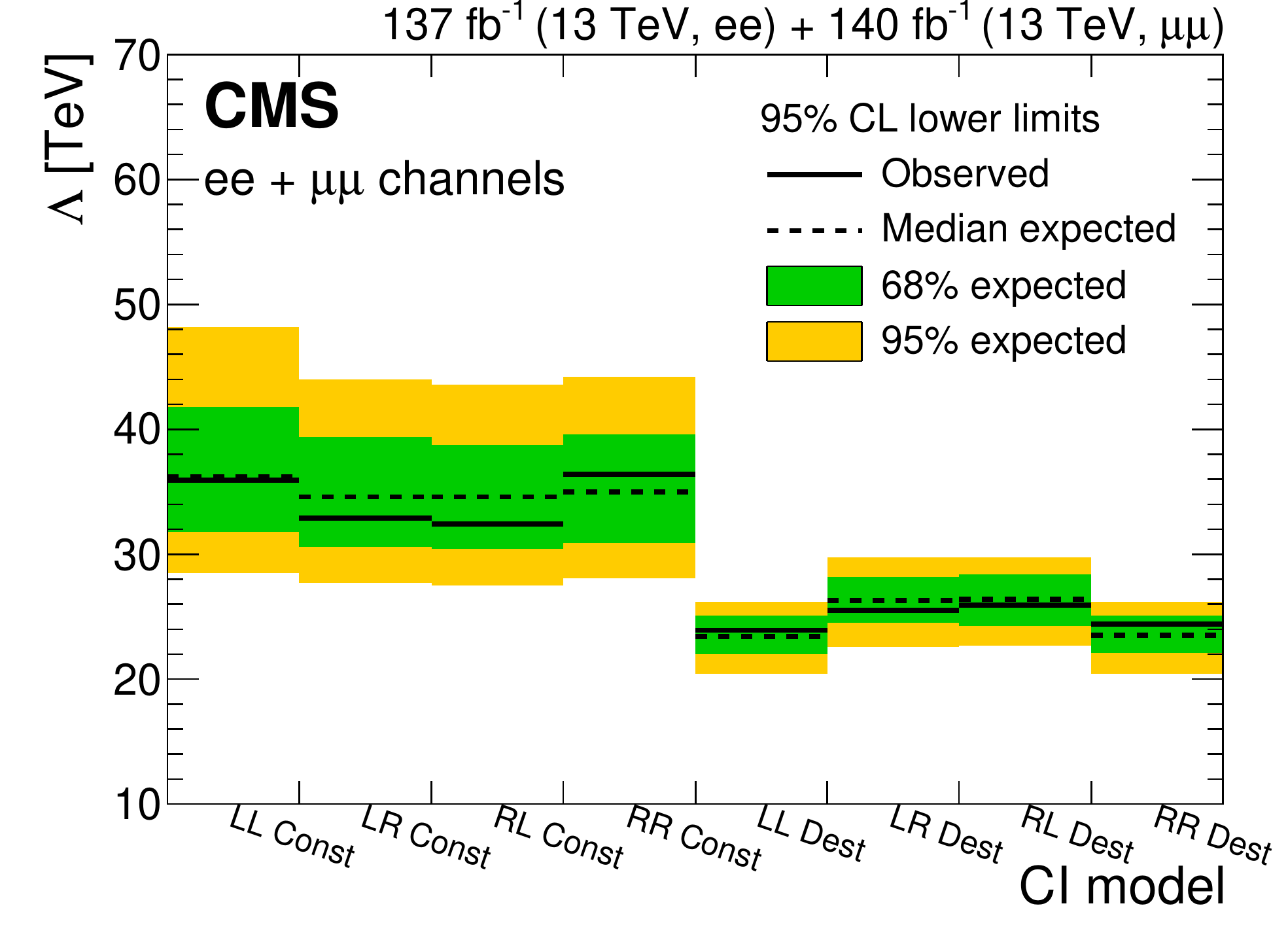}
  \caption{
    Dilepton lower exclusion limits at 95\% \CL on the CI scale ($\Lambda$) for the eight CI models considered, for (upper left) the dielectron channel, (upper right) the dimuon channel, and (lower) their combination. The limits are obtained for $\mll>400\GeV$.
  }
  \label{fig:ci-limits}
\end{figure}

The resulting limits for the dielectron, dimuon, and combined channels are shown in Fig.~\ref{fig:ci-limits}. The limits are given separately for the eight distinct CI models. In the electron channel, the observed limit is weaker than the expected limit by up to two $\sigma$. This is caused by the excess of electron events discussed in Section 8. The effect is more pronounced here than in the limits in the ADD model as the shape of the mass distribution for the CI model makes it more sensitive to the mass range in which the excess is present. The lower limit on $\Lambda$ ranges from 23.9 to 36.4\TeV, depending on the model, an improvement over previous CMS results by 3.5--4.5\TeV~\cite{Sirunyan:2018ipj}. Using the same framework of contact interactions, ATLAS has set limits of up to 35.8 TeV~\cite{Aad:2020otl}. Signal yields in the LR and RL models are reduced (enhanced) compared to LL and RR in the constructive (destructive) case.  However, due to the improved signal-to-background ratio in the negative \costheta bin for these models, the reduced sensitivity is mostly recovered in the constructive case, while in the destructive case the increased sensitivity compared to LL and RR is increased even further. 
The splitting of the event sample into two \costheta ranges improves the limits by $\approx$1\TeV in the case of destructive interference and $\approx$3\TeV in the case of constructive interference.

\subsection{Search for lepton flavor universality violation}
To search for signs of lepton flavor universality violation at high dilepton mass, we consider the ratio \Rmumuee of the differential dilepton production cross sections in the muon and electron channels. The reconstructed distributions are distorted compared to particle level by bin-to-bin migration caused by mass scale and resolution effects, and by the detector acceptance and lepton efficiencies. 

The mass distributions in data are unfolded after subtracting all backgrounds except for DY. For this purpose, the bin-to-bin migration is quantified as response matrices, i.e. 2D-histograms of the generated versus reconstructed dilepton mass in an event. They are obtained from the DY simulation for events passing the full analysis selection. The response matrices are then inverted and applied to the reconstructed mass spectra to unfold them to particle level using the
 \textsc{RooUnfold}~\cite{Prosper:1306523} utility. Since detailed studies found the off-diagonal elements of the response matrices to be small, no regularization is necessary. No significant dependence of the unfolded result on the characteristics of the DY simulation was observed. In the dimuon channel, the size of the unfolding correction rises from around 2\% at low mass to around 10\% at 3 TeV, caused by the worsening \mll resolution. The size of the unfolding corrections is 3-5\% in the dielectron channel, where the resolution is largely independent of \mll.

As the interest is in departures from the SM at high mass, we assume, as in the model of Ref.~\cite{Greljo:2017vvb}, that departures are negligible in the mass region 200--400\GeV.  Thus, to 
correct for the differences in acceptance and efficiency between the dielectron and dimuon channels in this mass region, \Rmumuee is normalized to unity here. To account for the remaining mass dependence of the difference between the channels, \Rmumuee is measured in simulated DY events after application of the full event selection, the unfolding procedure, and the normalization of the ratio at low mass. The inverse of this ratio is then used to correct \Rmumuee in data, which is obtained from the unfolded mass distributions. The size of this correction is up to 5\% for events with central leptons and as large as 20\% for events with forward leptons.

Several uncertainties, with the most significant originating from the PDFs, cancel in the flavor ratio. Uncertainties in the modeling of the mass resolution and scale are propagated through the unfolding procedure by obtaining response matrices for which the reconstructed mass values are smeared or shifted by the uncertainty. The uncertainties in the unfolded mass spectrum are then calculated as the differences between the mass spectra obtained with the different response matrices. The uncertainties in the detector acceptance and lepton efficiencies are propagated to the final flavor ratio via the acceptance and efficiency corrections.

\begin{figure}[h]
  \centering
  \includegraphics[width=0.48\textwidth]{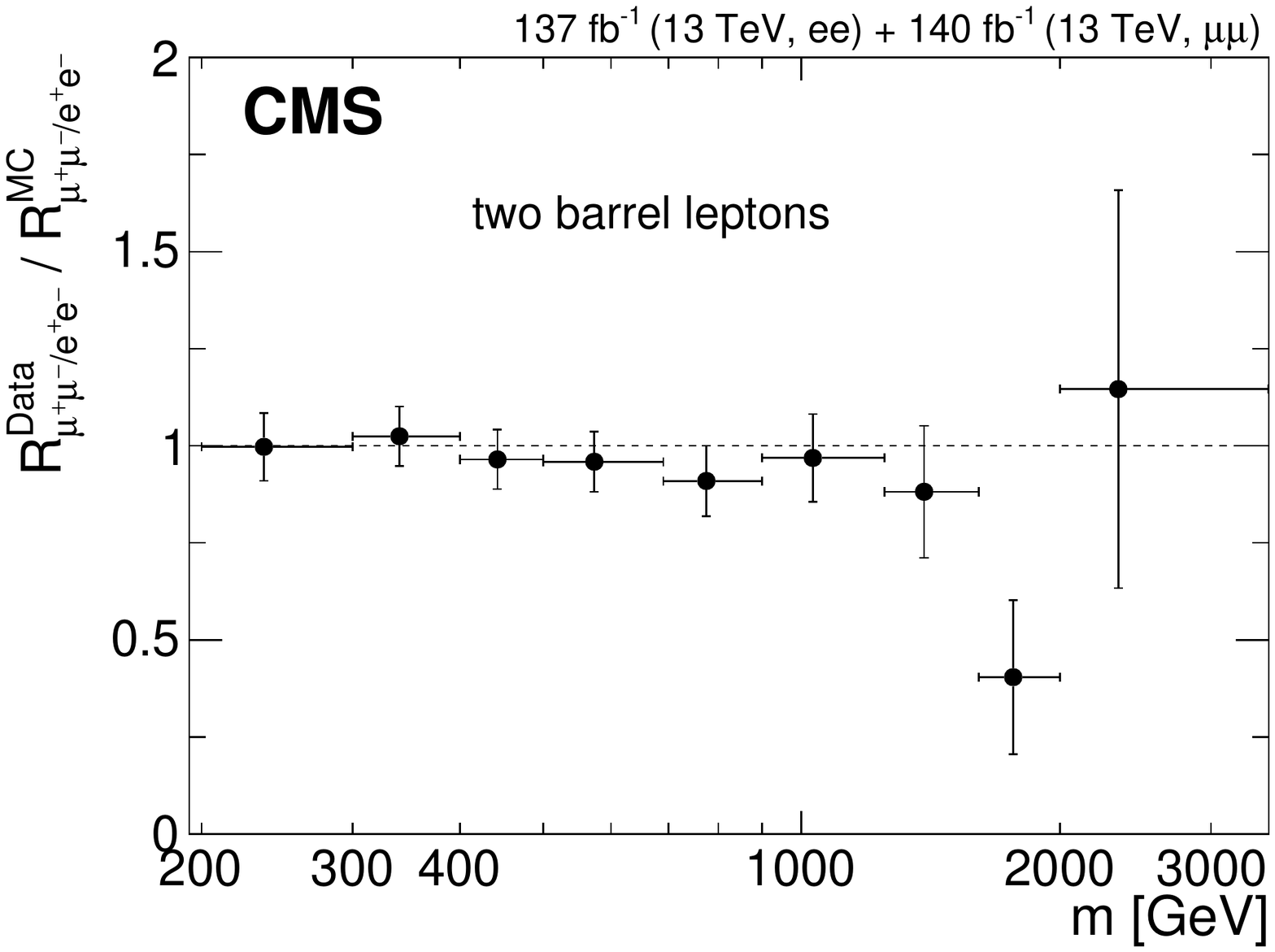}
  \includegraphics[width=0.48\textwidth]{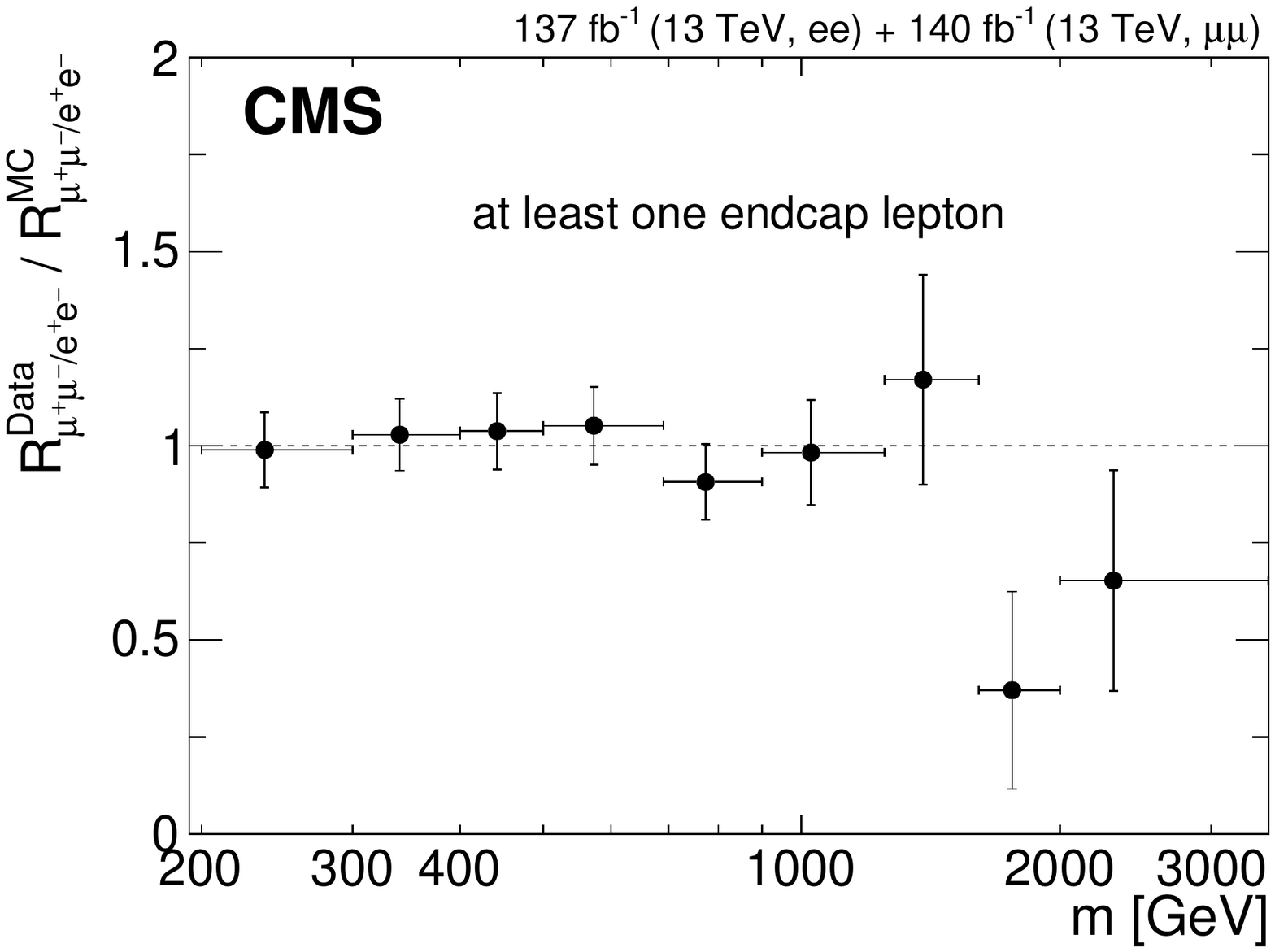}\\
    \includegraphics[width=0.48\textwidth]{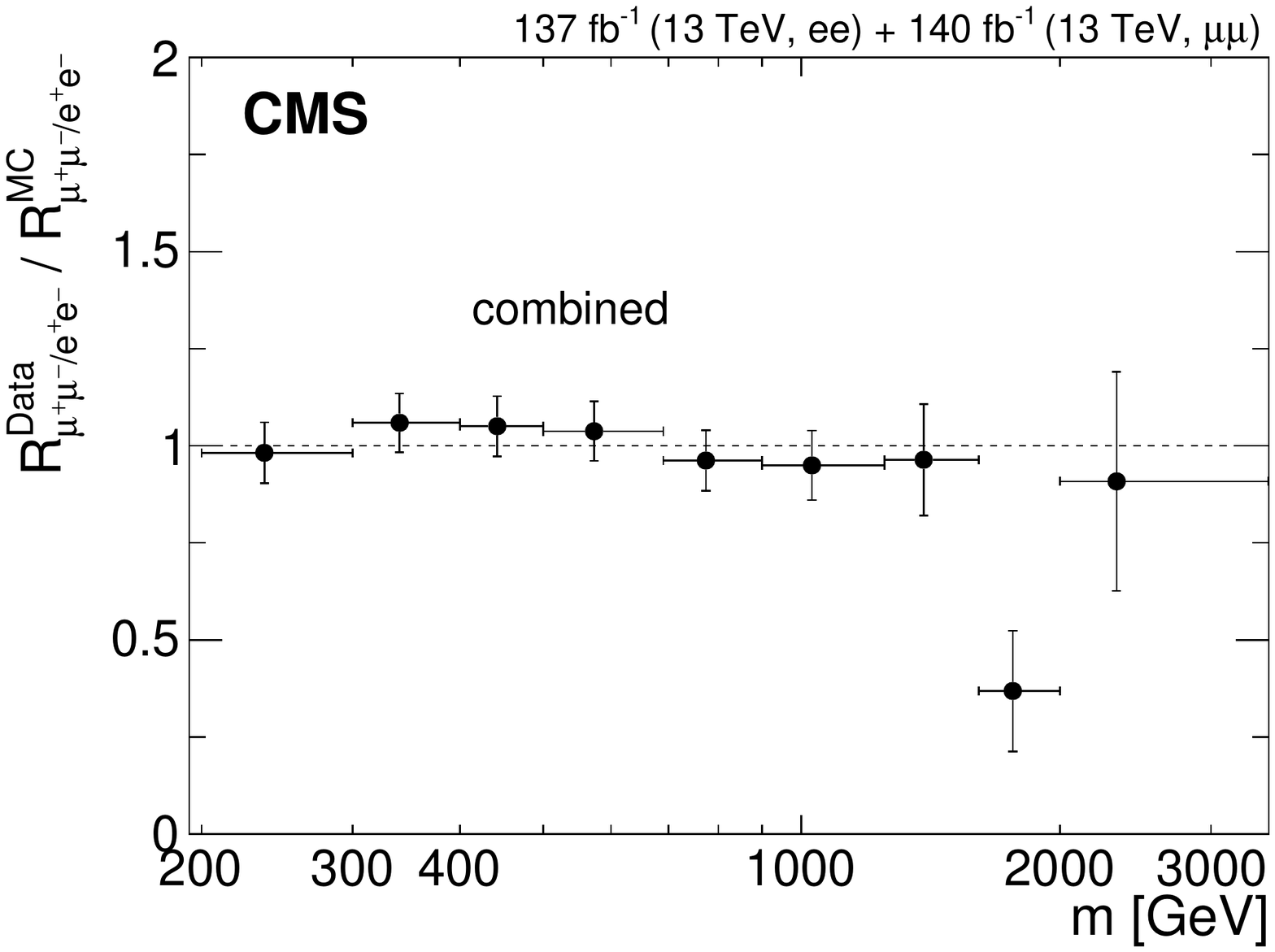}
  \caption{
Ratio of the differential dilepton production cross section in the dimuon and dielectron channels \Rmumuee, as a function of \mll for (upper left) events with two barrel leptons, (upper right) at least one lepton in the endcaps, and (lower) their combination. The ratio is obtained after correcting the reconstructed mass spectra to particle level. The error bars include both statistical and systematic uncertainties.}
  \label{fig:flavorratio}
\end{figure}

The resulting ratio \Rmumuee as a function of dilepton mass is shown in Fig.~\ref{fig:flavorratio}, for events with both leptons in the barrel region, events with at least one lepton in the endcaps, and their combination. Lepton flavor universality implies that this ratio is unity. Good agreement with this expectation is observed up to 1.5\TeV. At very high masses, the statistical uncertainties are large. Here, some deviations from unity are observed, caused by the slight excess in the dielectron channel discussed above. A $\chi^2$ test for the mass range above 400\GeV is performed. The resulting $\chi^2$/dof values are 11.2/7 for the events with two barrel leptons, 9.4/7 for those with at least one lepton in the endcaps, and 17.9/7 for the combined distribution. These correspond to one-sided $p$-values of 0.130 and 0.225, and 0.012, respectively.

As the flavor ratio unfolded to the particle level has been measured, these results can serve as a basis to test models that predict deviations from lepton flavor universality.
 
\section{Summary}
A search for resonant and nonresonant new phenomena in the dilepton invariant mass spectrum in proton-proton collisions at $\sqrt{s} = 13\TeV$ corresponding to an integrated luminosity of up to 140\fbinv has been presented. High-mass dielectron and dimuon events were reconstructed and selected with algorithms optimized for electrons and muons with high transverse momenta. Standard model (SM) backgrounds were primarily estimated from simulation, with the dominant Drell--Yan background corrected to the highest order calculations available, including the contribution from photon-induced processes. When searching for resonant signals, the background normalizations were obtained from sidebands in the data, while for non-resonant signals, the background was normalized to the data in a control region around the \PZ boson peak. No significant deviation from SM expectation is observed. 

Upper limits are set on the ratio of the product of the production cross section and the branching fraction in a dilepton channel of a new resonance with an intrinsic width of up to 10\% to that of the SM \PZ boson at 95\% confidence level. The limits are interpreted in the context of a sequential SM (SSM) and a superstring-inspired model that predict spin-1 resonances. Lower mass limits of 5.15 (4.56)\TeV are set in the \ZPSSM (\ZPPSI) models. The observed limit on narrow spin-1 resonances is translated into limits on generalized couplings of the \PZpr to up and down quarks in several classes of new physics models. For spin-2 graviton resonances in the Randall--Sundrum model of extra dimensions, lower limits on the graviton mass of 2.47--4.78\TeV are set for values of the coupling parameter $k/\overline{M}_{\mathrm{Pl}}$ between 0.01 and 0.1. The lower mass limits for spin-1 and spin-2 resonances are the most stringent to date.

For spin-1 resonances that act as a mediator between SM particles and dark matter (DM), exclusion limits are set in the mass plane of the mediator and DM particles. For large values of $m_{\mathrm{DM}}$, mediator masses below 1.92 (4.64)\TeV are excluded in a model where the mediator is a vector (axial vector) with small (large) coupling to leptons. For $m_{\mathrm{DM}} = 0$, these limits are reduced to 1.04 and 3.41\TeV, respectively. 

Two models of nonresonant signatures have been considered. In case of a four-fermion contact interaction, lower limits on the ultraviolet cutoff parameter $\Lambda$ range from 23.8 to 36.4\TeV depending on the helicity structure of the interaction and the sign of its interference with the SM Drell--Yan background. In the Arkani-Hamed, Dimopoulos, and Dvali model of large extra dimensions, lower limits on the ultraviolet cutoff ranging from 5.9 to 8.9\TeV are set, depending on the parameter convention.

The dimuon and dielectron invariant mass spectra are corrected for the detector effects and, for the first time in this kind of analysis, compared at the TeV scale. No significant deviation from lepton flavor universality is observed. 
\label{sec:summary}

\begin{acknowledgments}
  We congratulate our colleagues in the CERN accelerator departments for the excellent performance of the LHC and thank the technical and administrative staffs at CERN and at other CMS institutes for their contributions to the success of the CMS effort. In addition, we gratefully acknowledge the computing centers and personnel of the Worldwide LHC Computing Grid and other centers for delivering so effectively the computing infrastructure essential to our analyses. Finally, we acknowledge the enduring support for the construction and operation of the LHC, the CMS detector, and the supporting computing infrastructure provided by the following funding agencies: BMBWF and FWF (Austria); FNRS and FWO (Belgium); CNPq, CAPES, FAPERJ, FAPERGS, and FAPESP (Brazil); MES (Bulgaria); CERN; CAS, MoST, and NSFC (China); COLCIENCIAS (Colombia); MSES and CSF (Croatia); RIF (Cyprus); SENESCYT (Ecuador); MoER, ERC PUT and ERDF (Estonia); Academy of Finland, MEC, and HIP (Finland); CEA and CNRS/IN2P3 (France); BMBF, DFG, and HGF (Germany); GSRT (Greece); NKFIA (Hungary); DAE and DST (India); IPM (Iran); SFI (Ireland); INFN (Italy); MSIP and NRF (Republic of Korea); MES (Latvia); LAS (Lithuania); MOE and UM (Malaysia); BUAP, CINVESTAV, CONACYT, LNS, SEP, and UASLP-FAI (Mexico); MOS (Montenegro); MBIE (New Zealand); PAEC (Pakistan); MSHE and NSC (Poland); FCT (Portugal); JINR (Dubna); MON, RosAtom, RAS, RFBR, and NRC KI (Russia); MESTD (Serbia); SEIDI, CPAN, PCTI, and FEDER (Spain); MOSTR (Sri Lanka); Swiss Funding Agencies (Switzerland); MST (Taipei); ThEPCenter, IPST, STAR, and NSTDA (Thailand); TUBITAK and TAEK (Turkey); NASU (Ukraine); STFC (United Kingdom); DOE and NSF (USA).
  
  \hyphenation{Rachada-pisek} Individuals have received support from the Marie-Curie program and the European Research Council and Horizon 2020 Grant, contract Nos.\ 675440, 724704, 752730, and 765710 (European Union); the Leventis Foundation; the Alfred P.\ Sloan Foundation; the Alexander von Humboldt Foundation; the Belgian Federal Science Policy Office; the Fonds pour la Formation \`a la Recherche dans l'Industrie et dans l'Agriculture (FRIA-Belgium); the Agentschap voor Innovatie door Wetenschap en Technologie (IWT-Belgium); the F.R.S.-FNRS and FWO (Belgium) under the ``Excellence of Science -- EOS" -- be.h project n.\ 30820817; the Beijing Municipal Science \& Technology Commission, No. Z191100007219010; the Ministry of Education, Youth and Sports (MEYS) of the Czech Republic; the Deutsche Forschungsgemeinschaft (DFG), under Germany's Excellence Strategy -- EXC 2121 ``Quantum Universe" -- 390833306, and under project number 400140256 - GRK2497; the Lend\"ulet (``Momentum") Program and the J\'anos Bolyai Research Scholarship of the Hungarian Academy of Sciences, the New National Excellence Program \'UNKP, the NKFIA research grants 123842, 123959, 124845, 124850, 125105, 128713, 128786, and 129058 (Hungary); the Council of Science and Industrial Research, India; the Ministry of Science and Higher Education and the National Science Center, contracts Opus 2014/15/B/ST2/03998 and 2015/19/B/ST2/02861 (Poland); the National Priorities Research Program by Qatar National Research Fund; the Ministry of Science and Higher Education, project no. 0723-2020-0041 (Russia); the Programa Estatal de Fomento de la Investigaci{\'o}n Cient{\'i}fica y T{\'e}cnica de Excelencia Mar\'{\i}a de Maeztu, grant MDM-2015-0509 and the Programa Severo Ochoa del Principado de Asturias; the Thalis and Aristeia programs cofinanced by EU-ESF and the Greek NSRF; the Rachadapisek Sompot Fund for Postdoctoral Fellowship, Chulalongkorn University and the Chulalongkorn Academic into Its 2nd Century Project Advancement Project (Thailand); the Kavli Foundation; the Nvidia Corporation; the SuperMicro Corporation; the Welch Foundation, contract C-1845; and the Weston Havens Foundation (USA).\end{acknowledgments}

\bibliography{auto_generated}

\cleardoublepage \appendix\section{The CMS Collaboration \label{app:collab}}\begin{sloppypar}\hyphenpenalty=5000\widowpenalty=500\clubpenalty=5000\input{EXO-19-019-authorlist.tex}\end{sloppypar}
\end{document}

%% file: EXO-19-019-authorlist.tex
\vskip\cmsinstskip
\textbf{Yerevan Physics Institute, Yerevan, Armenia}\\*[0pt]
A.M.~Sirunyan$^{\textrm{\dag}}$, A.~Tumasyan
\vskip\cmsinstskip
\textbf{Institut f\"{u}r Hochenergiephysik, Wien, Austria}\\*[0pt]
W.~Adam, J.W.~Andrejkovic, T.~Bergauer, S.~Chatterjee, M.~Dragicevic, A.~Escalante~Del~Valle, R.~Fr\"{u}hwirth\cmsAuthorMark{1}, M.~Jeitler\cmsAuthorMark{1}, N.~Krammer, L.~Lechner, D.~Liko, I.~Mikulec, F.M.~Pitters, J.~Schieck\cmsAuthorMark{1}, R.~Sch\"{o}fbeck, M.~Spanring, S.~Templ, W.~Waltenberger, C.-E.~Wulz\cmsAuthorMark{1}
% ...
\vskip\cmsinstskip
\dag: Deceased\\
1:  Also at Vienna University of Technology, Vienna, Austria\\
% ...
